\documentclass[aip,reprint,twocolumn,floatfix]{revtex4-1}
\usepackage[english]{babel}
\usepackage{natbib,hyperref}
\usepackage{soul}
\usepackage{tabularx}
\usepackage{amsmath}
\usepackage{amssymb}
\usepackage[T1]{fontenc}
\usepackage[utf8]{inputenc}

\usepackage{xcolor}
\usepackage{graphicx}

\usepackage{xspace}
\newcommand{\RRef}[1]{Ref.~\cite{#1}}

\definecolor{capri}{rgb}{0.0, 0.75, 1.0}
\definecolor{aqua}{rgb}{0.0, 1.0, 1.0}

\graphicspath{{./figs/}}

\usepackage{titlesec}
\definecolor{darkblue}{rgb}{0.0, 0.0, 0.5}  
\titleformat{\section}
  {\color{darkblue}\normalfont\large\bfseries}
  {\thesection}{1em}{}
\titleformat{\subsection}
  {\color{darkblue}\normalfont\normalsize\bfseries}
  {\thesubsection}{1em}{}
\titleformat{\subsubsection}
  {\color{darkblue}\normalfont\normalsize\bfseries}
  {\thesubsubsection}{1em}{}
\titleformat{\paragraph}
  {\color{darkblue}\normalfont\normalsize\bfseries}
  {\theparagraph}{1em}{}

\usepackage[acronym]{glossaries}
\usepackage{hyperref}
\makeglossaries
\newacronym{2dpcfs}{2D-PCFs}{two-dimensional pair correlation functions}
\newacronym{acsfs}{ACSFs}{atom-centered symmetry functions}
\newacronym{aevs}{AEVs}{atomic environment vectors}
\newacronym{aipimd}{AI-PIMD}{{\it ab initio} path integral molecular dynamics}
\newacronym{aimd}{AIMD}{{\it ab initio} molecular dynamics}
\newacronym{asec}{ASEC}{average solvent environment configuration}
\newacronym{bpnn}{BPNN}{Behler--Parrinello neural network}
\newacronym{cnnps}{C-NNPs}{committee neural network potentials}
\newacronym{ccsd_t}{CCSD(T)}{coupled cluster singles doubles with perturbative triples}
\newacronym{cent}{CENT}{charge equilibration via neural net technique}
\newacronym{cosmo}{COSMO}{conductor-like screening model}
\newacronym{dft}{DFT}{density functional theory}
\newacronym{dnn}{DNN}{deep neural network}
\newacronym{fes}{FES}{free energy surface}
\newacronym{ff}{FF}{force field}
\newacronym{ffs}{FFs}{force fields}
\newacronym{gcnn}{GCNN}{graph convolutional neural network}
\newacronym{gnn}{GNN}{graph neural network}
\newacronym{gnns}{GNNs}{graph neural networks}
\newacronym{gpr}{GPR}{Gaussian process regression}
\newacronym{hdnnp}{HD-NNP}{high-dimensional neural network potentials}
\newacronym{krr}{KRR}{kernel ridge regression}
\newacronym{lotf}{LOTF}{learning-on-the-fly}
\newacronym{mad}{MAD}{mean absolute deviation}
\newacronym{mbe}{MBE}{many-body expansion}
\newacronym{mctdh}{MCTDH}{multi configuration time-dependent Hartree}
\newacronym{mc}{MC}{Monte Carlo}
\newacronym{md}{MD}{molecular dynamics}
\newacronym{metad}{MetaD}{metadynamics}
\newacronym{mlamd}{MLaMD}{machine learning-accelerated molecular dynamics}
\newacronym{mlp}{MLP}{machine-learned potential}
\newacronym{mlps}{MLPs}{machine-learned potentials}
\newacronym{ml}{ML}{machine learning}
\newacronym{mm-gbsa}{MM-GBSA}{molecular mechanics-generalized Born surface area}
\newacronym{mm-pbsa}{MM-PBSA}{molecular mechanics-Poisson Boltzmann surface area}
\newacronym{mm}{MM}{molecular mechanics}
\newacronym{mpnns}{MPNNs}{message-passing neural networks}
\newacronym{nn}{NN}{neural network}
\newacronym{nns}{NNs}{neural networks}
\newacronym{nqes}{NQEs}{nuclear quantum effects}
\newacronym{oniom}{ONIOM}{our own N-layered integrated molecular orbital + molecular mechanics}
\newacronym{orr}{ORR}{oxygen reduction reaction}
\newacronym{pb}{PB}{Poisson--Boltzmann}
\newacronym{pcm}{PCM}{polarizable continuum model}
\newacronym{pes}{PES}{potential energy surface}
\newacronym{pimc}{PIMC}{path integral Monte Carlo}
\newacronym{pimd}{PIMD}{path integral molecular dynamics}
\newacronym{qbc}{QbC}{query-by-committee}
\newacronym{qmmm}{QM/MM}{quantum mechanics/molecular mechanics}
\newacronym{rdf}{RDF}{radial distribution function}
\newacronym{rdfs}{RDFs}{radial distribution functions}
\newacronym{remd}{REMD}{replica exchange molecular dynamics}
\newacronym{rism}{RISM}{reference interaction site model}
\newacronym{rmse}{RMSE}{root mean squared error}
\newacronym{sdfs}{SDFs}{spatial distribution functions}
\newacronym{smd}{SMD}{solvation model based on density}
\newacronym{soap}{SOAP}{smooth overlap of atomic positions}
\newacronym{ts}{TS}{transition state}
\newacronym{vad}{VAD}{vacuum-accessible domain}
\newacronym{wtmtd}{WT-MTD}{well-tempered metadynamics}

\hypersetup{
   linkcolor=black,
   citecolor=black,
 }

\begin{document}

\title{
Machine-Learned Potentials for Solvation Modeling
}
\date{\today}

\author{Roopshree Banchode}
\affiliation{\'{E}cole Centrale School of Engineering, Mahindra University, Hyderabad 500043, India}

\author{Surajit Das}
\affiliation{Tata Institute of Fundamental Research, Hyderabad 500046, India}

\author{Shampa Raghunathan}
\email{shampa.raghunathan@mahindrauniversity.edu.in (Corresponding author)}
\affiliation{\'{E}cole Centrale School of Engineering, Mahindra University, Hyderabad 500043, India}

\author{Raghunathan Ramakrishnan}
\email{ramakrishnan@tifrh.res.in (Corresponding author)}
\affiliation{Tata Institute of Fundamental Research, Hyderabad 500046, India}

\begin{abstract}
Solvent environments play a central role in determining molecular structure, energetics, reactivity, and interfacial phenomena. 
However, modeling solvation from first principles remains difficult due to the complex interplay of interactions and unfavorable computational scaling of first-principles treatment with system size. 
Machine-learned potentials (MLPs) have recently emerged as efficient surrogates for quantum chemistry methods, offering first-principles accuracy at greatly reduced computational cost.
MLPs approximate the underlying potential energy surface, enabling efficient computation of energies and forces in solvated systems. They are also capable of accounting for effects such as hydrogen bonding, long-range polarization, and conformational changes. 
This review surveys the development and application of MLPs in solvation modeling. 
We summarize the theoretical basis of MLP-based energy and force predictions and present a classification of MLPs based on training targets, model types, and design choices related to architectures, descriptors, and training protocols. 
Integration into established solvation workflows is discussed, with case studies spanning small molecules, interfaces, and reactive systems. We conclude by outlining open challenges and future directions toward transferable, robust, and physically grounded MLPs for solvation-aware atomistic modeling.
\end{abstract}


\maketitle

\makeatletter
\def\l@subsubsection#1#2{}
\makeatother
\setcounter{tocdepth}{2} 
\tableofcontents

\section{Introduction\label{sec:introduction}}

Solvation is central to numerous phenomena in chemistry, biology, physics, and materials science. 
This process influences molecular stability, governs reaction mechanisms, and impacts the behavior of systems at interfaces~\cite{hirata2003molecular,mennucci2008continuum,ben2013solvation}. 
Solvent effects are deeply embedded in the mechanisms of many essential processes, ranging from redox processes in electrochemical materials to conformational changes in biomolecules.

Yet, modeling solvation with atomistic resolution remains a considerable challenge, especially when one seeks the accuracy of first-principles electronic structure methods. 
Classical approaches, including continuum solvation models and explicit solvent \gls*{md} simulations, offer valuable insights but often demand compromises between computational cost, interpretability, and accuracy.

Recently, \gls*{ml} based strategies referred as 
machine-learned atomistic potentials (MLAPs), 
machine-learned interatomic potentials (MLIPs), 
machine-learned force fields (MLFF)
or briefly, \gls*{mlps} have emerged as a promising alternative to traditional first-principles modeling. 
Such approaches aim to model the \gls*{pes} of a system with first-principles accuracy, while enabling scalable atomistic simulations.

\begin{figure*}[htpb]
\centering
\includegraphics[width=1.0\linewidth,angle=0]{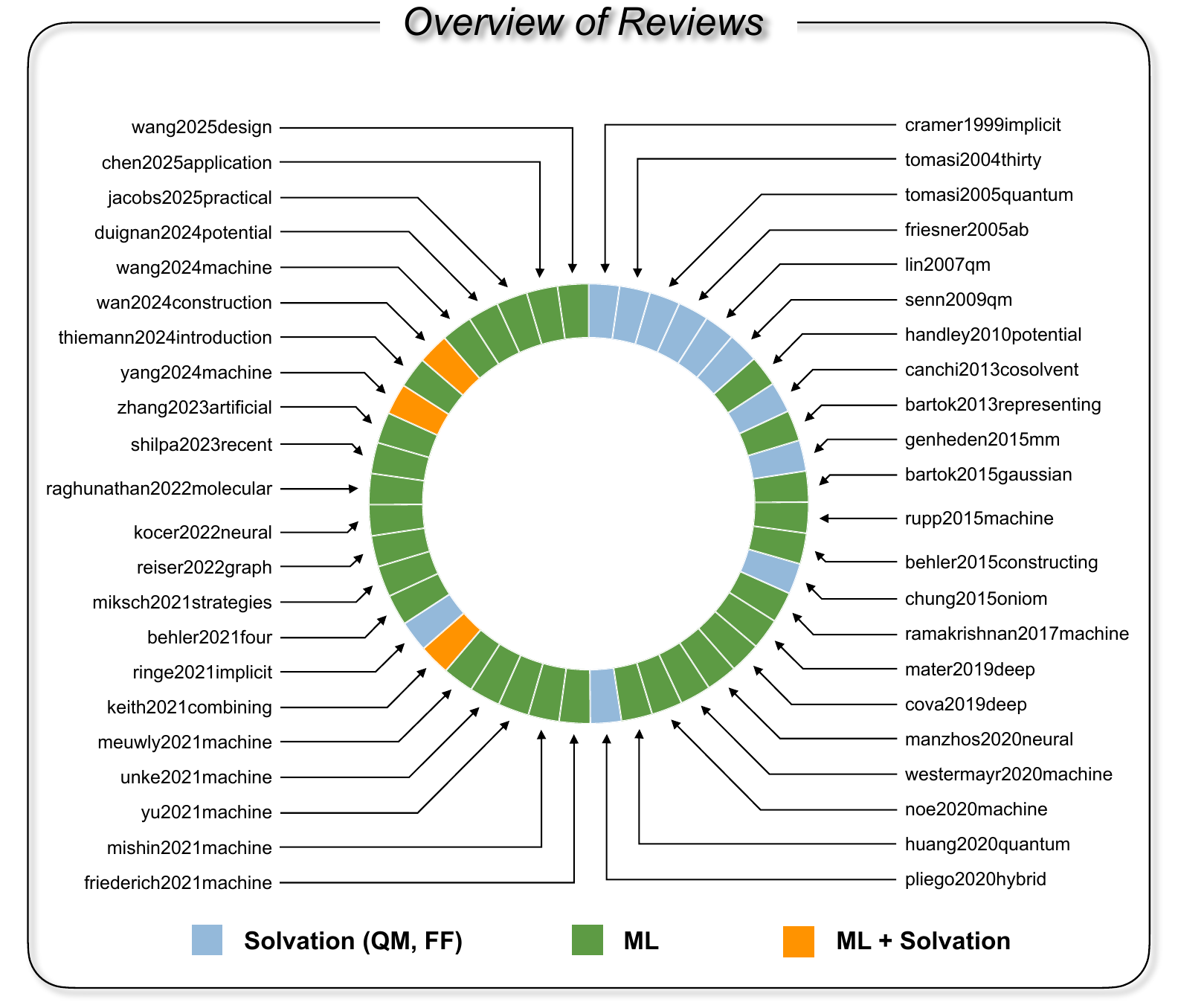}
\caption{
 Distribution of review articles cited in the present work, categorized by thematic focus (labeled using BibTeX-style citation keys, {\it lastnameYYYYfirstword}): traditional solvation modeling based on {\it ab initio} and force-field methods (blue), general ML strategies for molecular and materials modeling (green), and ML reviews with explicit discussions on solvation applications (orange). 
 As of yet, there is no focused review solely on ``MLPs for solvation modeling''. 
 Three reviews~\cite{wan2024construction,yang2024machine,keith2021combining} that include dedicated sections on solvation modeling are highlighted in orange.
 This classification highlights the emerging emphasis on MLPs in the context of solvation modeling.
}
\label{fig:overview}
\end{figure*}

This review is intended for researchers engaged in
\gls*{mlp}-based atomistic modeling, seeking a broader or alternative perspective of the field, focusing on solvation modeling.  
We cover a range of MLP approaches, including neural network–based and kernel-based models, not only in terms of their formal mathematical structure but also in their application to \gls*{pes} modeling and the direct quantitative prediction of numerous solvation-aware properties. 
While these models differ in algorithmic architecture or computer implementation, we emphasize their shared design principles, overlapping terminology, and common challenges. 
Here, we follow the approach of Gilmer et al.~\cite{gilmer2017neural}, 
who elegantly categorized several \gls*{nn} models into the common framework, \gls*{mpnns}. 
Furthermore, we discuss widely used practical strategies for enhancing accuracy without incurring the full cost of high-level quantum calculations. Additionally, we emphasize the growing importance of symmetry principles (such as invariance and equivariance) in \gls*{mlp}-based modeling, surveying these concepts from the perspective of computational chemists. 
Figure~\ref{fig:overview} summarizes the distribution of review articles cited in this work, grouped by thematic focus to highlight the extent of coverage on solvation-specific \gls*{ml} applications.

We provide a comprehensive account of the development of \gls*{mlps}, their deployment in molecular simulations, and their adaptation to the unique challenges of solvation modeling. 
Solvation is a vibrant setting to explore the capabilities and limitations of traditional and \gls*{ml}-based modeling approaches due to the diversity of physical effects involved, ranging from polarization and hydrogen bonding to long-range screening. 
By surveying a wide range of recent works on solvation-related problems, we aim to highlight conceptual and methodological connections that transcend model class and help bridge communities focused on 
structure, dynamics, and reactivities of molecules or materials.

The remainder of this section aims to motivate the importance of solvation (Section \ref{ssec:why_solvation_matters}), review traditional modeling paradigms (Section \ref{ssec:traditional_paradigms}), and introduce the emerging scope of \gls*{mlps} 
for capturing solvation effects (Section \ref{ssec:mlp_scope}).

\subsection{What is solvation and why it matters?\label{ssec:why_solvation_matters}}

Solvation refers to the process by which solvent molecules surround and interact with solute species, forming a composite phase characterized by distinct structural, energetic, and dynamic properties~\cite{canuto2010solvation,yoshida2017molecular}. 
These interactions take many forms, such as electrostatic forces, hydrogen bondings, dispersion effects, solvent-induced polarization, and hydrophobic interactions~\cite{ren2012biomolecular}. 
Depending on the nature of the solute and solvent, solvation can stabilize intermediates, shift transition state energies (i.e., reaction barriers), and alter spectroscopic or thermodynamic properties of molecules. 
A simple example is the dissolution of table salt (NaCl) in water, which is a common yet chemically rich 
process shaped entirely by solvation.

The influence of solvation extends across both natural systems and engineered materials. 
In biological contexts, for instance, water comprises roughly 65--90\% of an organism's mass~\cite{dargaville2022water}. 
Meanwhile, cosolvents like urea and trimethylamine N-oxide (TMAO) play important roles in helping marine life adapt to osmotic stress~\cite{yancey1982living,schroer2011exploring}. 
Solvent dynamics also govern fundamental biochemical events, such as proton transfer and hydrogen exchange reactions~\cite{esaki1985proton,dlugosz2005effects}. 
In synthetic chemistry, deep eutectic solvents are emerging as environmentally friendly alternatives for polymer production, offering low melting points and biodegradable properties that support low-waste or solvent-free synthesis~\cite{weerasinghe2024deep}.

The examples mentioned above raise several foundational questions: 
What are the driving forces behind solvent-mediated molecular processes?
When do solute–solvent interactions dominate, and when are solvent–solvent correlations equally significant as solute-solvent interactions? 
Can such effects be categorized using thermodynamics alone, or do structural motifs also play a decisive role? 
These are not new questions, but represent some of the problems that are being actively addressed 
by using experimental and computational techniques~\cite{ando2021molecular,foumthuim2023solvent,jaworek2021boosting,buntkowsky2018properties,ahanger2024small,devereux2020polarizable,jaganade2020urea,raghunathan2020urea,raghunathan2018role,devereux2014novel}.

 Consider the case of galanin, a 29-residue neuropeptide that adopts different conformations depending on the solvent environment as revealed through NMR spectroscopy and \gls*{md} simulations~\cite{de1992molecular}. 
 The structure of bulk liquid water has been intensively studied using neutron scattering and classical molecular models~\cite{soper1982hydrogen,head2002water}, while the phase-specific behavior of water across gas, liquid, and solid regimes has been elucidated using both kinetic experiments and atomistic models such as TIP4P and TIP5P~\cite{berendsen1987missing,jorgensen1983comparison,mahoney2000five,horn2004development,abascal2005general,vega2005melting,abascal2005potential}.

Even with decades of research and collective knowledge on this topic, a complete understanding of how solvation microstructure affects macroscopic thermodynamics and kinetics is lacking. 
Solvent effects span broad length and time scales and involve chemical effects ranging from local hydrogen bonding patterns to long-range polarization and reorganization. 
Hence, accurate depiction of solvation requires theoretical models that can simultaneously resolve molecular geometry and electronic response. 
Recent progress in high-performance computing, coupled with advances in atomistic modeling techniques, is helping to close this gap. 
While approaches based on \gls*{ml} have opened up new avenues~\cite{meuwly2021machine,zhang2021phase,magduau2023machine,ramakrishnan2017machine,von2020exploring,mater2019deep,keith2021combining,cova2019deep,ceriotti2021introduction,noe2020machine,westermayr2020machine}, the broader challenge remains: developing frameworks that are not only data-driven, but also grounded in physical insight and transferable across diverse solvation environments.

\begin{figure}[ht]
\centering
\includegraphics[width=\linewidth,angle=0]{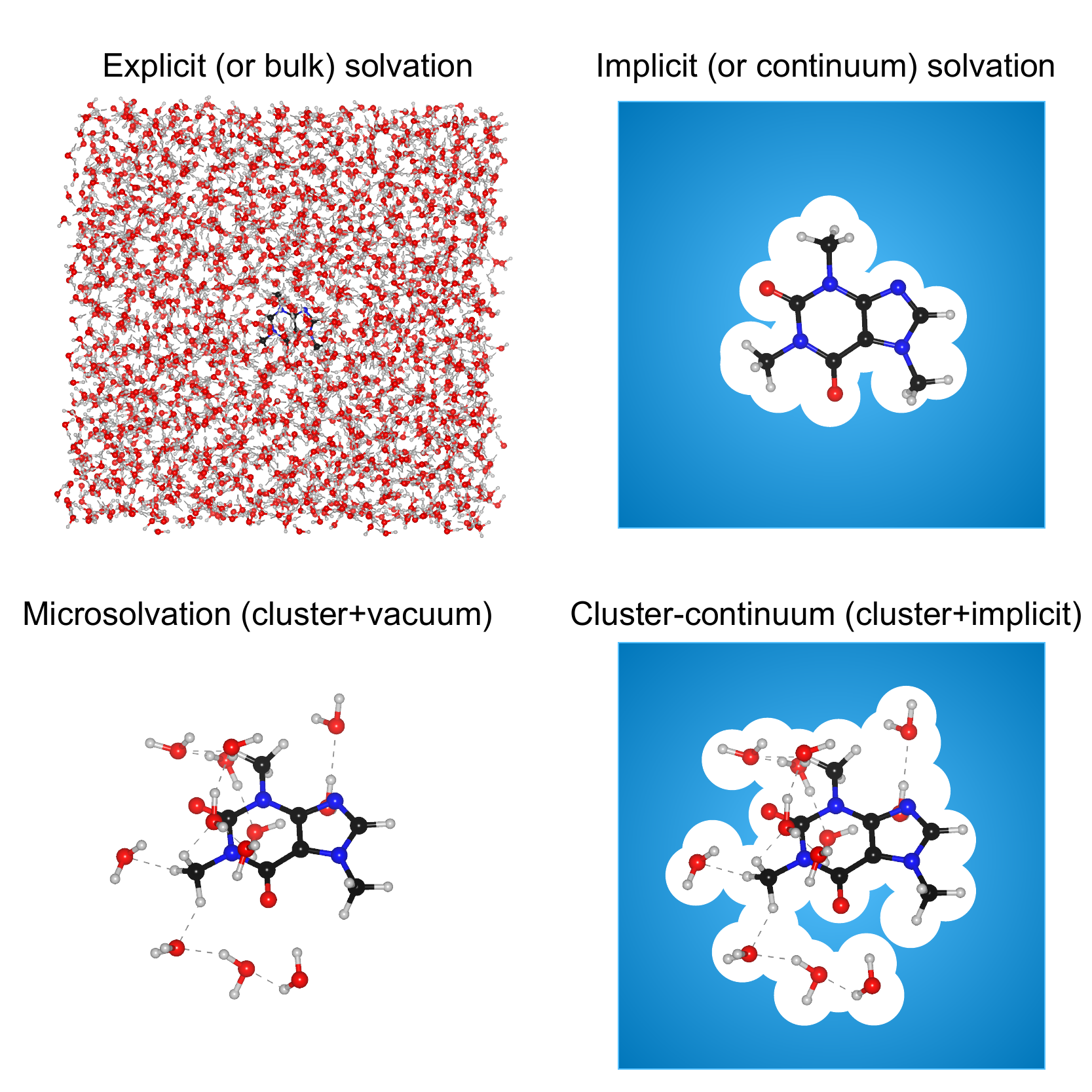}
\caption{
Four levels of solvation modeling are illustrated using caffeine as the example solute:
(1) {\it Explicit solvation}, where the solute is fully immersed in bulk water, typically modeled with periodic boundary conditions; here, a 10~\AA{} cubic water box was used for illustration;
(2) {\it Implicit solvation}, where the solute resides in a molecule-shaped cavity within a polarizable dielectric continuum;
(3) {\it Microsolvation}, a hybrid approach in which the solute is surrounded by a small number of explicit solvent molecules, with vacuum beyond; and
(4) {\it Cluster–continuum}, another hybrid approach where the solute–solvent cluster is embedded within a continuum model, capturing both local explicit interactions and long-range dielectric effects.
}
\label{fig:solvation_paradigm}
\end{figure}

\subsection{Traditional solvation modeling paradigms\label{ssec:traditional_paradigms}}

Solvation modeling, such as free-energy calculations, has long been approached using both quantum mechanical and classical techniques, typically falling into three broad categories: explicit, implicit, and hybrid solvation models~\cite{chipot2007free}, as illustrated in Figure~\ref{fig:solvation_paradigm}. Each of these frameworks offers a different balance between physical realism and computational cost, especially when it comes to accurately capturing intermolecular interactions, which is a long-standing challenge in solvation modeling~\cite{tomasi2005quantum,marenich2009universal,cramer1999implicit,skyner2015review,ringe2021implicit}.

In {\it explicit solvent models}, each solvent molecule is modeled individually, enabling a detailed description of both solute–solvent and solvent–solvent interactions. 
These models are commonly implemented using \gls*{md} or \gls*{mc} simulations to model the bulk solvation environment, where the solute is embedded in a solvent box comprising a few hundred to thousands of solvent molecules~\cite{foumthuim2020can,foumthuim2023solvent}. 
When using first-principles methods, a carefully selected unit cell configuration can be
treated within the periodic \gls*{dft} formalism. While
such explicit treatments offer a high degree of realism, they become computationally intractable with the increase in the number of solvent molecules.
Whereas \gls*{ffs} formalism enables simulations of large solvation boxes, the resulting free energy estimates can be highly sensitive to the selected \gls*{ffs}~\cite{shirts2003extremely}. 
For example, simulations of protein stability in osmolyte solutions, such as urea and TMAO, have shown notable variations depending on the \gls*{ffs} employed in \gls*{md} simulations~\cite{canchi2013cosolvent,ganguly2016hydrophobic}.

By contrast, {\it implicit solvation models} avoid treating individual solvent molecules altogether. Instead, the electrostatic potential provided by the solvent environment is approximated as a continuous polarizable medium~\cite{tomasi2004thirty}, thus significantly reducing the computational effort, facilitating rapid and qualitative predictions of solvation-dependent properties and energies. 
Common frameworks include 
the \gls*{pcm}~\cite{tomasi2005quantum}, 
the \gls*{cosmo}~\cite{klamt1993cosmo}, and universal models~\cite{cramer2008universal}
that can describe diverse chemical species (neutral, charged, polar, nonpolar, organics, inorganics) without requiring system-specific parameter tuning such as the \gls*{smd}~\cite{marenich2009universal}. More physically grounded approaches, such as 
the \gls*{rism}~\cite{hirata1981extended,hirata1987new} and its three-dimensional extension 
(3D-\gls*{rism})~\cite{sindhikara2013analysis,imai2004solvation}, attempt to recover solvent structure through statistical mechanics. 
In molecular simulations, energy-based methods like \gls*{mm-gbsa} and \gls*{mm-pbsa} also fall into the implicit category that can provide fast free energy estimates from snapshots of MD trajectories~\cite{genheden2015mm}. Still, these models often miss essential features such as directional hydrogen bonding, ion pair specificity, or the dynamic rearrangement of solvation shells.

In many applications, where the interest is primarily in the role of solute-solvent
interactions within the first solvation shell, {\it hybrid solvation strategies} provide a tradeoff between computational cost and realistic modeling. 
In {\it microsolvation}, a small number of solvent molecules are placed directly around the 
solute, while the entire solute-solvent cluster is treated at the first-principles level~\cite{basdogan2020advances,pliego2020hybrid}. 
This approach is also known as the cluster-vacuum approach, as the entire cluster is modeled in vacuum, which lacks long-range solute-solvent interactions. 

A popular additivity-based microsolvation scheme is based on the 
\gls*{oniom} approach~\cite{chung2015oniom,dapprich1999new}. In this approach,
the chemically interesting part (or layer) of the system (e.g., the solute) 
is modeled with a highly accurate (albeit expensive) method, such as correlated wavefunction methods, and the surrounding solvent layer is modeled using \gls*{dft}. 
To account for a large cluster size, one can also model the solute layer with \gls*{dft}
and solvent layer with \gls*{mm} (i.e., with a classical \gls*{ff}). 
In the first-generation \gls*{oniom} method~\cite{maseras1995imomm}, the total energy of the cluster is approximated as
\begin{equation}
    E_{\rm ONIOM}^{\rm cluster} \approx  E_{\rm QM}^{\rm solute}
    + \left( E_{\rm MM}^{\rm cluster}  - E_{\rm MM}^{\rm solute} \right) 
    \label{ref:oniom}
\end{equation}
where the solute is treated with QM, and the MM treatment captures the full cluster
minus the solute, hence the solute-solvent interaction is captured at the MM level. 
In Eq.~\ref{ref:oniom}, one can introduce
\begin{equation}
E_{\rm MM}^{\rm cluster} = E_{\rm MM}^{\rm solute} + E_{\rm MM}^{\rm solvent}
+ E_{\rm MM}^{\rm interaction}
    \label{ref:cluster}
\end{equation}
to arrive at
\begin{equation}
    E_{\rm ONIOM}^{\rm cluster} \approx  E_{\rm QM}^{\rm solute}
    + E_{\rm MM}^{\rm solvent}
+ E_{\rm MM}^{\rm interaction} 
    \label{ref:oniom2}
\end{equation}

In the \gls*{qmmm}~\cite{lin2007qm,senn2009qm,friesner2005ab,warshel2006electrostatic} the interaction energy is not estimated with MM but
through mechanical and electrostatic embedding strategies to model the boundary between the two subsystems (i.e., the QM-MM boundary) 
\begin{equation}
    E_{\rm QM/MM}^{\rm cluster} \approx  E_{\rm QM}^{\rm solute}
    + E_{\rm MM}^{\rm solvent}
+ E_{\rm QM-MM}^{\rm interaction} 
    \label{ref:qmmm}
\end{equation}

While the basic \gls*{oniom} approach involves three separate energy evaluations per structure, 
\gls*{qmmm} methods, especially when based on polarizable or self-consistent electrostatic embedding, may require multiple iterative passes even for single-point energy calculations, due to the coupling between QM and MM regions. Overall, both \gls*{oniom} and \gls*{qmmm} schemes allow for the use of a large number of solvent molecules to account for
dynamic, explicit solvent interactions while keeping the total computational cost manageable.

In {\it cluster–continuum models}, a small number of solvent molecules near the solute is treated explicitly either at the {\it ab initio} (e.g., with \gls*{dft}) or \gls*{oniom} or \gls*{qmmm} methods, while the remaining bulk solvent is described as a polarizable dielectric continuum~\cite{pliego2001cluster,tomanik2020solvation,simm2020systematic,pliego2020hybrid} (see Figure~\ref{fig:solvation_paradigm}). Both microsolvation and cluster-continuum methods offer tractable yet chemically informed solvation modeling strategies, combining the accuracy of first-principles methods with the scalability of empirical and continuum models.

\subsection{Scope of MLPs for solvation modeling\label{ssec:mlp_scope}}


Solvents play a central role in shaping chemical reactivity at the molecular level, primarily through electrostatic and non-covalent interactions such as hydrogen bonding, dipole–dipole forces, and van der Waals (vdW) interactions. These effects can selectively stabilize different species along the reaction coordinate, reactants, intermediates, or products, alter activation barriers, and ultimately influence both the rate and yield of reactions.
Recent advances have shown that \gls*{mlps} can capture these solvent-mediated effects with near-quantum accuracy and significantly reduced computational cost, enabling realistic modeling of solvation environments in complex chemical processes~\cite{anmol2024unveiling,zhang2024modelling,gastegger2021machine,chen2023accelerating}.

The structural and energetic landscape of a reactive system can shift dramatically depending on the nature of the solvent. Polarity, in particular, can strongly affect the synchronicity of bond-making and bond-breaking events, the lifetime of intermediates, and the accessibility of alternative reaction pathways. For example, in the case of glycosylation, Vitartas et al.~\cite{vitartas2025active} showed that non-polar solvents favor an S$_\text{N}$2 mechanism, while polar solvents shift the reaction toward an S$_\text{N}$1 pathway involving distinct reactive species. More broadly, polar environments support long-range charge stabilization and solute–solvent coupling, whereas non-polar media typically allow only localized interactions~\cite{celerse2024capturing}. These observations highlight the solvent’s critical role in 
modulating short-range electronic effects and long-range structural organization.

To understand these complex phenomena at the molecular and electronic scale, one must accurately model the underlying \gls*{pes}. 
An accurate framework for simulating solvent effects with quantum mechanical accuracy, capturing features like electronic polarization, dynamic hydrogen bonding, and charge transfer, is \gls*{aimd}~\cite{marx2000ab,marx2009ab,tuckerman1996ab,carloni2002role}. 
However, its computational complexity restricts its application domain to small systems and short timescales. 
\gls*{mlps} offer a practical solution by reproducing the \gls*{pes} with near-\gls*{aimd} accuracy, enabling longer simulations, enhanced sampling of the configuration space, and facilitating the calculation of statistically converged thermodynamic observables. 
The data-driven nature of \gls*{mlps} makes them particularly well-suited for capturing the subtle, many-body interactions that arise during the dynamics of solvated systems.
When combined with active learning, these models can efficiently explore configuration space, focusing training efforts on chemically relevant regions while maintaining fidelity to the electronic structure.

In recent years, \gls*{mlps}, especially those built on deep \gls*{nn}s and active learning frameworks, have gained popularity in computational chemistry~\cite{behler2016perspective,gastegger2021machine,zhang2018potential,mishin2021machine,friederich2021machine,duignan2024potential}. These models offer a scalable way to approximate \gls*{pes}s with near first-principles accuracy at a fraction of the computational cost. Training on energies and forces allows for the construction of \gls*{pes}s that not only accelerate \gls*{md} simulations but also facilitate applications in spectroscopy, thermodynamics, and chemical reaction dynamics~\cite{meuwly2021machine}. \gls*{mlps} are also increasingly replacing classical \gls*{ff}s in the \gls*{mm} region of hybrid QM/MM simulations. 
Beyond non-reactive systems, MLPs have also been developed to capture reactive processes, with models such as ReaxFF enabling the description of bond breaking and formation within complex chemical environments~\cite{yang2024machine}.

\begin{figure*}[htpb]
\centering
\includegraphics[width=1.0\linewidth,angle=0]{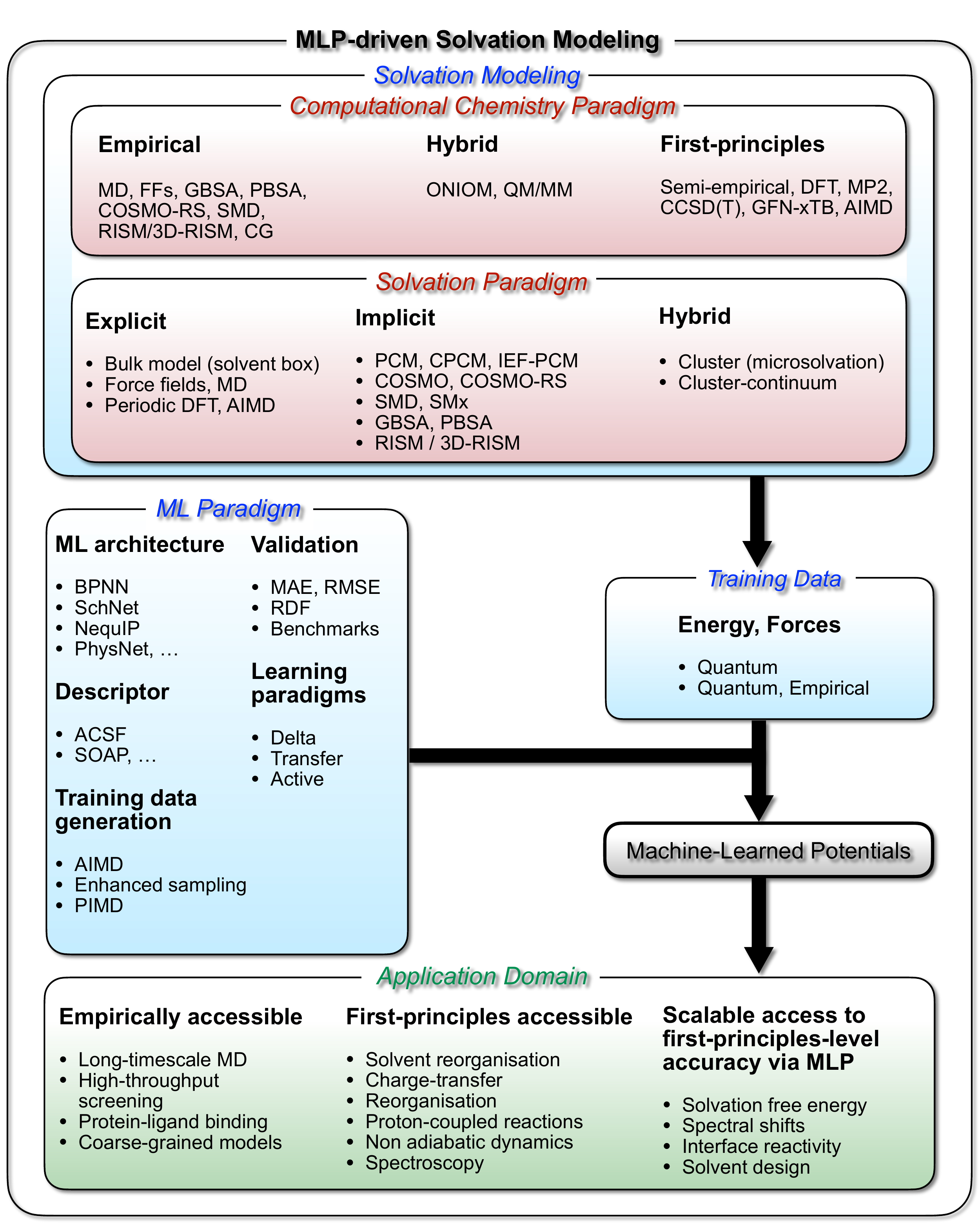}
\caption{Concept map of MLP-driven solvation modeling, as presented in this review.
Solvation modeling is organized around computational and solvation paradigms and interfaces with machine learning to produce rapid, accurate, and transferable data-driven force fields.}
\label{fig:concept_map}
\end{figure*}

At the heart of any \gls*{mlp} lies a descriptor, which is a numerical representation of the atomic environment, 
and a mathematical model that maps atomistic configurations to scale, vector, and/or tensor-valued molecular properties~\cite{thiemann2024introduction}. The performance of a given \gls*{mlp} hinges on the choice of descriptor and modeling strategy, which can vary widely across \gls*{nn} and kernel-based frameworks~\cite{pinheiro2021choosing}. Common approaches include symmetry functions \cite{behler2021machine} and graph-based encodings. Equally important is the training dataset: it must be diverse and representative enough to capture the relevant configurational space. While high-level electronic structure data remains the gold standard, generating such datasets, especially for systems involving transition metals, enzyme active sites, or complex solvent networks, remains computationally intractable.

Recent efforts have focused on standardizing workflows for building and validating these models across different domains, ranging from catalysis and enzymology to materials discovery. By enabling accurate, transferable modeling of both intramolecular interactions and solvent effects, MLPs are opening up new avenues for understanding and predicting solvation-driven reactivity. This review highlights key developments on this topic, particularly emphasizing how MLPs help bridge the gap between electronic structure theory and atomistic solvation modeling.

\vspace{1em}
\noindent
In this review, we aim to provide a broad yet focused account of how MLPs are advancing solvation modeling. 
Section~\ref{sec:theory} lays the theoretical foundation, covering various ingredients of MLPs such as training objectives, energy and force formulations, many-body expansions, and the role of $\Delta$-machine learning in improving accuracy. In Section~\ref{subsec:classifyMLPs}, we present a taxonomy of MLP architectures that span neural network, kernel-based, linear, and gradient-domain models, and examine how their design choices relate to the specific challenges of solvation. 
We also explore how MLPs are incorporated into practical modeling workflows, from training strategies to software deployment.
Section~\ref{sec:catalogue} highlights selected case studies that showcase the use of 
MLPs to construct PES of solvated species, and to directly predict solvent-modulated properties.  
Further, we highlight the essential role of sampling strategies in the pipeline of solvation-aware
MLP modeling. 
Section~\ref{sec:challenge} turns to key open challenges, including limitations in model transferability, gaps in benchmarking practices, and the inherent complexity of real solvation environments. 
Finally, Section~\ref{sec:outlook} offers a forward-looking perspective on how the field might evolve toward unified and transferable MLP frameworks for solvation modeling.

To orient the reader to the broader structure and themes of this review, Figure~\ref{fig:concept_map} presents a concept map summarizing how MLPs interface with solvation modeling. This schematic serves as a roadmap for the sections that follow, outlining the relationships among simulation techniques, solvation strategies, and MLP formalisms discussed throughout the review.

\section{Physical Foundations of MLPs\label{sec:theory}}

MLPs are reshaping the landscape of atomistic simulations by offering a practical way to approximate quantum mechanical PESs with remarkable accuracy and far lower computational cost. Traditional quantum chemistry methods, such as DFT and wavefunction-based approaches, deliver high fidelity but become prohibitively expensive as system size grows, limiting their use in large and/or long-timescale MD simulations. Whereas classical FFs are computationally efficient but often fall short in accuracy and transferability, especially for systems that are chemically diverse or reactive.

MLPs offer a compelling middle ground. By learning the PES directly from {\it ab initio} reference data, they achieve near-DFT accuracy while retaining the speed of empirical models. This combination has made it possible to carry out atomistic MD simulations of complex systems, ranging from solvated biomolecules to condensed-phase materials, at spatial and temporal scales that are prohibitively expensive for conventional modeling.

At their core, MLPs are model trained on {\it ab initio} data on both energies and forces across a wide range of atomic environments. 
In materials modeling, it is also common to include stress tensor contributions in the training loss to better capture elastic and mechanical responses~\cite{ocampo2024adaptive}. This combined-training paradigm mirrors practices in computational spectroscopy, where accurate PES construction typically involves fitting to high-level \textit{ab initio} data, including energies, gradients, and, in some cases, even Hessians~\cite{carbonniere2003least,carbonniere2004construction}.

The rest of this section lays out the theoretical underpinnings of these models, including their training, loss function formulations, the architectural strategies they employ, and establishes their connection to classical PES fitting and many-body expansions.

\subsection{From PES fitting to MLPs}
Traditionally, \gls*{pes}s are constructed using Taylor-like expansions around a reference geometry, typically expressed in chemically meaningful internal coordinates that are functions of interatomic distances, bond angles, and dihedral angles, or in terms of dimensionless normal coordinates~\cite{jasien1988general,schatz2000fitting}. For a $D$-dimensional system, the total potential energy (with the minimum energy set to zero) is written as a multivariate polynomial in the $D$-dimensional displacement vector $\mathbf{x}$
\begin{equation}
V_{\text{Taylor}}(\mathbf{x}) = \sum_{i,j} k_{ij} x_i x_j + \sum_{i,j,k} k_{ijk} x_i x_j x_k + \cdots
\end{equation}
Here, $x_i$ represents the displacement along coordinate $i$, and the coefficients $k_{ij}$, $k_{ijk}$, etc., are fitted to reference {\it ab initio} data. The expansion captures harmonic and anharmonic contributions via quadratic, cubic, quartic, and higher-order terms. 
This expansion can also be viewed schematically as a \emph{sum of products of monomials}
\begin{equation}
V_{\text{Taylor}}(\mathbf{x}) = \sum_{f} c_f \prod_{i=1}^{D} x_i^{p_{if}}
\label{eq:taylorsop}
\end{equation}
where $f$ indexes the terms in the expansion (i.e., the number of force field terms), 
$p_{if}$ is the power (or exponent) of $x_i$ in terms of $f$; and $c_f$ is the corresponding coefficient. 
Typically, each term involves only a small subset of the $D$ coordinates, and the total degree of interaction is bounded: 
\(\sum_{i=1}^{D} p_{if} \leq d_{\text{max}}\), 
where $d_{\text{max}}$ is the maximum order of the polynomial expansion (e.g., 2 for quadratic, 3 for cubic).

PESs can also be fitted as a sum of $N$ neuron activations~\cite{manzhos2020neural,manzhos2006using}
\begin{equation}
V_{\text{NN}}(\mathbf{x}) = \sum_{q=1}^{N} c_q \, \sigma\left( \sum_{p=1}^{D} w_{qp} x_p + b_q \right)
\end{equation}
where $\sigma(\cdot)$ is the activation function; $c_q, w_{qp}$, and $b_q$ are learnable parameters. 
When using an exponential activation $\sigma(y) = e^y$, the potential takes on a sum-of-products form, structurally similar to the Taylor expansion~\cite{manzhos2006using}
\begin{equation}
V_{\text{NN}}(\mathbf{x}) = \sum_{q=1}^{N} c_q' \prod_{p=1}^{D} \phi_{qp}(x_p)
\label{eq:nnsop}
\end{equation}
The coefficients $c_q'$ and weights $w_{qp}$ are learned during training. This representation effectively decomposes the potential into a \emph{sum of products of univariate exponential functions}, denoted as $\phi_{qp}(x_p) = e^{w_{qp} x_p}$.

Like the polynomial expansion in Eq.~\ref{eq:taylorsop}, the \gls*{nn}-potential form in Eq.~\ref{eq:nnsop} also decomposes the PES into a sum of products. This structural similarity makes both representations suitable for efficient evaluation of quantum observables by reducing high-dimensional integrals into products of one-dimensional ones, which can accelerate methods such as the \gls*{mctdh} approach~\cite{manzhos2020neural}. 
This sum-of-products perspective also parallels the design of the atomic cluster expansion (ACE) framework, which employs a systematic product-basis ansatz to represent atomic environments accurately~\cite{drautz2019atomic}. In this sense, ACE serves as a physically grounded and interpretable analogue of the NN–based product decomposition.

In both approaches to PES fitting, the coefficients, $c_f$ in Eq.~\ref{eq:taylorsop} and $c_q'$ in Eq.~\ref{eq:nnsop}, are obtained by minimizing a loss function that measures the squared deviation between the fitted and reference (e.g., \textit{ab initio}) energies over a set of $N_t$ training configurations
\begin{equation}
    \mathcal{L}_{\text{PES}} = \sum_{n=1}^{N_t} \left( E_n^{\rm fit} - E_n^{\rm ref} \right)^2
\end{equation}
In the case of the Taylor expansion, this fitting reduces to an exact linear least squares problem. In contrast, for \gls*{nn} potentials, the energy is a nonlinear function of the parameters due to the use of activation functions and learned weights. As a result, the optimization is performed using iterative, gradient-based methods (e.g., stochastic gradient descent or Levenberg--Marquardt), and the associated loss landscape may exhibit multiple local minima.

Modern MLP methods provide a significantly more flexible and expressive alternative to traditional PES representations. These models learn complex, high-dimensional 
potential energy function, 
$E\left( \lbrace  {\bf r}_j \rbrace;  \lbrace Z_j \rbrace \right)$, directly from Cartesian coordinates, 
$\lbrace {\bf r}_j \rbrace $, 
and atomic numbers, $\lbrace Z_j \rbrace$, without relying on predefined functional forms. The inclusion of nuclear charges enables training across chemical space, allowing a single model to generalize across different elements and molecular compositions~\cite{rupp2012fast}. A common training objective is to minimize an energy-only loss function
\begin{equation}
\mathcal{L}_{\text{energy}} = \sum_{n=1}^{N_t} \left( E_n^{\text{ML}} - E_n^{\text{ref}} \right)^2,
\label{eq:energyonlyloss}
\end{equation}
where $E_n^{\rm ML}$ and $E_n^{\rm ref}$ denote the predicted and reference total energies for configuration $n$.

While total energy is the most commonly used global property in supervised PES learning, other scalar molecular properties, $P\left( \lbrace  {\bf r}_j \rbrace;  \lbrace Z_j \rbrace \right)$, such as dipole moments, HOMO-LUMO gaps, or solvation free energies, can also be learned using the same framework. In such cases, the loss function takes the general form
\begin{equation}
\mathcal{L}_{\text{property}} = \sum_{n=1}^{N_t} \left( P_n^{\text{ML}} - P_n^{\text{ref}} \right)^2,
\end{equation}
where $P_n^{\text{ML}}$ and $P_n^{\text{ref}}$ denote the predicted and reference values of a given property for configuration $n$. Although this approach has been applied successfully in various molecular property prediction tasks~\cite{ramakrishnan2015many}, it often lacks a clear theoretical justification. Unlike potential energy, which is a well-defined function of nuclear coordinates via the Born--Oppenheimer approximation, there is often no fundamental reason to expect a smooth or unique mapping from structure to global properties such as redox potential or reactivity, particularly when solvent effects, electronic correlation, or conformational variability play a significant role.

If the learned energy function is smooth and differentiable, atomic forces can be obtained as analytical gradients of the predicted energy. In principle, an accurately learned PES determines the corresponding \gls*{ff}, making this energy-only strategy viable, at least in regions of configuration space that are well sampled. However, in practice, limited sampling or model inaccuracies can lead to poor force predictions, even when the energy fit appears accurate. 

To overcome the limitations of energy-only training, most modern MLP frameworks employ a combined loss function that incorporates a force component~\cite{christensen2020role,unke2021machine,yue2021short,tokita2023train,batzner20223}
\begin{equation}
\mathcal{L}_{\text{force}}= \sum_{n=1}^{N_t}
\sum_{i \in n} \left\| \mathbf{F}_{i,n}^{\mathrm{ML}} - \mathbf{F}_{i,n}^{\mathrm{ref}} \right\|_2^2,
\label{eq:forceonlyloss}
\end{equation}
where $\mathbf{F}_{i,n}^{\mathrm{ML}}$ and $\mathbf{F}_{i,n}^{\mathrm{ref}}$ are the predicted and reference 
force vectors on atom $i$ in configuration $n$.
The force error is computed using the Euclidean norm, where \(\|{\bf x}\|_2 = \sqrt{ \sum_i x_i^2} \).

This force-based term is typically combined with the energy loss to form a joint training objective
\begin{equation}
\mathcal{L}_{\text{combined}} = 
w_E \mathcal{L}_{\text{energy}} + w_F \mathcal{L}_{\text{force}},
\label{eq:combinedloss}
\end{equation}
where the weights $w_E$ and $w_F$ control the trade-off between energy and force accuracy during training. 
In some cases, additional loss terms are included, such as partial charges, dipole moments, or 
L$_2$ regularization, particularly in charge-aware or kernel-based models~\cite{christensen2020role}. 
Typically, $w_F\propto w_E/(3N)$, as for every energy label, the loss function includes $3N$ force labels.

\subsection{Many-body expansion of energy\label{subsec:mbe}}

The total potential energy of a molecular or condensed-phase system can be formally decomposed using the \gls*{mbe}, which expresses the energy as a hierarchy of contributions from isolated atoms, pairs, triplets, and higher-order clusters\cite{varandas1977many,richard2014aiming,heindel2021molecular,jindal2022capturing}

\begin{equation}
E_{\text{total}} = \sum_i \varepsilon_i + \sum_{i<j} \varepsilon_{ij} + \sum_{i<j<k} \varepsilon_{ijk} + \cdots,
\end{equation}
where,
\( \varepsilon_i \) is the one-body energy of atom \( i \) in isolation,
\( \varepsilon_{ij} \) is the two-body correction for interactions between atoms \( i \) and \( j \),
\( \varepsilon_{ijk} \) is the three-body correction involving atoms \( i, j\) and \( k \),
and so forth.

Each higher-order term is defined recursively by subtracting all lower-order contributions from the total energy of the corresponding cluster
$\varepsilon_i = E_i$,
$\varepsilon_{ij} = E_{ij} - \varepsilon_i - \varepsilon_j$, 
$\varepsilon_{ijk} = E_{ijk} - \sum_{\alpha<\beta} \varepsilon_{\alpha\beta} - \sum_{\alpha} \varepsilon_{\alpha}$, and so on. Here, \( E_{ij} \) and \( E_{ijk} \) represent the total quantum-mechanical energies of dimers and trimers, respectively, computed from isolated sub-cluster calculations.

Once the energy is decomposed in this way, atomic forces can be derived by analytic differentiation~\cite{demerdash2016convergence,kang2023global}
\begin{equation}
\mathbf{F}_i =
- \sum_{j} \nabla_{\mathbf{r}_i} \varepsilon_j 
- \sum_{j<k} \nabla_{\mathbf{r}_i} \varepsilon_{jk} 
- \sum_{j<k<l} \nabla_{\mathbf{r}_i} \varepsilon_{jkl} 
\cdots,
\end{equation}
with only those terms contributing that contain atom \( i \) in their index set. 

\subsection{Force prediction: Direct and energy-based}

\begin{figure}[ht]
\centering
\includegraphics[width=\linewidth,angle=0]{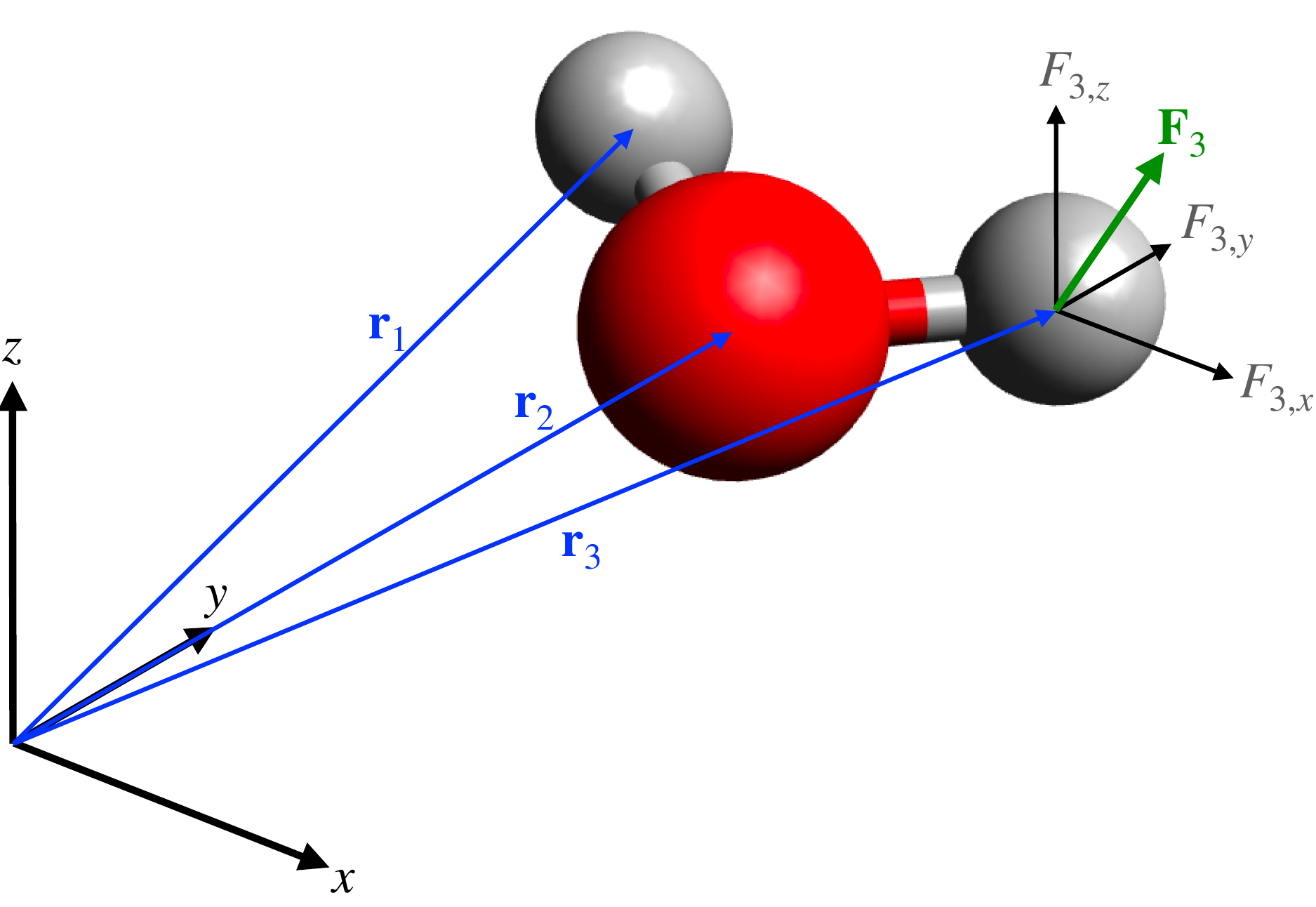}
\caption{
Illustration of a water molecule showing atomic position vectors, $\lbrace {\bf r}_i \rbrace$, and the Cartesian 
components of the force vector, ${\bf F}_3$, acting on a selected atom. All vectors are shown in a fixed
coordinate system. 
}
\label{fig:water_force}
\end{figure}

The central goal of MLPs is to accurately reproduce the \gls*{pes} that governs atomic interactions, enabling efficient and reliable \gls*{md} simulations. Since MD requires integrating Newton’s equations of motion, MLPs must not only predict the total energy as a scalar function of atomic positions for a given chemical composition, but also provide accurate energy gradients
\begin{equation}
\mathbf{F}_i (\lbrace  {\bf r}_j \rbrace;  \lbrace Z_j \rbrace ) = -\nabla_{\mathbf{r}_i} E(\lbrace  {\bf r}_j \rbrace;  \lbrace Z_j \rbrace ),
\label{eq:force_energy_based}
\end{equation}
where \(\mathbf{F}_i\) is the force acting on atom \(i\), computed as the negative gradient of the total energy with respect to its position vector \(\mathbf{r}_i\); Figure~\ref{fig:water_force} gives the definitions of the vector quantities in Eq.~\ref{eq:force_energy_based}.
Here, \( \nabla_{\mathbf{r}_i} \) denotes the gradient with respect to the Cartesian coordinates of atom \( i \), expressed as
\begin{equation}
\nabla_{\mathbf{r}_i} E = \left(
\hat{\mathbf{x}} \frac{\partial }{\partial x_i} + 
\hat{\mathbf{y}} \frac{\partial }{\partial y_i} + 
\hat{\mathbf{z}} \frac{\partial }{\partial z_i} \right) E,
\end{equation}
where \( \hat{\mathbf{x}}, \hat{\mathbf{y}}, {\rm and}~\hat{\mathbf{z}} \) are unit vectors in the \( x \), \( y \), and \( z \) directions, respectively, and \( (x_i, y_i, z_i) \) are the Cartesian coordinates of atom \( i \). This vector gives the local slope of the \gls*{pes} in each spatial direction for that atom, 
holding all other atomic positions \( \{\mathbf{r}_j\}_{j \neq i} \) and all nuclear charges \( \{Z_j\} \) fixed.
To enable simultaneous learning of energies and forces, these models are commonly trained using a combined loss function that balances errors in both quantities (see Eq.~\ref{eq:combinedloss}).

A central design principle in many MLP architectures, such as the \gls*{bpnn}~\cite{behler2007generalized}, is to decompose the total potential energy into a sum of \emph{quasi-atomic} contributions
\begin{equation}
    E_{\rm total}^{\rm ML}(\{\mathbf{r}_j\};  \lbrace Z_j \rbrace)  =  \sum_{j=1}^N E_j^{\rm ML}(\mathbf{r}_j; Z_j),
    \label{eq:energysumpool}
\end{equation}
where each atomic energy term \(E_j^{\rm ML}\) is modeled as a function of the local environment of atom \(j\).

This decomposition serves several important purposes. First, it enforces \emph{size-extensivity} by ensuring that the total energy scales linearly with system size. Second, it promotes \emph{transferability}, as the local energy mappings can be applied to systems of varying size and composition. This formulation also closely mirrors the structure of the many-body expansion (see Section~\ref{subsec:mbe}).
While the locality of atomic energy mappings (Eq.~\ref{eq:energysumpool}) ensures size extensivity and offers the possibility of transferability across systems of varying size and composition, achieving practical transferability ultimately depends on the diversity and representativeness of the training data. As highlighted in recent studies~\cite{goodwin2024transferability}, even models based on local or equivariant architectures require carefully curated datasets to generalize reliably to unseen chemical compositions or mixtures.

While the force vector acting on atom \( i \) is formally defined as the gradient of the total energy, it is computed by differentiating each atomic energy term with respect to \( {\bf r}_i \)
\begin{equation}
{\bf F}_i^{\rm ML} =
 -\nabla_{{\bf r}_i} E_{\text{total}}^{\rm ML} = - \sum_j
  \nabla_{{\bf r}_i} E^{\rm ML}_j.
\end{equation}
Although the total energy is expressed as a sum over atomic contributions, each \(E_j^{\rm ML}\) typically depends on the positions of neighboring atoms. As a result, the gradient with respect to \({\bf r}_i\) can have nonzero contributions even for \(j \ne i\):
\begin{equation}
    \nabla_{\mathbf{r}_i} E_j^{\rm ML} \ne 0 \quad \text{for } i \ne j.
\end{equation}
These derivatives are computed \emph{analytically}, either through explicit formulae or automatic differentiation, ensuring that the resulting \gls*{ff} is smooth, conservative, and numerically stable.

To reduce computational cost, a finite cutoff radius is introduced to limit the range of interactions. 
In practice, cutoff distances between 5 and 10~\AA~ are commonly employed, as this range has been shown to capture most intermolecular interactions effectively.~\cite{kovacs2023evaluation}
However, the optimal cutoff value can vary depending on the molecular system and the underlying architecture.
Within this approximation, the force on atom \(i\) is computed by considering only nearby atoms:
\begin{equation}
    {\bf F}_i^{\rm ML} \approx - \sum_{j \in \mathcal{N}_i}  \nabla_{\mathbf{r}_i} E_j^{\rm ML},
\end{equation}
where \(\mathcal{N}_i = \{ j \mid {\bf r}_j \in \mathrm{env}(i) \}\) denotes the local environment of atom \(i\) within the cutoff.
This locality assumption, grounded in the \emph{nearsightedness of electronic matter}~\cite{prodan2005nearsightedness,glielmo2017accurate,poltavsky2021machine,zeng2022nearsighted}, enables linear-scaling evaluations and makes MLPs suitable for large-scale simulations.

Modern \gls*{gnn} architectures operate on molecular graphs and learn embeddings for each atom by propagating information through message passing. Some GNNs, such as SchNet~\cite{schutt2018schnet}, predict global molecular properties as a sum of learned atom-wise contributions, using a sum pooling operation to ensure extensivity
\begin{equation}
  E^{\rm ML}_{\text{total}} = \sum_i \varepsilon({\bf h}_i),  
  \label{eq:energysumpoolgnn}
\end{equation}
where \( {\bf h}_i \) is the atomic embedding and \( \varepsilon \) is a shared output \gls*{nn} layer applied to each atom. In contrast, Buterez~\textit{et al.}~\cite{buterez2023modelling} introduced an attention-based pooling mechanism that computes a dynamically weighted mean of atomic embeddings. This approach enables the model to focus on atoms most relevant to the target property, improving performance on both localized and global quantities.
In these approaches, the total energy is treated as a holistic, learned function of atomic positions and identities, \(E^{\rm ML}(\lbrace {\bf r}_i \rbrace)\), rather than a sum over atomic energies. Forces are then obtained as analytical gradients of this total energy
\begin{equation}
  {\bf F}_i^{\rm ML} =  -\nabla_{\mathbf{r}_i} E_{\rm total}^{\rm ML},
  \label{eq:force_energybased_ML}
\end{equation}
ensuring that predictions remain physically consistent with the underlying \gls*{pes}. 

Table~\ref{tab:mlp_summary} categorizes various MLP architectures based on whether atomic forces are computed as the sum of gradients of per-atom energy contributions, \( -\sum_j \nabla_{{\bf r}_i} E_j^{\rm ML} \), or as the gradient of the total energy, \( -\nabla_{{\bf r}_i} E_{\rm total}^{\rm ML} \).

While most MLPs compute forces as analytic gradients of a scalar energy function, an alternative strategy is to train models that directly learn and predict atomic force vectors without explicitly learning the energy. These 
models, which can further be classified as \emph{gradient-domain} or 
\emph{force-only} are particularly attractive when accurate force labels, such as those from AIMD or geometry optimization trajectories, are available. Such models are trained using a force-based loss function that penalizes the difference between predicted and reference forces (see Eq.~\ref{eq:forceonlyloss}). Force-only approaches are instrumental in applications such as vibrational analysis, solvation-induced perturbations, or geometry optimization in complex environments.

\subsection{Invariant and conservative forces}

An essential physical constraint for isolated systems is that the total force must vanish at all configurations
\begin{equation}
    \sum_i {\bf F}_i = {\bf 0}.
\end{equation}
This condition reflects conservation of linear momentum and follows from the translational invariance of the potential energy function. 
If the total energy \( E(\{{\bf r}_i\}) \) depends only on internal coordinates (e.g., interatomic distances) and not on absolute positions, then it is not affected by a uniform translation of all atomic positions. 
Formally, under a global shift \({\bf r}_i \rightarrow {\bf r}_i + {\bf s} \), the energy remains unchanged
\begin{equation}
    E(\{{\bf r}_i + {\bf s}\}) = E(\{{\bf r}_i\}) \, \Rightarrow \, \sum_i \nabla_{{\bf r}_i} E = \nabla_{{\bf R}_{\mathrm{CM}}} E = {\bf 0},
\end{equation}
where \({\bf R}_{\mathrm{CM}} = \frac{1}{M} \sum_i m_i {\bf r}_i\) is the center of mass of the system with total mass \( M = \sum_i m_i \). 

Hence, in energy-based models that respect translational invariance, the total force is zero by construction. If this condition is violated, the system will experience a net translation of its center of mass during time evolution, leading to a nonphysical drift in \gls*{md} simulations.

Similarly, conservation of angular momentum imposes that the total torque on the system must vanish
\begin{equation}
    \sum_i \mathbf{r}_i \times \mathbf{F}_i = \mathbf{0}.
\end{equation}
This condition arises from the rotational invariance of the potential energy. Under an infinitesimal rotation of all atomic positions, the displacement of each atom is given by 
$\delta \mathbf{r}_i = \boldsymbol{\delta \theta} \times \mathbf{r}_i$,
where \( \boldsymbol{\delta \theta} \) is an infinitesimal rotation vector. If the energy depends only on internal coordinates (e.g., distances and angles), it remains invariant under such a transformation. The resulting variation in energy is
\begin{equation}
    \delta E = \sum_i \nabla_{\mathbf{r}_i} E \cdot (\boldsymbol{\delta \theta} \times \mathbf{r}_i) = \boldsymbol{\delta \theta} \cdot \sum_i \mathbf{r}_i \times \mathbf{F}_i.
\end{equation}
Since this must vanish for arbitrary \( \boldsymbol{\delta \theta} \), it follows that
\begin{equation}
\sum_i \mathbf{r}_i \times \mathbf{F}_i = \mathbf{0}.
\end{equation}
Violation of this condition leads to non-conservation of angular momentum and can cause artificial rotational drift in MD simulations.

Importantly, while translational and rotational invariance ensure the correct global behavior of a force field, they do not guarantee that the force field is \emph{conservative}~\cite{chmiela2019towards,dupuy2024exciting,cova2019deep,williams2024stable}. A vector-valued force field \( {\bf F}({\bf r}) \) is conservative if and only if it is the gradient of a scalar potential (as in Eq.~\ref{eq:force_energy_based}).
A necessary local condition for this to hold is that the curl of the force vanishes~\cite{goldstein2011classical}
\begin{equation}
    \nabla \times {\bf F}({\bf r}) = {\bf 0}.
\end{equation}
If this condition is violated
\begin{equation}
    \nabla \times \mathbf{F}_i \neq {\bf 0} \quad \Rightarrow \quad \mathbf{F}_i \neq -\nabla_{\mathbf{r}_i} E,
\end{equation}
the forces do not correspond to any well-defined PES. Such non-conservative fields can cause energy drift in MD or yield unphysical free energy profiles in thermodynamic applications.
This zero-curl condition applies pointwise throughout configuration space and ensures local integrability of the \gls*{ff}. In contrast to the global force and torque constraints, which govern system-wide symmetries, the curl condition governs the \emph{local} consistency of the vector field. Violation of this condition implies that the \gls*{ff} cannot be derived from any scalar energy function, even if global symmetries are satisfied.

To illustrate why conservative force fields satisfy the zero-curl relation \( \nabla \times \mathbf{F} = \nabla \times (-\nabla V) = \mathbf{0} \), consider a scalar potential energy function \( V(x, y, z) \). The corresponding force field is
\begin{eqnarray}
    \mathbf{F}  & = & -\nabla V(x,y,z) \nonumber \\ 
             & = & -\left( \hat{\mathbf{x}} g_1(x,y,z) + \hat{\mathbf{y}} g_2(x,y,z) + \hat{\mathbf{z}} g_3(x,y,z) \right),
\end{eqnarray}
where \( g_1 = \partial V / \partial x \), \( g_2 = \partial V / \partial y \), and \( g_3 = \partial V / \partial z \). Taking the curl
\begin{eqnarray}
\nabla \times \mathbf{F} & =&  -\nabla \times \nabla V \nonumber \\ 
& = &  
- \left(
\frac{\partial g_3}{\partial y} - \frac{\partial g_2}{\partial z}, 
\frac{\partial g_1}{\partial z} - \frac{\partial g_3}{\partial x}, 
\frac{\partial g_2}{\partial x} - \frac{\partial g_1}{\partial y}
\right),
\end{eqnarray}
and using the equality of mixed second derivatives
\begin{equation}
\frac{\partial^2 V}{\partial y \partial z} = \frac{\partial^2 V}{\partial z \partial y}, \quad
\frac{\partial^2 V}{\partial z \partial x} = \frac{\partial^2 V}{\partial x \partial z}, \quad
\frac{\partial^2 V}{\partial x \partial y} = \frac{\partial^2 V}{\partial y \partial x},
\end{equation}
each component of the curl vanishes. Therefore, any sufficiently smooth scalar potential yields a conservative force field with \( \nabla \times \mathbf{F} = \mathbf{0} \).

\vspace{1em}
Together, the conditions of zero total force, zero total torque, and zero curl ensure that the \gls*{ff} is physically consistent at both global and local levels. These constraints are automatically satisfied in conservative, energy-based MLPs that use translationally and rotationally invariant inputs (e.g., interatomic distances and angles), but may require explicit enforcement in force-only models.
Descriptor-based models enforce symmetry constraints through explicitly constructed scalar features, while GNN-based models achieve this via rotationally equivariant architectures that preserve physical symmetries by design.

\subsection{\texorpdfstring{$\Delta$-}-machine learning\label{sec:delta-ML}}

$\Delta$-machine learning ($\Delta$-ML) is a hybrid modeling strategy where an MLP can be trained to correct a lower-level baseline method~\cite{ramakrishnan2015big}
\begin{eqnarray}
E^{\text{target}}(\{\mathbf{r}_j\}; \{ Z_j \} ) & = & E^{\text{base}}(\{\mathbf{r}_j\}; \{ Z_j \}) + \nonumber \\
& & \Delta E^{\mathrm{ML}}(\{\mathbf{r}_j\}; \{ Z_j \}),
\end{eqnarray}
where $E^{\text{base}}$ is typically a lower-level but qualitatively accurate energy, 
and $\Delta E^{\mathrm{ML}}$ is the machine-learned correction trained on the difference between high-level and baseline energies.

\begin{figure}[ht]
\centering
\includegraphics[width=\linewidth,angle=0]{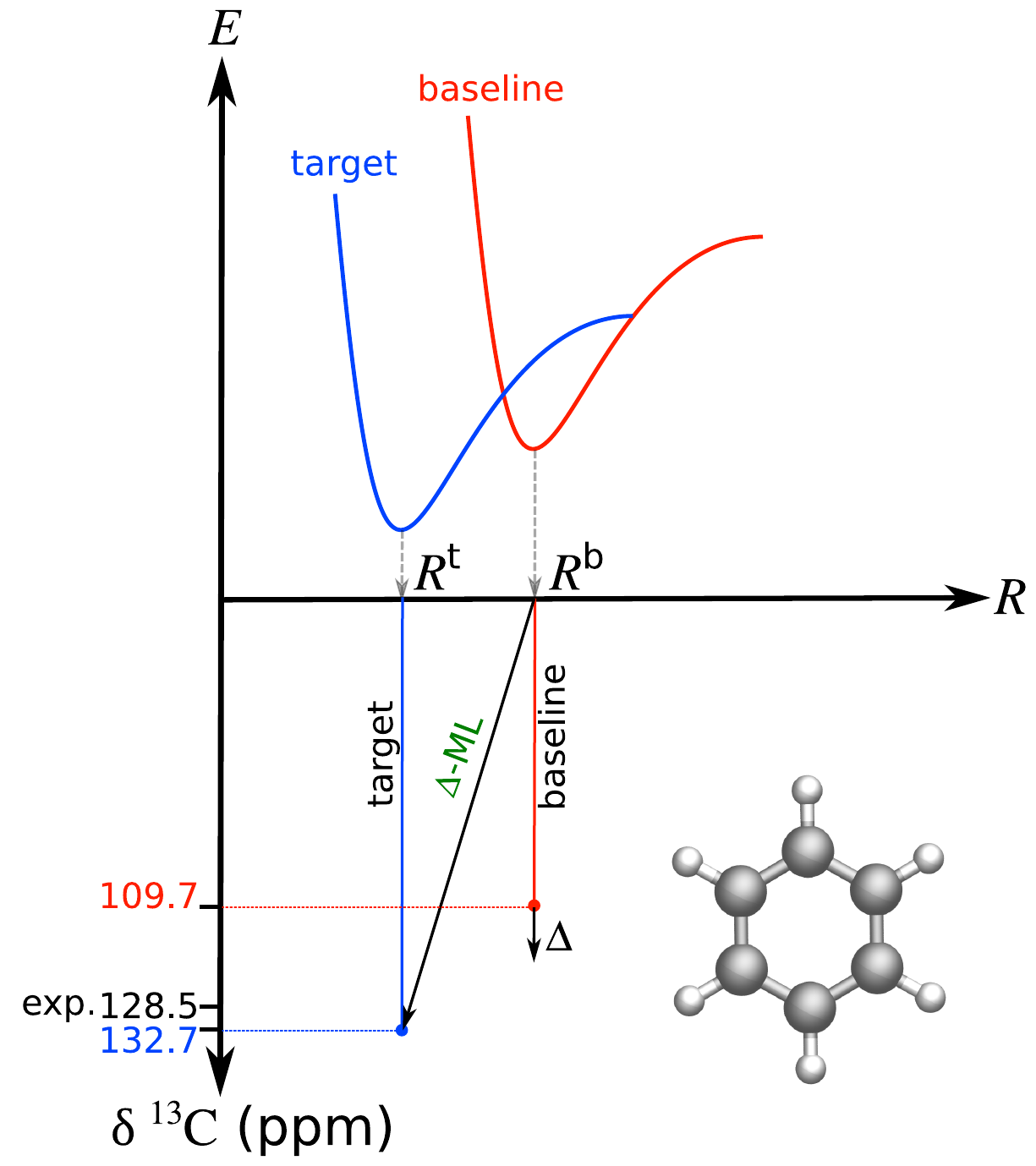}
\caption{
Illustration of $\Delta$-ML for predicting NMR chemical shifts in a prototypical molecule, adapted from Ref.~\cite{gupta2021revving}. The model is trained to learn the difference between a baseline prediction (DFT with a small basis set using semiempirical geometries) and a higher-level reference (DFT with a large basis set using geometries optimized at a comparable level).
}
\label{fig:deltaML}
\end{figure}

A main advantage of this approach is that the baseline and target methods need not use the same geometries. 
In practice, the baseline-level geometry is often used as input for both \( E^{\text{base}} \) and \( \Delta E^{\mathrm{ML}} \), even though the reference (target) energy corresponds to a higher-level optimized structure. This flexibility allows the correction model to be trained on baseline geometries alone. As a result, for new predictions, only the baseline geometry is needed, avoiding expensive high-level structure optimization. This strategy has been successfully applied in many studies to predict quantum chemical properties across chemical space with improved accuracy and reduced cost~\cite{ramakrishnan2017machine, gupta2021revving, tripathy2024chemical,unzueta2021predicting}. Figure~\ref{fig:deltaML} illustrates the application of $\Delta$-ML to predict NMR chemical shifts using baseline and target quantum chemical levels.

If the ML model is differentiable, atomic forces can be derived by analytic differentiation
\begin{equation}
\mathbf{F}_i = -\nabla_{\mathbf{r}_i} E^{\text{target}} = -\nabla_{\mathbf{r}_i} E^{\text{base}} - \nabla_{\mathbf{r}_i} \Delta E^{\mathrm{ML}}.
\end{equation}
This ensures that the predicted forces are conservative (i.e., derived from a scalar potential), and that energy conservation is preserved during \gls*{md} simulations.

Alternatively, force labels can be used to train the force correction directly
\begin{equation}
\Delta \mathbf{F}_i^{\text{ML}} = \mathbf{F}_i^{\text{target}} - \mathbf{F}_i^{\text{base}},
\end{equation}
so that the learned model predicts the force discrepancy rather than total energy \cite{pattnaik2020machine}.

As in the energy-based formulation, the delta-force approach offers flexibility in the choice of geometries used for training. The correction model \( \Delta \mathbf{F}_i^{\text{ML}} \) can be trained on forces evaluated at baseline-level geometries, even if the target-level forces originate from a different, higher-level theory. In such cases, both \( \mathbf{F}_i^{\text{base}} \) and \( \mathbf{F}_i^{\text{target}} \) are computed at the same baseline geometry, and the model learns to predict their difference. At inference time, only the baseline geometry is required to generate corrected forces, enabling efficient and accurate predictions without the need for expensive high-level structural optimization.

This flexibility in geometry choice also enables the seamless integration of $\Delta$-ML corrections into MD simulations. Since both the baseline method and the ML correction can be evaluated at the same geometry, MD can be performed using baseline-level trajectories while applying learned corrections at each step. At every timestep, the total force is computed as
\begin{equation}
\mathbf{F}_i^{\text{target}} = \mathbf{F}_i^{\text{base}} + \Delta \mathbf{F}_i^{\text{ML}}(\{\mathbf{r}_j^{\rm base}\}),
\end{equation}
or, equivalently, the corrected energy is used to derive forces via analytic differentiation
\begin{equation}
\mathbf{F}_i^{\text{target}} = -\nabla_{\mathbf{r}_i} \left[ E^{\text{base}}(\{\mathbf{r}_j^{\rm base}\}) + \Delta E^{\text{ML}}(\{\mathbf{r}_j^{\rm base}\}) \right].
\end{equation}
In either case, the dynamics require only the baseline-level geometries as input, allowing efficient, energy-conserving MD simulations with accuracy approaching that of the target method.

\vspace{1em}
\noindent
Although $\Delta$-ML, MBEs, and hybrid schemes such as ONIOM or QM/MM (see Section\ref{ssec:traditional_paradigms}) differ in implementation and domain of application, they are unified by a common modeling philosophy: a lower-cost approximation is systematically corrected to approach a higher-accuracy reference. This {\it divide-and-correct paradigm} manifests itself in different ways. $\Delta$-ML models are trained to predict the difference between a baseline method and a target-level label. In ONIOM and QM/MM, a system is partitioned into regions treated at different levels of theory. Similarly, in MBE frameworks, lower-order interaction terms (e.g., 1-body and 2-body) can be computed at high-level quantum chemical accuracy, while higher-order terms can be approximated using low-level methods or ML models.

While ONIOM partitions the system spatially, MBE partitions it by interaction order. Both approaches apply layered accuracy where it matters most, and $\Delta$-ML can be viewed as an ML analogue of ONIOM, replacing explicit high-level calculations with data-driven surrogates. This analogy clarifies the conceptual role of $\Delta$-ML in the broader modeling landscape.

Such additive schemes also underpin cluster expansion (CE) techniques for materials modeling~\cite{sanchez1984generalized}, where energies of alloy configurations or defects are expressed as sums over contributions from pairs, triplets, and higher-order clusters. ML methods have recently been used to improve the fitting and generalization of CE models~\cite{xie2024machine}.

The scope of such additive schemes is far-reaching. For instance, they can be used to correct density functional tight-binding (DFTB)-predicted phonon band structures of materials using DFT-level phonons at the $\Gamma$-point~\cite{cook2020reduced}, or to perform basis set extrapolation by combining results from low-cost and high-accuracy basis sets to approximate the complete basis set (CBS) limit~\cite{truhlar1998basis}. $\Delta$-ML models have also been developed to learn basis set convergence patterns directly, enabling extrapolation to CBS-quality modeling~\cite{holm2023single}.


\begin{figure*}[!htpb]
\centering
\includegraphics[width=1.0\linewidth,angle=0]{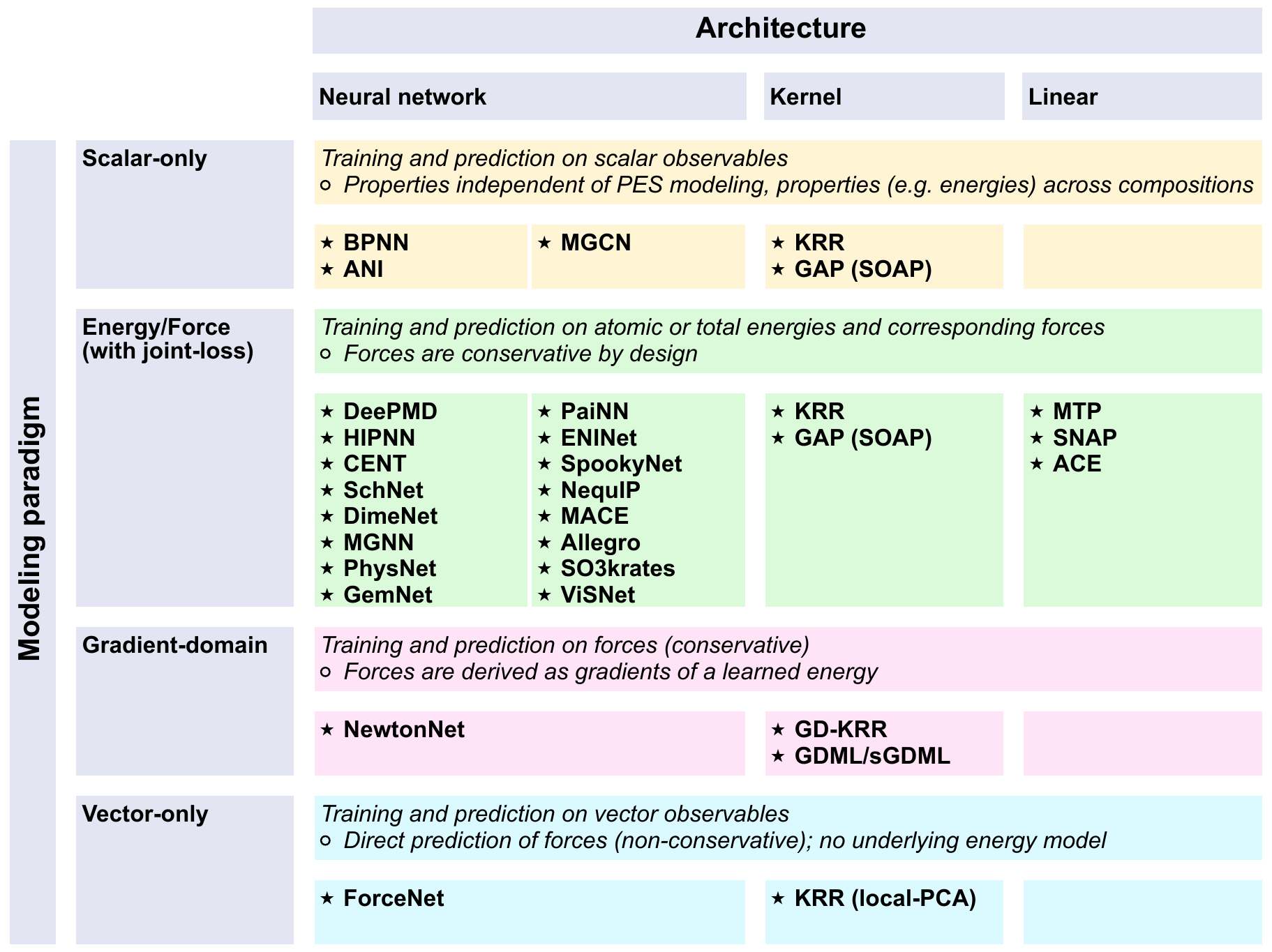}
\caption{
A logical classification of machine-learned potentials (MLPs) 
organized by modeling paradigm (rows) and architectural class (columns) 
that distinguishes conservative, force-only, and scalar models while grouping 
them under neural, kernel, or linear frameworks. All acronyms used here are defined in the main text.
}
\label{fig:taxonomy}
\end{figure*}
\section{Taxonomy, Techniques and Performances\label{subsec:classifyMLPs}}

MLPs differ in how they decompose the total energy into atomic contributions, represent atomic environments, and enforce physical symmetries. For solvation modeling, desirable features include the ability to capture both local and long-range interactions, accommodate dynamic and heterogeneous solvent shells, and preserve rotational, translational, and permutational invariance. This section summarizes the key architectural families employed in MLP-based solvation modeling.

Several studies have proposed taxonomies for MLPs based on descriptors, symmetry constraints, and physical priors~\cite{ocampo2024adaptive,thiemann2024introduction,yang2025efficient,reiser2022graph,duval2023hitchhiker,wang2023graph,behler2021four,omranpour2025machine,yang2024machine,kocer2022neural,wang2024machine}. To provide a unified perspective on this landscape, we present in Fig.~\ref{fig:taxonomy} a formal taxonomy organized along two orthogonal axes:
\begin{itemize}
    \item the \emph{modeling paradigm}, which defines what quantities are learned and how they are predicted,
    \item the \emph{architectural class}, which reflects the inductive biases and representational design of the model.
\end{itemize}

This classification distinguishes scalar-only models, conservative energy–force models, gradient-domain force-only models, and vector-only models, all of which bypass energy learning from one another. 
Simultaneously, it groups models by architectural aspects, such as descriptor-based \gls*{nn}s, graph message-passing networks, kernel methods, and linear models. Organizing models along these conceptual axes clarifies how their design choices impact physical consistency, transferability, and suitability for solvation modeling.

Although Figure~\ref{fig:taxonomy} presents a logically grounded classification, a different taxonomy is adopted in the review to better align with the historical development of models, architectural similarity, training strategies, and pedagogical clarity. 

This alternative structure also avoids sparsely populated or underdefined regions in the $4\times3$ classification (e.g., vector-only linear models). Importantly, the absence of models in some combinations of target-labels and architecture is not due to mathematical impossibility, but often reflects practical factors such as historical priorities, data availability, or the coupling between certain model types and accessible software infrastructures. 

Moreover, the boundaries between some categories in the principled classification shown in Figure~\ref{fig:taxonomy} are inherently fluid. For instance, scalar-only models can be viewed as limiting cases of energy–force models when trained solely on scalar quantities. Many models originally developed for scalar-valued energy prediction can also be extended to energy–force or vector prediction tasks, depending on the availability of labels and the choice of loss functions. While such distinctions are conceptually sound, they often blur in practice, as most MLP architectures can be flexibly trained with or without force supervision. These considerations motivate a departure from the strict quadrant-style classification presented in Figure~\ref{fig:taxonomy} in favor of a more narrative-driven taxonomy.

Accordingly, we distinguish five major categories of MLPs:
\begin{enumerate}
    \item \textbf{Neural network–based MLPs (NN-MLPs)}. These are further subdivided based on how atomic environments are represented and learned:
    \begin{itemize}
        \item (a) \textbf{Descriptor-based architectures} (Desc-NN-MLPs), which use fixed or learnable atom-centered descriptors;
        \item (b) \textbf{End-to-end networks with local message passing} (MP-LNN-MLPs), which operate directly on coordinates with locally defined neighborhoods;
        \item (c) \textbf{End-to-end networks with graph-based message passing} (MP-GNN-MLPs), which leverage atomistic graphs and learned edge-wise interactions.
    \end{itemize}
    \item \textbf{Kernel-based MLPs (Kernel-MLPs)}, which use \gls*{krr}, \gls*{gpr}, or symmetry-adapted kernels to learn energies or forces.
    \item \textbf{Linear MLPs (Linear-MLPs)}, which use fixed functional forms (e.g., polynomials or bispectrum components) and linear regression for fitting.
    \item \textbf{Gradient-domain MLPs (GD-MLPs)}, which learn conservative \gls*{ff}s by training on forces and defining them as gradients of a learned scalar energy.
    \item \textbf{Force-only MLPs (F-MLPs)}, which bypass energy learning entirely and directly regress atomic forces from force labels. These models are not guaranteed to be energy-conserving but offer computational advantages in training and inference.
\end{enumerate}

While both descriptor-based and end-to-end NN-MLPs share a standard downstream structure that maps learned atomic embeddings through a readout layer to predict atomic energies, the key distinction lies in their input representations. Descriptor-based models operate on fixed, handcrafted features, whereas end-to-end architectures learn atomic representations directly from atomic numbers and positions via trainable message-passing or convolutional layers.

Gradient-domain MLPs can themselves be further subdivided into \gls*{nn}–based and kernel-based models. 

A fundamental distinction across MLP architectures lies in how physical symmetries—such as translational, rotational, and permutational invariance—are enforced. In \textit{descriptor-based MLPs} (both NN and kernel-based), these symmetries are typically encoded into the descriptors themselves, such as \gls*{acsfs}, Coulomb matrices, or the \gls*{soap}. These descriptors are designed to be invariant under transformations, ensuring that scalar outputs (e.g., energy) remain unchanged. In contrast, \textit{end-to-end message-passing NN-MLPs} must encode these symmetry constraints architecturally. Thus, symmetry preservation shifts from descriptor-level design to architectural-level design as one moves from traditional descriptor-based models to modern end-to-end models.

We do not attempt a parallel taxonomy of descriptors here, as other studies have already covered this topic extensively~\cite{raghunathan2022molecular,yi2023towards,lange2024comparative}. Table~\ref{tab:mlp_categories} provides a high-level comparison of representative MLP classes, organized by descriptor formalism and architectural design. 
While this taxonomy abstracts away many finer distinctions, such as whether long-range interactions are learned or explicitly modeled, it serves as a conceptual guide to selecting and comparing MLP architectures.

Throughout this work, ``MLP'' refers to machine-learned interatomic potentials, encompassing a broad class of models that use \gls*{nn}s, kernel methods, or linear regressions to approximate \gls*{pes}s. To avoid confusion with the standard abbreviation for \emph{multi-layer perceptrons} in neural network architectures, we explicitly refer to such components as \emph{dense layers}, \emph{fully connected layers}, or \emph{shared hidden layers} (i.e., combinations of a few intermediate layers applied across atoms) where appropriate.

\begin{table*}[!htpb]
\centering
\caption{
Comparison of major categories of machine-learned interatomic potentials (MLIPs) in atomistic modeling. Models are grouped by descriptor type (explicit vs.\ learned), architecture (feedforward, graph-based, kernel, or tensor-based), and learning paradigm (energy-based, force-based, or hybrid).}
\label{tab:mlp_categories}
\begin{tabular}{l l l l l}
\hline 
MLP Type & Examples & Descriptor & Architecture & Category \\
\hline 
Descriptor-based NN        & BPNN, ANI                 & Explicit               & Feedforward NN               & NN-MLP \\
End-to-end local NN        & DeePMD, PhysNet           & Learned                & Deep NN                      & NN-MLP \\
Message-passing GNN        & SchNet, DimeNet           & Learned                & Invariant GNN                & NN-MLP \\
Equivariant MP-GNN         & NequIP, MACE      & Learned                & Equivariant GNN              & NN-MLP \\
Kernel-based (Energy)      & GAP, FCHL                 & Explicit               & KRR      & Kernel-MLP \\
Tensor/polynomial models   & MTP, ACE                  & Explicit               & Tensor contraction & Linear-MLP \\
Gradient-domain kernel     & GDML, FCHL19      & Explicit               & KRR (force)    & GD-MLP \\
Gradient-domain NN         & NewtonNet                 & Learned                & Equivariant GNN              & GD-MLP \\
Force-only kernel          & KRR (with local-PCA)                   & Explicit               & KRR (force-only) & F-MLP \\
Force-only NN              & ForceNet                  & Learned                & GNN (force-only)             & F-MLP \\
\hline  
\end{tabular}
\end{table*}

\subsection{Neural network models}

NN-MLPs represent one of the most widely used and actively developed classes of ML models. They predict total energies, forces on atoms, or other properties by learning mappings directly from atomic coordinates and chemical species through trainable neural architectures. Depending on how atomic environments are represented and processed, NN-MLPs can be further divided into descriptor-based models and end-to-end models based on \gls*{mpnns} with learned internal features.

In the broader sense defined by Behler~\cite{behler2021four}, an NN-MLP from any of these categories can be termed a \gls*{hdnnp} if it: 
(i) explicitly depends on all atomic degrees of freedom, 
(ii) preserves the required symmetry invariances such as translational, rotational, and permutational, and 
(iii) enables scalable calculation with an increase in the size of the atomistic system. 
In essence, \gls*{hdnnp}s construct a unique mapping from atomic configurations to a high-dimensional \gls*{pes}.

This subsection provides an overview of the key NN-MLP architectures used in solvation modeling, with attention to their strengths and limitations for different solvation environments.

\subsubsection{Descriptor-based NN-MLPs (Desc-NN-MLPs)}

Descriptor-based NN-MLPs are among the earliest and most widely used approaches for ML-based PES modeling. 
These models rely on the assumption that the total molecular energy can be decomposed into atomic contributions, each determined by the atom's local chemical environment. 
This environment is encoded using symmetry-preserving descriptors that are invariant to translation, rotation, and permutation of like atoms, such as \gls*{acsfs}~\cite{behler2007generalized,behler2011atom}, Coulomb matrix~\cite{rupp2012fast}, and \gls*{soap}~\cite{bartok2013representing}.

\begin{figure*}[ht]
\centering
\includegraphics[width=\linewidth,angle=0]{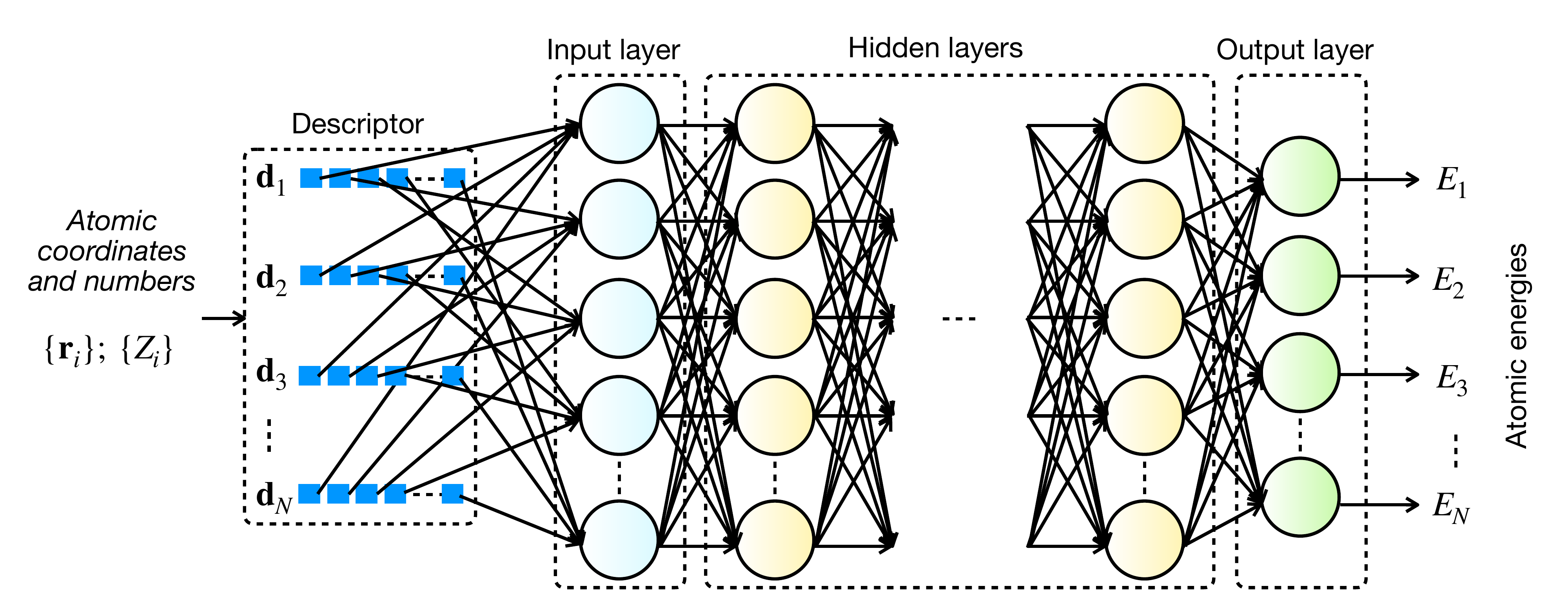}
\caption{
Schematic of a DeepNN architecture for predicting atomic energy contributions, as employed in models such as BPNN and DeepMD. Each atom is represented by a descriptor vector, which is passed through a feedforward neural network composed of an input layer, multiple hidden layers, and an output layer that returns the atomic energies. The same network is applied to all atoms of a given element type, and the resulting atomic energies are summed to obtain the total energy of the system. In BPNN, descriptors are predefined symmetry functions based on local atomic environments, whereas in DeepMD, descriptors are learned from the relative positions of neighboring atoms within a locally defined reference frame.
}
\label{fig:DeepNN}
\end{figure*}

Once the descriptors \( \mathbf{d}_i \) are computed as numerical vectors, they are passed into species-specific feedforward NNs to predict atomic energies as shown in Figure~\ref{fig:DeepNN}. The total energy is obtained by summing over atoms as in Eq.~\ref{eq:energysumpool} or Eq.~\ref{eq:energysumpoolgnn}.

For a shallow feedforward \gls*{nn} with a single hidden layer, the atomic energy \( E_i \) is computed from the input descriptor vector \( \mathbf{d}_i \in \mathbb{R}^F \), where $\mathbb{R}^F$ denotes $F$-dimensional real vector space ($F$ is used to indicate feature vector length). 

Let \( \mathbf{W}^{(1)} \in \mathbb{R}^{F \times M} \) be the weight matrix from the input to the hidden layer, \( \mathbf{b}^{(1)} \in \mathbb{R}^M \) the hidden layer bias vector ($M$ is the number of neurons in the hidden layer), and \( f^{(1)}(\cdot) \) the elementwise activation function (e.g., Tanh, ReLU, sigmoid, or SoftMax). 
The hidden layer activation for atom \( i \) is given by
\begin{equation}
\mathbf{h}_i = f^{(1)}\left( \mathbf{W}^{(1)} \mathbf{d}_i + \mathbf{b}^{(1)} \right),
\label{eq:shallowactivation}
\end{equation}

Let \( \mathbf{w}^{(2)} \in \mathbb{R}^{M} \) be the output weight vector connecting the hidden layer to the output, and \( b^{(2)} \in \mathbb{R} \) the scalar output bias, then
and the atomic energy is computed as
\begin{equation}
E_i = (\mathbf{w}^{(2)})^\top \mathbf{h}_i + b^{(2)}.
\end{equation}

The vector \( \mathbf{h}_i \), denoted as the hidden-layer activation for atom \( i \), is a learned latent representation, also referred to as an \textit{embedding} of the input geometry. It captures chemically relevant features of the local environment in a continuous, trainable space.

Alternatively, if \( \mathbf{H} \in \mathbb{R}^{N \times M} \) is the matrix of hidden layer outputs (with each row \( \mathbf{h}_i^\top \) corresponding to atom \( i \)), the vector of atomic energies can be written compactly as
\begin{equation}
\mathbf{E} = \mathbf{H} \mathbf{w}^{(2)} + \mathbf{b}^{(2)} ,
\end{equation}
where 
\( \mathbf{E} = [E_1, E_2, \dots, E_N]^\top \in \mathbb{R}^N \) denote the vector of atomic energies for all \( N \) atoms in the system, the total energy is ${\rm sum}( \mathbf{E})$.
Note that the output weight \( \mathbf{w}^{(2)} \in \mathbb{R}^M \) is a vector because it is shared across all atoms; the same set of weights is used to map each hidden representation \( \mathbf{h}_i \) to a scalar atomic energy \( E_i \), ensuring consistency, efficiency, and permutation invariance across the system.

This formulation generalizes to deep neural networks (DeepNNs) with \( L \) layers. Let \( \mathbf{h}^{(0)}_i = \mathbf{d}_i \in \mathbb{R}^{N_0} \) be the input descriptor for atom \( i \), and let each hidden layer compute its output as
\begin{equation}
\mathbf{h}^{(l)}_i = f^{(l)}\left( \mathbf{W}^{(l)} \mathbf{h}^{(l-1)}_i + \mathbf{b}^{(l)} \right), \quad l = 1, \dots, L-1,
\label{eq:deepactivation}
\end{equation}
where
\( \mathbf{W}^{(l)} \in \mathbb{R}^{N_l \times N_{l-1}} \) is the weight matrix for layer \( l \), 
\( \mathbf{b}^{(l)} \in \mathbb{R}^{N_l} \) is the bias vector, 
\( f^{(l)} \) is the elementwise activation function for layer \( l \), and 
\( \mathbf{h}^{(l)}_i \in \mathbb{R}^{N_l} \) is the output of layer \( l \).
A prototypical DeepNN architecture is shown in Figure~\ref{fig:DeepNN} and its schematic representation
is presented in Figure~\ref{fig:DeepNN_HIPNN}.

The final output energy is computed as
\begin{equation}
E_i = (\mathbf{w}^{(L)})^\top \mathbf{h}^{(L-1)}_i + b^{(L)},
\label{eq:deenatomicenergy}
\end{equation}
where \( \mathbf{w}^{(L)} \in \mathbb{R}^{N_{L-1}} \) and \( b^{(L)} \in \mathbb{R} \) are the output-layer weights and biases, respectively.

\paragraph*{BPNN: Behler--Parrinello Neural Networks}
BPNNs are descriptor-based MLPs based on DeepNN architecture
to decompose the total molecular energy into atomic contributions predicted by species-specific NNs~\cite{behler2007generalized}. The input descriptors \( \mathbf{d}_i \) consist of \gls*{acsfs}, which encode the radial and angular environment around atom \( i \) in a symmetry-invariant form.

The simplest ACSF corresponds to a smooth coordination number
\begin{equation}
G_i^{1} = \sum_j f_c(R_{ij}),
\end{equation}
where \( R_{ij} \) is the interatomic distance and \( f_c(R_{ij}) \) is a cutoff function that smoothly decays to zero at a predefined cutoff radius \( R_c \)
\begin{equation}
f_c(R_{ij}) = 
\begin{cases}
0.5 \left[ \cos\left( \frac{\pi R_{ij}}{R_c} \right) + 1 \right], & R_{ij} \leq R_c, \\
0, & R_{ij} > R_c.
\end{cases}
\end{equation}

Radial symmetry functions probe the distribution of neighbors at different distances
\begin{equation}
G_i^{2} = \sum_{j \neq i} e^{-\eta (R_{ij} - R_s)^2} f_c(R_{ij}),
\end{equation}
where \( \eta \) and \( R_s \) control the width and center of the Gaussian.

Angular symmetry functions encode three-body correlations
\begin{eqnarray}
G_i^{3} &=& 2^{1 - \zeta} \sum_{j,k \neq i}^{j < k} 
\left(1 + \lambda \cos \theta_{ijk} \right)^\zeta f_c(R_{ij}) f_c(R_{ik}) f_c(R_{jk}) \times \nonumber \\
&& e^{-\eta \left(R_{ij}^2 + R_{ik}^2 + R_{jk}^2\right)},
\end{eqnarray}
where \( \theta_{ijk} \) is the angle between atoms \( i, j, k \); \( \lambda = \pm 1 \) selects angular phase, and \( \zeta \) controls angular resolution.

These ACSFs are evaluated over neighbors \( j \), \( k \) within the cutoff radius. A variety of hyperparameters \( \eta, R_s, \zeta, \lambda \) are used to generate a high-dimensional descriptor vector \( \mathbf{d}_i \). Typical ranges include \( R_s \in [0.5, 6.0]~\text{\AA} \), \( \eta \in [0.001, 10]~\text{\AA}^{-2} \), and \( \zeta \in [1, 10] \). The radial and angular channels produce tens to hundreds of features, depending on grid resolution.

\paragraph*{ANI}
The ANI (Accurate NeurAl networK engINe for Molecular Energies, or briefly ANI) framework builds upon the BPNN architecture by introducing \gls*{aevs} as input descriptors~\cite{smith2017ani}. While retaining the decomposition of total energy into atomic contributions via species-specific NNs, ANI replaces ACSFs with a modified variant tailored for extensibility and transferability across diverse chemical environments.
The radial part of AEVs uses Gaussian functions centered at \( R_s \) with width \( \eta \), multiplied by a cutoff function
\begin{equation}
G_{i}^{\mathrm{rad.}} = \sum_{j \in s} e^{-\eta (R_{ij} - R_s)^2} f_c(R_{ij}),
\end{equation}
similar in form to ACSFs but parameterized with a fixed \( \eta \) and a dense grid of \( R_s \) values to ensure smooth feature resolution. The angular AEVs incorporate angle shifts \( \theta_s \), radial shifts \( R_s \), and exponent tuning via \( \zeta \), allowing selective probing of angular regions
\begin{eqnarray}
G_{i}^{\mathrm{ang}} &=& \sum_{j,k} \left(1 + \cos(\theta_{ijk} - \theta_s)\right)^\zeta  f_c(R_{ij}) f_c(R_{ik}) \times \nonumber  \\
&&e^{-\eta \left( \frac{R_{ij} + R_{ik}}{2} - R_s \right)^2}.
\end{eqnarray}
Unlike ACSFs, which aggregate neighbor contributions more generically, AEVs are explicitly structured to resolve radial and angular interactions by atom and bond types, resulting in a fixed-length, chemically discriminating feature vector optimized for transferability and GPU efficiency.
ANI models (e.g., ANI-1, ANI-1x, ANI-2x) are trained on extensive datasets constructed from normal-mode sampled conformations of small organic molecules (e.g., from GDB-11 with up to 8 heavy atoms), using DFT energies ($\omega$B97X/6-31G($d$)) as reference. 
ANI-1ccx~\cite{smith2020ani} is an NN-MLP trained on a diverse subset of molecular geometries with coupled-cluster (CCSD(T)/CBS) reference data, designed to achieve high accuracy in quantum property prediction for organic molecules.

\subsubsection*{Electrostatics-aware descriptor-based models: CENT, QeqNN, and QRNN}

\textbf{CENT:}
The \gls*{cent}~\cite{ghasemi2015interatomic} extends descriptor-based MLPs by predicting electronegativities \( \chi_i \) and computing charges \( q_i \) via a global charge equilibration (Qeq) scheme. The total energy is
\begin{equation}
U_{\text{tot}} = \sum_i \left( E_i^0 + \chi_i q_i + \frac{1}{2} J_{ii} q_i^2 \right) + \frac{1}{2} \iint \frac{\rho(\mathbf{r}) \rho(\mathbf{r'})}{|\mathbf{r} - \mathbf{r'}|} d\mathbf{r} d\mathbf{r'},
\end{equation}
with \( \rho_i(\mathbf{r}) = q_i \left( \frac{1}{\alpha_i^3 \pi^{3/2}} \right) \exp\left( -\frac{|\mathbf{r} - \mathbf{r}_i|^2}{\alpha_i^2} \right) \). Charges are obtained by solving
\begin{equation}
\sum_j A_{ij} q_j + \chi_i = 0, \quad
A_{ij} =
\begin{cases}
J_{ii} + \frac{2\gamma_{ii}}{\sqrt{\pi}}, & i = j, \\
\frac{\text{erf}(\gamma_{ij} r_{ij})}{r_{ij}}, & i \neq j,
\end{cases}
\label{eq:qeq}
\end{equation}
where \( \gamma_{ij} = 1/\sqrt{\alpha_i^2 + \alpha_j^2} \). CENT uses ACSF-like descriptors and a compact 51-3-3-1 neural network to predict \( \chi_i \), achieving high accuracy for neutral and ionized NaCl clusters.

\noindent 
\textbf{QeqNN:}
QeqNN~\cite{ko2021fourth} enhances CENT by adding a short-range atomic energy model to the global electrostatics. It uses a similar Qeq procedure (as in Eq.~\ref{eq:qeq}) but includes charges in the atomic energy network
\begin{align}
E_{\text{Qeq}} &= E_{\text{elec}} + \sum_i \left( \chi_i q_i + \frac{1}{2} J_i q_i^2 \right), \\
E_{\text{elec}} &= \sum_{i<j} \frac{\text{erf}(r_{ij}/\sqrt{2}\gamma_{ij})}{r_{ij}} q_i q_j + \sum_i \frac{q_i^2}{2\sigma_i \sqrt{\pi}}, \\
E_{\text{total}} &= E_{\text{elec}} + \sum_i E_i(\mathbf{G}_i, q_i).
\end{align}
Here, \( \mathbf{G}_i \) represents an ACSF-style descriptor encoding the local environment of atom \( i \), used for both the electronegativity prediction and the short-range atomic energy model. 

\noindent 
\textbf{QRNN:}
QRNN~\cite{jacobson2022transferable} removes the Qeq solve step (as in Eq.~\ref{eq:qeq}) by using a learned recursive update for the charges
\begin{equation}
q_i^{(l+1)} = f_{\rm update}(q_i^{(l)}, \boldsymbol{h}_i, \boldsymbol{R}_i),
\end{equation}
where \( f_{\rm update} \) is a shared \gls*{nn}. After a few iterations, 
the final charges \( q_i^{(L-1)} \) are used in the energy prediction
\begin{equation}
E_{\text{total}} = \sum_i E_i(\mathbf{G}_i, q_i^{(L-1)}).
\end{equation}
Unlike QeqNN, QRNN uses AEVs as input descriptors, following the ANI framework.

\textbf{CENT $\rightarrow$ QeqNN $\rightarrow$ QRNN:}
CENT introduced neural charge prediction for long-range electrostatics; QeqNN combined it with short-range energy learning; QRNN further enhanced efficiency via learned recursive charge updates. Together, these models bring charge-aware, physics-informed predictions into the descriptor-based NN-MLP framework.

These Desc-NN-MLPs exemplify a growing class~\cite{song2024charge,kocer2025iterative} of \textit{hybrid} models that embed physical priors (such as electrostatics, polarization, and global charge conservation) into trainable architectures. Unlike traditional descriptor-based models that rely solely on local environments, these models achieve enhanced generalization and accuracy by coupling local learning with physically grounded long-range interactions. 

As such, they bridge the gap between hand-designed physical models and fully end-to-end learned representations, while underscoring accurate treatment of charge transfer, polarization, and long-range electrostatics that remain a persistent challenge, even in descriptor-free, end-to-end architectures.

\subsubsection{MP-LNN-MLPs: End-to-end NN-MLPs with local message passing}

Unlike traditional Desc-NN-MLPs, end-to-end architectures with local message passing (MP-LNN-MLPs) construct atomic representations directly from Cartesian coordinates and atomic numbers, without relying on pre-defined descriptors. These models operate in a purely data-driven fashion, encoding the local chemical environment of each atom using relative atomic positions within a cutoff radius. Interactions are typically modeled through distance-sensitive filters, without explicit graph connectivity or attention-based pooling.

A key architectural feature of several models in this category, such as deep potential molecular dynamics (referred to as DPMD or DeePMD)~\cite{zhang2018deep}, is the use of a \textit{local coordinate frame}. For each atom, a symmetry-preserving transformation maps its neighbors into a canonical local reference frame, enabling the model to maintain rotational, translational, and permutational invariance while learning environment-sensitive features. Other models, such as HIPNN~\cite{lubbers2018hierarchical,chigaev2023lightweight}, achieve similar invariance using hierarchical message passing schemes built on radial basis expansions and smooth cutoff functions.

These models are particularly well-suited for large-scale MD simulations, where locality and linear scaling are critical. Although they typically lack explicit modeling of long-range electrostatics or directional information, such effects can be implicitly learned from data, making MP-LNN-MLPs competitive surrogates for {\it ab initio} MD in solvated or condensed-phase systems.

In models such as DeePMD and HIPNN, the atomic feature vectors \( \mathbf{x}_i \) are learnable latent embeddings that are updated during training. This contrasts with descriptor-based models like BPNN or ANI, where static descriptors \( \mathbf{d}_i \), computed from local geometry (e.g., ACSFs), serve as direct input.

The feature vector \( \mathbf{x}_i \), commonly used in end-to-end models, corresponds to the output of the final hidden layer (see Eq.~\ref{eq:deepactivation}) in the DeepNN architecture
\begin{equation}
\mathbf{x}_i = \mathbf{h}_i^{(L-1)}.
\end{equation}
The final atomic energy is then computed analogously to Eq.~\ref{eq:deenatomicenergy}
\begin{equation}
E_i = (\mathbf{w}^{(L)})^\top \mathbf{x}_i + b^{(L)},
\label{eq:deepenergyfeature}
\end{equation}
where \( \mathbf{w}^{(L)} \in \mathbb{R}^{N_{L-1}} \) and \( b^{(L)} \in \mathbb{R} \) are the output-layer weights and biases, respectively.

\paragraph*{DeePMD: Deep potential molecular dynamics}
DeePMD~\cite{zhang2018deep} is designed to perform AIMD-quality simulations at reduced computational cost. Rather than relying on hand-crafted descriptors, DeePMD learns atomic energies \( E_i \) directly from Cartesian coordinates via symmetry-preserving coordinate transformations and DeepNNs.

Each atom's environment is encoded in a local coordinate frame, using scaled relative positions of neighbors within a cutoff \( R_c \). 
The input feature vectors, $\lbrace \mathbf{x}_i \rbrace$, are based on the inverse interatomic distances
\( \{ 1/r_{ij} \} \)  
to preserve translational, rotational, and permutational symmetries. 
These features are processed by a DeepNN (typically with five hidden layers, e.g., 240--120--60--30--10 neurons) to yield atomic contributions, $E_i$, to the total energy.

The network is trained using a composite loss function involving energy, force, and stress (virial) errors
\begin{equation}
\mathcal{L}  = w_E \mathcal{L}_{\rm energy} +  w_F \mathcal{L}_{\rm force} +  w_s \mathcal{L}_{\rm stress}, 
\end{equation}
where the stress term is essential for accurate modeling of periodic systems under pressure or strain, it is typically omitted for isolated molecules.

DeePMD has demonstrated high fidelity across both periodic (e.g., liquid water, ice) and finite systems (e.g., benzene, aspirin), accurately reproducing AIMD-level energies, forces, and structural observables such as radial and angular distribution functions. The method scales linearly with system size and enables long-timescale simulations of large systems.

Most implementations employ the DeepPot-SE (Smooth Edition) descriptor~\cite{zhang2018end}, which enhances differentiability and smoothness of the local environment representation. DeePMD has been widely used in explicit solvation studies, including simulations of bulk water, electrolyte solutions, and catalytic metal–water interfaces. It also underpins large-scale active learning workflows such as DP-GEN~\cite{zhang2020dp}, and offers a balance of scalability, physical fidelity, and AIMD-level accuracy well-suited to modeling condensed-phase systems.

\begin{figure*}[ht]
\centering
\includegraphics[width=\linewidth,angle=0]{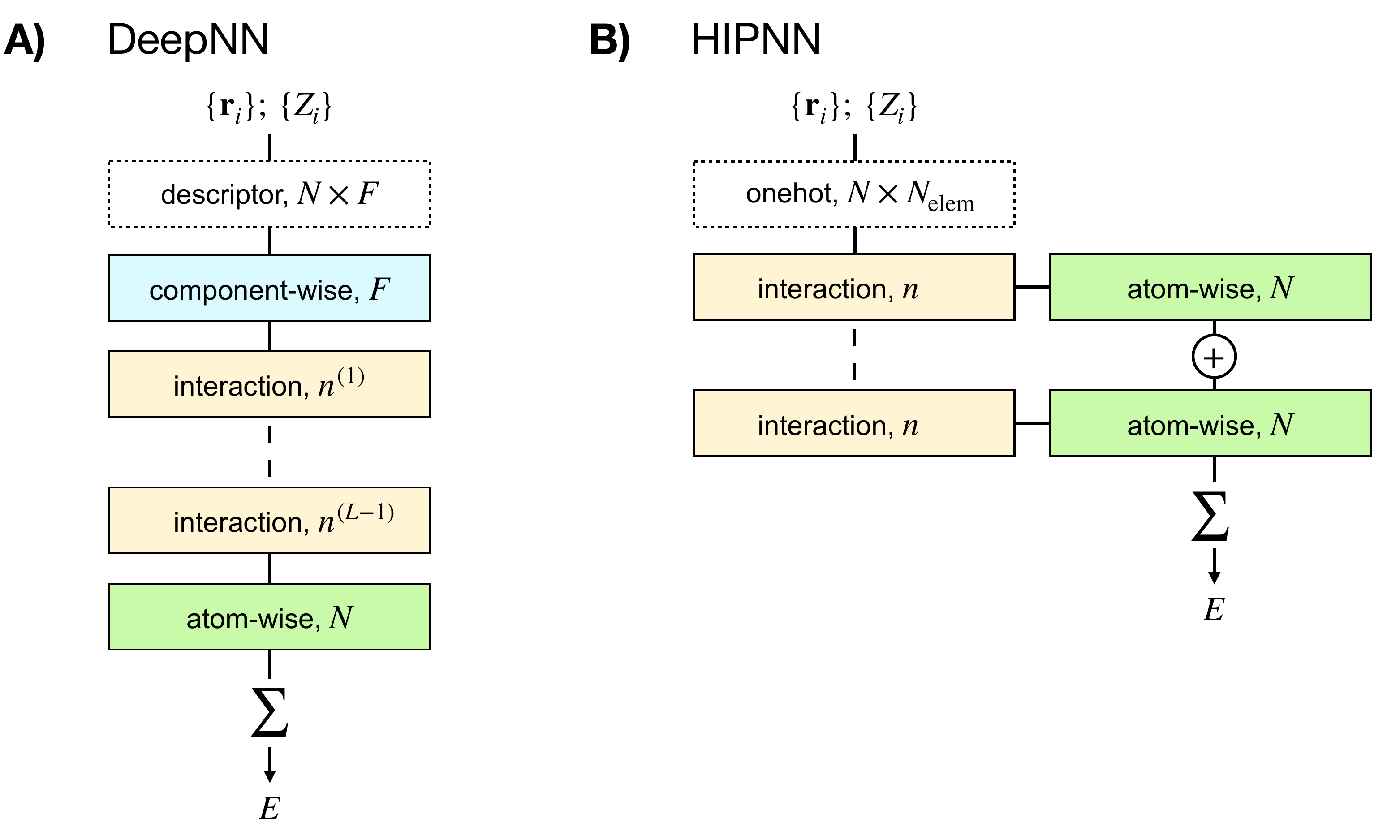}
\caption{
Schematic comparison of descriptor-based deep neural network architectures.  
\textbf{a)} A standard DeepNN model (e.g., BPNN), where each atom’s descriptor is processed through shared layers and a single atom-wise output to yield the total energy via sum pooling.  
\textbf{b)} The HIPNN architecture, which introduces multiple readout layers at selected interaction layers. Each readout produces an atom-wise partial energy contribution, and their sum defines the final atomic energies used to compute the total energy via sum pooling. Boxes with dashed lines denote input descriptors or initialized embeddings, with their dimensionality. Boxes with solid lines indicate shallow or fully connected neural layers, with the number of neurons or output dimensionality indicated.
}
\label{fig:DeepNN_HIPNN}
\end{figure*}

\paragraph*{HIPNN: Hierarchically interacting particle neural network}

HIPNN~\cite{lubbers2018hierarchical,chigaev2023lightweight} is designed to predict molecular energies and forces using a hierarchical architecture that captures many-body correlations without relying on explicit graph representations. It operates directly on atomic coordinates and chemical identities, making it part of the MP-LNN-MLP family.

Instead of constructing molecular graphs, HIPNN defines interatomic interactions via smooth, distance-based filters. 
Each atom is assigned a learnable feature vector \( {\bf x}_i \), which is iteratively refined through interaction layers. These layers aggregate information from neighboring atoms within a cutoff radius using sensitivity functions such as a Gaussian in inverse distance
\begin{equation}
s_\nu(r_{ij}) = \exp\left( -\frac{(\frac{1}{r_{ij}} - \frac{1}{\mu_\nu})^2}{2 \sigma_\nu^{-2}} \right) \phi_{\text{cut}}(r_{ij}),
\end{equation}
with
\begin{equation}
\phi_{\text{cut}}(r_{ij}) = 
\begin{cases}
\frac{1}{2} \left[ \cos\left( \frac{1}{2} \frac{r_{ij}}{R_c} \right) \right]^2, & r_{ij} \leq R_c, \\
0, & r_{ij} > R_c,
\end{cases}
\end{equation}
where \( r_{ij} \) is the pairwise interatomic distance, and \( \mu_\nu, \sigma_\nu \) are learnable parameters.

Both HIPNN and DeePMD model the total molecular energy as a sum over atom-wise contributions, but differ in how these atomic energies are constructed.
In DeePMD, each atomic energy is predicted from a single learned feature vector \( \mathbf{x}_i \) using a single output layer as in Eq.~\ref{eq:deepenergyfeature}.
In contrast, HIPNN decomposes the atomic energy into a {\it hierarchy of contributions} from multiple intermediate layers of the network, capturing different levels of interaction complexity
\begin{equation}
E_i^{(\ell_n)} = \mathbf{w}^{(n)} \cdot \mathbf{x}_i^{(\ell_n)} + b^{(n)}; 
\, 
E_i^{\text{HIPNN}} = \sum_n E_i^{(\ell_n)}.
\end{equation}
Here, \( \mathbf{x}_i^{(\ell_n)} \) denotes the learned representation of atom \( i \) from a selected hidden layer \( \ell_n \), and the corresponding partial energy contribution \( E_i^{(\ell_n)} \) reflects interactions of increasing complexity (e.g., 2-body, 3-body, etc.). In HIPNN, only a subset of hidden layers, typically chosen at regular intervals (e.g., layers 1, 3, 5), are used for energy prediction. These layers are manually designated to capture progressively higher-order interactions, and an independent readout head is positioned at the end of each layer. Architecturally, these readouts are placed like \textit{T-junctions}, branching off from the main hidden layer stack to compute partial atomic energy contributions. This hierarchical summation enhances both accuracy and interpretability by explicitly modeling many-body interactions at multiple levels of resolution. The architecture of HIPNN is compared with that of a
DeepNN in Figure~\ref{fig:DeepNN_HIPNN}.

\subsubsection{End-to-End NN-MLPs with Graph-Based Message Passing (MP-GNN-MLPs)\label{sec:MP-GNN-MLPs}}

Unlike MP-LNN-MLPs, which use local, coordinate-based neighborhoods, MP-GNN-MLPs represent molecules as graphs with atoms as nodes and interatomic relationships (e.g., distances, bonds, angles) as edges~\cite{wang2023graph,reiser2022graph}. These models update atomic or edge features iteratively through trainable message-passing layers. A unified framework for such models was introduced by Gilmer {\it et al.}~\cite{gilmer2017neural}, who formulated \gls*{mpnns} as a general class of graph-based neural architectures for molecular property prediction. Their work established message passing as a foundational mechanism for end-to-end learning on molecular graphs, inspiring numerous subsequent architectures covered in this section.

Each layer of a message-passing neural network updates atomic features via local interactions. A general formulation applicable to the fourteen GNN-MLPs discussed in Section~\ref{sec:MP-GNN-MLPs} is
\begin{equation}
\mathbf{h}_i^{(l+1)} = f_{\mathrm{update}}\left(
\mathbf{h}_i^{(l)},\ 
\sum_{j \in \mathcal{N}(i)} 
f_{\mathrm{msg}}\left( \mathbf{S}_{i\mathcal{N}(i)}^{(l)} \right)
\right),
\label{eq:mp_update_compact}
\end{equation}
where \( \mathbf{S}_{i\mathcal{N}(i)}^{(l)} \) is the \textit{neighbor-informed local structural message} ({N-msg}), a shorthand we introduce in this work to denote the structural and feature context required to compute messages from neighbors \( j \in \mathcal{N}(i) \). To the best of our knowledge, this terminology is not standard in existing 
literature, but we introduce it here as we attempt to an architectural comparison across various models. 

Depending on the specific model, \( \mathbf{S}_{i\mathcal{N}(i)}^{(l)} \) may include:
\begin{itemize}
  \item Node features \( \mathbf{h}_i^{(l)} \),
  \item Edge features \( \mathbf{e}_{ij} \), such as distances, angles, or learned filters,
  \item Relative positions \( \mathbf{r}_{ij} \), angular descriptors \( \theta_{ijk} \), or higher-order geometric terms,
  \item Basis function expansions (e.g., radial, spherical, or tensor products).
\end{itemize}
This abstraction accommodates a wide range of architectures, ranging from pairwise models (e.g., SchNet) to tensor-valued equivariant models (e.g., NequIP, MACE).

MP-GNN-MLPs fall into two categories based on symmetry treatment. \textbf{Invariant models:} SchNet, MGCN, DimeNet, MGNN, PhysNet, GemNet, and SpookyNet. These ensure scalar outputs remain unchanged under rotation, translation, and permutation. \textbf{Equivariant models:} PaiNN, ENINet, NequIP, MACE, Allegro, SO3krates, and ViSNet. These enforce $SE(3)$, $O(3)$, or $E(3)$ equivariance, so that tensorial outputs (e.g., forces, dipoles) rotate consistently with the input.

In all models, atomic energy contributions are typically obtained via a shared readout function \( \mathcal{R} \), using one of three strategies as:

\begin{enumerate}
    \item \textbf{Final-layer atomic readout:} 
    SchNet, MGCN, DimeNet, MGNN, PhysNet, GemNet, PaiNN, ENINet, SpookyNet, SO3krates, and ViSNet
    \begin{equation}
    E_i = \mathcal{R}(\mathbf{h}_i^{(L)}), \quad E_{\mathrm{tot}} = \sum_i E_i.
    \label{eq:readout_final}
    \end{equation}
    
    \item \textbf{Layerwise aggregation:} 
    NequIP and MACE
    \begin{equation}
    E_i = \sum_{l=1}^{L} \mathcal{R}^{(l)}(\mathbf{h}_i^{(l)}), \quad
    E_{\mathrm{tot}} = \sum_i E_i.
    \label{eq:readout_layerwise}
    \end{equation}
    
    \item \textbf{Final-layer edge readout:} 
    Allegro predicts energy from the final-layer edge features
    \begin{equation}
    E_{ij} = \mathcal{R}(x_{ij}^{(L)}), \quad
    E_{\mathrm{tot}} = \sum_{i,j \in \mathcal{N}(i)} \sigma_{Z_i,Z_j} E_{ij}.
    \end{equation}
\end{enumerate}

SchNet, MGCN, DimeNet, GemNet, and Allegro support scalar-only prediction, and the forces are obtained
as gradients. PaiNN, PhysNet, SpookyNet, ENINet, SO3krates, ViSNet NequIP, and MACE support direct
vector/tensor prediction. MGNN supports post hoc vector prediction. For all models, forces are computed as analytic gradients of the total energy as per Eq.~\ref{eq:force_energybased_ML}.

Table~\ref{tab:mpgnn-summary-readout} summarizes key architectural properties, including symmetry type, readout method, and ability to predict vector/tensor observables directly (beyond forces).

\begin{table}[!htpb]
\centering
\caption{Summary of readout type, symmetry treatment (invariance/equivariance under Euclidean group operations), and vector/tensor prediction capability for MP-GNN-MLPs.
Vec./Tens. denotes whether the model supports direct prediction of vector/tensor quantities via equivariant readouts. All models compute forces as energy gradients.
}
\label{tab:mpgnn-summary-readout}
\begin{tabular}{@{}llll@{}}
\hline
Model & Readout & Symmetry & Vec./Tens. \\
\hline
SchNet      & Final-atom    & Invariant       & No \\
MGCN        & Final-atom    & Invariant       & No \\
DimeNet     & Final-atom    & Invariant       & No \\
MGNN        & Final-atom    & Invariant       & Optional\\
PhysNet     & Final-atom    & Invariant       & Yes \\
GemNet      & Final-atom    & Invariant       & No \\
PaiNN       & Final-atom    & $SE(3)$         & Yes \\
ENINet      & Final-atom    & $O(3)$          & Yes \\
SpookyNet   & Final-atom    & Hybrid          & Yes \\
NequIP      & Layer-atom    & $E(3)$          & Yes \\
MACE        & Layer-atom    & $E(3)$          & Yes \\
Allegro     & Final-edge    & $E(3)$          & No \\
SO3krates   & Final-atom    & $SO(3)$         & Yes \\
ViSNet      & Final-atom    & $SO(3)$         & Yes \\
\hline
\end{tabular}
\end{table}

\begin{figure*}[ht]
\centering
\includegraphics[width=\linewidth,angle=0]{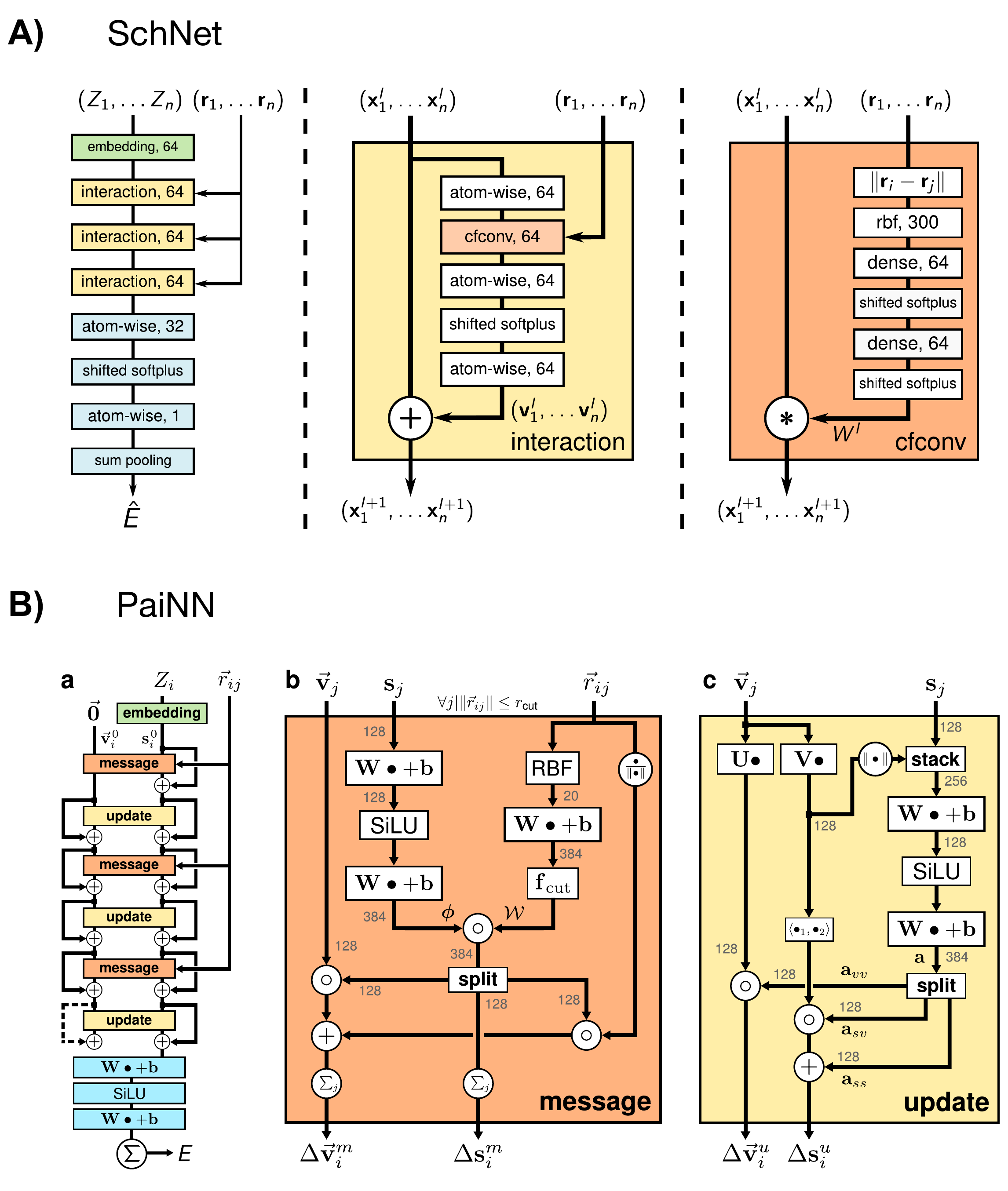}
\caption{
Schematic description of representative modern message-passing GNN architectures: 
A) SchNet architecture, reused from the arXiv preprint of \RRef{schutt2018schnet}. 
{\it The original figure is distributed under arXiv’s non-exclusive license to distribute}.
B) PaiNN architecture, reused from the arXiv preprint of \RRef{schutt2021equivariant}. 
{\it The original figure is distributed under arXiv’s non-exclusive license to distribute}.
}
\label{fig:schnet_painn}
\end{figure*}

A detailed treatment of symmetry-aware message passing, including irreducible representations, tensor product couplings, and group theory foundations, is provided in Section~\ref{symmMPNN}. 
Comparative architectural diagrams are shown in 
Fig.~\ref{fig:DeepNN_HIPNN} (MP-LNN-MLPs)
and
Fig.~\ref{fig:schnet_painn} (MP-GNN-MLPs), 
illustrating the growing architectural complexity in modern GNN models as they incorporate angular resolution, equivariance, and long-range attention mechanisms.

\paragraph*{SchNet: Continuous-filter convolutional neural network}

SchNet~\cite{schutt2018schnet} is a rotationally and translationally invariant MP-GNN designed to predict scalar molecular properties such as potential energy surfaces (PESs), atomic forces, and dipole moments. The model extends the
deep tensor neural network (DTNN) approach~\cite{schutt2017quantum} by operating directly on atomic numbers \( \{Z_i\} \) and coordinates \( \{\mathbf{r}_i\} \), using continuous-filter convolutional (cfconv) layers to represent interatomic interactions in a fully differentiable and symmetry-aware framework.
Although the SchNet study~\cite{schutt2018schnet} does not explicitly classify the model as an
MPNN, its architecture (based on continuous-filter convolutions) naturally implements message and update functions, making it fully compatible with the MP-GNN-MLP framework.

Each atom \( i \) is initialized with a learned embedding \( \mathbf{h}_i^{(0)} = \mathbf{a}_{Z_i} \in \mathbb{R}^F \). Pairwise distances \( r_{ij} = \|\mathbf{r}_j - \mathbf{r}_i\| \) are expanded using radial basis functions
\begin{equation}
\phi_k(r_{ij}) = \exp\left[-\gamma (r_{ij} - \mu_k)^2\right], \quad k = 1, \dots, K.
\end{equation}

Over \( L \) layers, atomic features are updated using the general message-passing form of Eq.~\ref{eq:mp_update_compact}, with the neighbor-informed local structural message (N-msg) defined as
\begin{equation}
\mathbf{S}_{i\mathcal{N}(i)}^{(l)} = \left\{ \mathbf{h}_j^{(l)} \circ \mathbf{W}_{ij}^{(l)} \right\}_{j \in \mathcal{N}(i)}, 
\quad \mathbf{W}_{ij}^{(l)} = \text{Dense}^{(l)}\left( \phi(r_{ij}) \right),
\end{equation}
where \( \circ \) denotes elementwise multiplication. This yields a smooth and differentiable interaction kernel for each pair \( (i,j) \), forming the basis of SchNet’s continuous-filter convolution.

A schematic diagram of the SchNet architecture is provided in Figure~\ref{fig:schnet_painn}.

\paragraph*{MGCN: Multilevel graph convolutional network}

MGCN~\cite{lu2019molecular} is a rotationally invariant MP-GNN that jointly updates atom (node) and bond (edge) features over multiple message-passing layers. Unlike conventional GNNs that rely solely on node-level updates, MGCN incorporates bond-level embeddings and geometric distance features to enrich structural representation.

Each atom \( i \) and edge \( (i, j) \) is initialized with learnable embeddings \( \mathbf{h}_i^{(0)} \in \mathbb{R}^F \) and \( \mathbf{e}_{ij}^{(0)} \in \mathbb{R}^E \), respectively. Interatomic distances are encoded using a radial basis expansion \( \phi(r_{ij}) \) as in SchNet.

Over \( L \) layers, MGCN alternates between updating atom and edge features using the general message-passing structure of Eq.~\ref{eq:mp_update_compact}. The neighbor-informed local structural message (N-msg) for atomic updates is defined as
\begin{equation}
\mathbf{S}_{i\mathcal{N}(i)}^{(l)} = \left\{ f_{\mathrm{msg}}^{\mathrm{node}}(\mathbf{h}_j^{(l)}, \mathbf{e}_{ij}^{(l)}, \phi(r_{ij})) \right\}_{j \in \mathcal{N}(i)}.
\end{equation}

Simultaneously, edge features are updated using atom features at the same layer
\begin{equation}
\mathbf{e}_{ij}^{(l+1)} = f_{\mathrm{msg}}^{\mathrm{edge}}(\mathbf{h}_i^{(l)}, \mathbf{h}_j^{(l)}, \mathbf{e}_{ij}^{(l)}).
\end{equation}

\paragraph*{DimeNet: Directional message passing neural network}

DimeNet~\cite{gasteiger2020directional} is a rotationally invariant MP-GNN that incorporates angular correlations through directional triplets \( (k \rightarrow j \rightarrow i) \). Unlike pairwise GNNs, DimeNet models three-body geometry by expanding bond angles \( \theta_{kji} \) and distances \( r_{ji} \) into radial and spherical basis functions.

Each directed edge \( (j \rightarrow i) \) is initialized with a message embedding \( \mathbf{m}_{ji}^{(0)} \), and angle-dependent context is encoded as
\begin{equation}
\mathbf{S}_{kji}^{(l)} = \left( \mathbf{m}_{kj}^{(l)}, \phi(r_{ji}), \psi(\theta_{kji}) \right),
\end{equation}
where \( \phi(r) \) and \( \psi(\theta) \) are radial and spherical basis functions, respectively.

Message passing proceeds over \( L \) layers. At each step, edge messages are updated using triplet interactions
\begin{equation}
\mathbf{m}_{ji}^{(l+1)} = f_{\mathrm{update}}\left( 
\mathbf{m}_{ji}^{(l)}, 
\sum_{\substack{k \in \mathcal{N}(j) \\ k \ne i}} f_{\mathrm{int}}(\mathbf{S}_{kji}^{(l)})
\right),
\end{equation}
followed by atomic updates
\begin{equation}
\mathbf{h}_i^{(l+1)} = f_{\mathrm{update}}\left( 
\mathbf{h}_i^{(l)}, 
\sum_{j \in \mathcal{N}(i)} \mathbf{m}_{ji}^{(l)} 
\right).
\end{equation}

DimeNet++~\cite{gasteiger2020fast} introduces architectural improvements, including Hadamard-based message interactions and reduced network depth via hierarchical projections.

Directionality in DimeNet is learned through basis-expanded angles, but the model remains rotationally invariant and does not support direct prediction of vector or tensor properties.

\paragraph*{MGNN: Moment graph neural network}

MGNN~\cite{chang2025mgnn} is a rotationally invariant MP-GNN that captures angular sensitivity using scalar-valued moment contractions over atomic triplets. Inspired by moment tensor potentials, it avoids explicit tensor propagation by encoding three-body geometry through moment scalars.

Each atom \( i \) is initialized with an embedding \( \mathbf{h}_i^{(0)} = \mathbf{a}_{Z_i} \in \mathbb{R}^F \), and pairwise distances \( r_{ij} = \|\mathbf{r}_j - \mathbf{r}_i\| \) are expanded using Chebyshev radial basis functions to generate edge features \( \mathbf{e}_{ij}^{(0)} \).

Over \( L \) layers, atomic features are updated using moment-informed messages defined over triplets \( (j, i, k) \). The neighbor-informed local structural message (N-msg) is
\begin{equation}
\mathbf{S}_{i\mathcal{T}(i)}^{(l)} = \left\{ 
f_{\mathrm{msg}}\left(
\mathbf{e}_{ij}^{(l)}, \mathbf{e}_{ik}^{(l)}, 
M^{(1)}_{jik}, M^{(2)}_{jik}
\right)
\right\}_{(j,k) \in \mathcal{T}(i)},
\end{equation}
where the moment contractions are
\begin{align}
M^{(1)}_{{jik}} &= \hat{\mathbf{r}}_{ij} \cdot \hat{\mathbf{r}}_{ik}, \\
M^{(2)}_{{jik}} &= (\hat{\mathbf{r}}_{ij} \otimes \hat{\mathbf{r}}_{ij}) : (\hat{\mathbf{r}}_{ik} \otimes \hat{\mathbf{r}}_{ik}),
\end{align}
which encode cosine and cosine-squared dependencies on triplet angles. Here, \( \hat{\mathbf{r}}_{ij} \) is the normalized direction vector and \( : \) denotes the double-dot product.

The atomic update follows Eq.~\ref{eq:mp_update_compact}, using messages derived from the above N-msg. MGNN supports optional vectorial outputs via post hoc projections of final-layer embeddings \( \mathbf{h}_i^{(L)} \), modulated by position vectors \( \mathbf{r}_i \).

\paragraph*{PhysNet: Physically informed neural network}

PhysNet~\cite{unke2019physnet} is a rotationally invariant MP-GNN that augments the SchNet framework with physics-motivated enhancements, including long-range electrostatics, dispersion corrections, and multitask supervision over energy, forces, dipoles, and partial charges.

Each atom \( i \) is initialized with an embedding \( \mathbf{h}_i^{(0)} = \mathbf{a}_{Z_i} \in \mathbb{R}^F \). Pairwise distances \( r_{ij} \) are expanded using a Gaussian radial basis \( \phi(r_{ij}) \), and the resulting features are passed through learned filters
\begin{equation}
\mathbf{W}_{ij}^{(l)} = \text{Dense}^{(l)}(\phi(r_{ij})).
\end{equation}

Over \( T \) layers, atomic embeddings are updated following Eq.~\ref{eq:mp_update_compact}, with the neighbor-informed local structural message (N-msg) defined as
\begin{equation}
\mathbf{S}_{i\mathcal{N}(i)}^{(l)} = \left\{ f_{\mathrm{msg}}(\mathbf{h}_j^{(l)}, \mathbf{W}_{ij}^{(l)}) \right\}_{j \in \mathcal{N}(i)}.
\end{equation}

After message passing, atomic energies are predicted via a shared readout \( \mathcal{R} \), and the total energy is augmented with physics-based terms
\begin{equation}
E_{\mathrm{tot}} = \sum_i \mathcal{R}(\mathbf{h}_i^{(T)}) + E_{\mathrm{elec}} + E_{\mathrm{disp}}.
\end{equation}

Electrostatic energy is computed from corrected partial charges \( \tilde{q}_i \) using a damped Coulomb kernel
\begin{align}
\tilde{q}_i &= q_i - \frac{1}{N} \left( \sum_j q_j - Q_{\text{tot}} \right), \\
E_{\mathrm{elec}} &= \sum_{i<j} \tilde{q}_i \tilde{q}_j \chi(r_{ij}), \\
\boldsymbol{\mu} &= \sum_i \tilde{q}_i \mathbf{r}_i.
\end{align}

PhysNet is trained using a composite loss over energy, forces, and dipole moment
\begin{equation}
\mathcal{L} = w_E \, \mathcal{L}_E + w_F \, \mathcal{L}_F + w_\mu \, \mathcal{L}_\mu + w_{\mathrm{nh}} \, \mathcal{L}_{\mathrm{nh}},
\end{equation}
where \( \mathcal{L}_{\mathrm{nh}} \) regularizes layerwise feature decay. The partial charges \( \tilde{q}_i \) are not trained directly but are inferred implicitly via their influence on \( E_{\mathrm{elec}} \) and \( \boldsymbol{\mu} \).

\paragraph*{GemNet: Geometric message passing neural network}

GemNet~\cite{gasteiger2021gemnet} is a rotationally consistent MP-GNN that extends DimeNet++ by incorporating two-hop message passing over quadruplets and richer geometric encodings such as dihedral angles. It enables scalar energy and force prediction using directional edge-based embeddings, without requiring full tensor equivariance.

Each atom pair \( (i,j) \) is represented by an edge embedding \( \mathbf{m}_{ij}^{(0)} \in \mathbb{R}^F \), initialized using a learned embedding of atomic types \( \{Z_i\} \) and relative geometry. Interatomic distances and angular features (bond and dihedral angles) are encoded using radial, circular, and spherical basis functions.

Over \( L \) layers, directional edge features are updated via a two-hop message-passing scheme. The neighbor-informed local structural message (N-msg) over quadruplets \( (c \rightarrow a, d \rightarrow b) \) is given by
\begin{equation}
\mathbf{S}_{ca\mathcal{N}(a)}^{(l)} = \left\{ 
f_{\mathrm{msg}}\left( \mathbf{m}_{db}^{(l)}, \mathbf{e}_{db}, \mathbf{e}_{ab}, \mathbf{e}_{ca}, \theta_{cabd} \right) 
\right\}_{b,d},
\end{equation}
where \( \mathbf{e}_{ij} \) are geometric basis encodings, and \( \theta_{cabd} \) is the dihedral angle across the quadruplet. Message updates follow
\begin{equation}
\mathbf{m}_{ca}^{(l+1)} = f_{\mathrm{update}}(\mathbf{m}_{ca}^{(l)}, \mathbf{S}_{ca\mathcal{N}(a)}^{(l)}).
\end{equation}
Final atomic features are constructed by aggregating incoming edge messages
\begin{equation}
\mathbf{h}_i^{(L)} = \sum_{j \in \mathcal{N}(i)} \mathbf{m}_{ji}^{(L)}.
\end{equation}

Energy is predicted via an atomwise readout \( \mathcal{R} \), and forces are computed from energy gradients. Training uses a combined energy–force loss. Variants include:
{GemNet-Q}: full model with quadruplet message passing,
{GemNet-T}: triplet-only variant without quadruplet interactions,
{GemNet-dQ/dT}: direct-force versions trained without an energy functional.

GemNet enforces \emph{rotational consistency} through directional geometry encodings, even though it does not maintain strict $SO(3)$ or $E(3)$ equivariance.

\paragraph*{PaiNN: Polarizable atom interaction neural network}

PaiNN~\cite{schutt2021equivariant} is an $SE(3)$-equivariant MP-GNN that predicts scalar and tensorial molecular properties by propagating coupled scalar and vector features. It extends the design of SchNet by incorporating directional information using continuous-filter convolutions and equivariant update blocks.

Each atom \( i \) is initialized with scalar features \( \mathbf{s}_i^{(0)} = \mathbf{a}_{Z_i} \in \mathbb{R}^F \) and vector features \( \mathbf{v}_i^{(0)} = \mathbf{0} \in \mathbb{R}^{F \times 3} \). Pairwise distances \( r_{ij} = \|\mathbf{r}_j - \mathbf{r}_i\| \) are expanded via radial basis functions \( \phi(r_{ij}) \), and directionality is encoded by normalized relative vectors \( \hat{\mathbf{r}}_{ij} \).

At each layer, scalar and vector features are updated using equivariant message-passing operations. The neighbor-informed message (N-msg) for scalar and vector channels is defined by
\begin{align}
\mathbf{S}_{i\mathcal{N}(i)}^{(l, \mathrm{scal})} &= \left\{ f_{\mathrm{msg}}^{\mathrm{scal}}(\mathbf{s}_j^{(l)}, \mathbf{W}_{ij}) \right\}_{j \in \mathcal{N}(i)}, \\
\mathbf{S}_{i\mathcal{N}(i)}^{(l, \mathrm{vec})}  &= \left\{ f_{\mathrm{msg}}^{\mathrm{vec}}(\mathbf{s}_j^{(l)}, \hat{\mathbf{r}}_{ij}, \mathbf{W}_{ij}) \right\}_{j \in \mathcal{N}(i)},
\end{align}
with learned filters \( \mathbf{W}_{ij} = \text{Dense}(\phi(r_{ij})) \).

The updates are then performed as
\begin{align}
\mathbf{s}_i^{(l+1)} &= f_{\mathrm{update}}^{\mathrm{scal}}(\mathbf{s}_i^{(l)}, \mathbf{v}_i^{(l)}, \mathbf{S}_{i\mathcal{N}(i)}^{(l, \mathrm{scal})}), \\
\mathbf{v}_i^{(l+1)} &= f_{\mathrm{update}}^{\mathrm{vec}}(\mathbf{v}_i^{(l)}, \mathbf{S}_{i\mathcal{N}(i)}^{(l, \mathrm{vec})}).
\end{align}

After \( L \) layers, atomic energies are predicted from the scalar features via a shared readout
\begin{equation}
E_i = \mathcal{R}(\mathbf{s}_i^{(L)}), \quad E_{\mathrm{tot}} = \sum_i E_i,
\end{equation}
and atomic forces are computed as analytic gradients.

PaiNN also supports tensorial property prediction. Dipole moments and polarizabilities are computed from scalar and vector features
\begin{align}
\boldsymbol{\mu} &= \sum_i \mathbf{v}_i + q_i \mathbf{r}_i, \\
\boldsymbol{\alpha} &= \sum_i \alpha_0(\mathbf{s}_i) \mathbf{I}_3 + \mathbf{v}_i \otimes \mathbf{r}_i + \mathbf{r}_i \otimes \mathbf{v}_i.
\end{align}

\paragraph*{ENINet: Equivariant N-body interaction network}

ENINet~\cite{mao2024molecule} is an $O(3)$-equivariant MP-GNN that encodes many-body geometric correlations using a line graph representation to model directional triplet interactions explicitly. It addresses the issue of message cancellation in equivariant GNNs with symmetric neighborhoods by leveraging explicit three-body angular context.

Each atom \( i \) is initialized with scalar and vector features \( \mathbf{h}_i^{(0)} = (\mathbf{h}_i^{\ell=0}, \mathbf{h}_i^{\ell=1}) \in \mathbb{R}^F \times \mathbb{R}^{F \times 3} \). Edges \( (i,j) \) and triplets \( (j \rightarrow i \rightarrow k) \in \mathcal{T}(i) \) are initialized with similarly structured features \( \mathbf{e}_{ij}^{(0)} = (\ell=0,1) \) and \( \mathbf{t}_{ji,ki}^{(0)} = (\ell=0,1) \).

Message passing proceeds hierarchically over \( L \) layers using directional neighborhood messages (N-msg). At each step:
\begin{enumerate}
    \item \textit{Triplet features} \( \mathbf{t}_{ji,ki}^{(l)} \) are computed from angular geometry (not shown explicitly).
    \item \textit{Edge updates} aggregate triplet features
    \begin{eqnarray}
    \mathbf{S}_{ij\mathcal{T}(i)}^{(l)} &=& \left\{ \mathbf{t}_{ji,ki}^{(l)} \right\}_{(j,k) \in \mathcal{T}(i)}, \nonumber \\
    \mathbf{e}_{ij}^{(l+1)} &=& f_{\mathrm{update}}^{\mathrm{edge}}(\mathbf{h}_i^{(l)}, \mathbf{e}_{ij}^{(l)}, \mathbf{S}_{ij\mathcal{T}(i)}^{(l)}).
    \end{eqnarray}
    \item \textit{Node updates} use edge-informed messages
    \begin{eqnarray}
    \mathbf{S}_{i\mathcal{N}(i)}^{(l)} &=& \left\{ \mathbf{e}_{ij}^{(l+1)} \right\}_{j \in \mathcal{N}(i)}, \nonumber \\
    \mathbf{h}_i^{(l+1)} &=& f_{\mathrm{update}}^{\mathrm{node}}(\mathbf{h}_i^{(l)}, \mathbf{S}_{i\mathcal{N}(i)}^{(l)}).
    \end{eqnarray}
\end{enumerate}

\paragraph*{SpookyNet: Physically informed neural network with nonlocal attention}

SpookyNet~\cite{unke2021spookynet} is a rotationally invariant MP-GNN that extends PhysNet by incorporating nonlocal attention and explicit conditioning on electronic state (charge \( Q \) and spin \( S \)). It is designed for systems with strong electronic delocalization, net charge, or spin effects.

Each atom \( i \) is initialized with a sum of embeddings for atomic number, charge, and spin
\begin{equation}
\mathbf{h}_i^{(0)} = \mathbf{a}_{Z_i} + \mathbf{a}_Q + \mathbf{a}_S,
\end{equation}
where \( \mathbf{a}_{Z_i} \in \mathbb{R}^F \), and the global \( \mathbf{a}_Q, \mathbf{a}_S \in \mathbb{R}^F \) are broadcast to all atoms via attention.

Over \( T \) layers, SpookyNet combines local geometric updates with nonlocal transformer-style attention. The neighbor-informed local structural message (N-msg) and update are given by
\begin{align}
\mathbf{S}_{i\mathcal{N}(i)}^{(t)} &= \left\{ f_{\mathrm{msg}}^{\mathrm{local}}(\mathbf{h}_j^{(t-1)}, \mathbf{r}_{ij}) \right\}_{j \in \mathcal{N}(i)}, \\
\mathbf{h}_i^{(t)} &= \mathbf{h}_i^{(t-1)} + f_{\mathrm{local}}(\mathbf{S}_{i\mathcal{N}(i)}^{(t)}) + f_{\mathrm{nonlocal}}(\{\mathbf{h}_j^{(t-1)}\}_{j=1}^N),
\end{align}
where \( f_{\mathrm{nonlocal}} \) is a global attention layer over all atoms.

After message passing, energy is predicted via a shared readout \( \mathcal{R} \) applied to \( \mathbf{h}_i^{(T)} \), and the total energy includes physical corrections
\begin{equation}
E_{\mathrm{tot}} = \sum_i \mathcal{R}(\mathbf{h}_i^{(T)}) + E_{\mathrm{rep}} + E_{\mathrm{elec}} + E_{\mathrm{vdW}}.
\end{equation}

Electrostatics are modeled with predicted partial charges \( q_i \), corrected to conserve total charge
\begin{align}
\tilde{q}_i &= q_i - \frac{1}{N} \left( \sum_j q_j - Q_{\text{tot}} \right), \\
E_{\mathrm{elec}} &= \sum_{i<j} \tilde{q}_i \tilde{q}_j \chi(r_{ij}), \\
\boldsymbol{\mu} &= \sum_i \tilde{q}_i \mathbf{r}_i.
\end{align}

SpookyNet is trained with a multitask loss
\begin{equation}
\mathcal{L} = w_E \, \mathcal{L}_E + w_F \, \mathcal{L}_F + w_q \, \mathcal{L}_q + w_\mu \, \mathcal{L}_\mu,
\end{equation}
where losses are computed over energy, force, charge, and dipole moment predictions. Like PhysNet, partial charges are not supervised directly, but learned implicitly via their role in the electrostatic and dipole terms.

The inclusion of global attention and electronic state embeddings allows SpookyNet to generalize effectively to charged and open-shell species.

\paragraph*{NequIP: Neural equivariant interatomic potentials}

NequIP~\cite{batzner20223} is an $E(3)$-equivariant MP-GNN that models atomic environments using geometric tensor features such as scalars, vectors, and higher-order irreducible representations (irreps) of the 3D rotation group \( O(3) \). It achieves high accuracy and data efficiency by explicitly enforcing symmetry across all layers.

Each atom \( i \) is initialized with a tensor-valued embedding \( \mathbf{h}_i^{(0)} \in \bigoplus_{\ell=0}^{\ell_{\max}} \mathbb{R}^{C_\ell} \otimes \mathcal{D}^\ell \), where \( \mathcal{D}^\ell \) denotes the irrep of \( O(3) \) at degree \( \ell \), and \( C_\ell \) is the number of channels.

Over \( L \) message-passing layers, features are updated via equivariant convolutions built from spherical harmonics \( Y_\ell(\hat{\mathbf{r}}_{ij}) \), radial functions \( R_\ell(r_{ij}) \), and Clebsch--Gordan contractions. The neighbor-informed local structural message (N-msg) at layer \( l \) is defined as
\begin{equation}
\mathbf{S}_{i\mathcal{N}(i)}^{(l)} = \left\{
f_{\mathrm{msg}}\left( \mathbf{h}_j^{(l)}, R_\ell(r_{ij}) Y_\ell(\hat{\mathbf{r}}_{ij}) \right)
\right\}_{j \in \mathcal{N}(i)},
\end{equation}
and used in the equivariant update
\begin{equation}
\mathbf{h}_i^{(l+1)} = f_{\mathrm{update}}\left( \mathbf{h}_i^{(l)}, \sum_{j \in \mathcal{N}(i)} \mathbf{S}_{ij}^{(l)} \right).
\end{equation}

Following the layerwise aggregation scheme of Eq.~\ref{eq:readout_layerwise}, atomic energies are computed from the scalar channels (\( \ell = 0 \))
\begin{equation}
E_i = \sum_{l=1}^{L} \mathcal{R}^{(l)}\left( \mathbf{h}_i^{(l, \ell=0)} \right), 
\end{equation}

\paragraph*{MACE: Multilayer atomic cluster expansion}

MACE~\cite{batatia2022mace} is an $E(3)$-equivariant MP-GNN that extends the ACE into a learnable neural network framework using body-ordered tensor interactions. It operates on tensor-valued atom features and enforces exact equivariance via spherical harmonics and Clebsch--Gordan products.

Each atom \( i \) is initialized with an embedding
\begin{equation}
\mathbf{h}_i^{(0)} \in \bigoplus_{\ell=0}^{\ell_{\max}} \mathbb{R}^{C_\ell} \otimes \mathcal{D}^\ell,
\end{equation}
where \( \mathcal{D}^\ell \) is the irreducible representation of the rotation group \( O(3) \), and \( C_\ell \) is the number of channels per order \( \ell \).

At each layer, atomic features are updated using a symmetric body-ordered message constructed over multi-atom clusters. The neighbor-informed structural message (N-msg) is defined as:
\begin{equation}
\mathbf{S}_{i\mathcal{N}(i)}^{(l)} = \sum_{k=1}^\nu \sum_{j_1, \dots, j_k \in \mathcal{N}(i)} f_k^{(l)}\left( \mathbf{h}_i^{(l)}, \mathbf{h}_{j_1}^{(l)}, \dots, \mathbf{h}_{j_k}^{(l)} \right),
\end{equation}
where \( f_k^{(l)} \) are equivariant functions acting on \( k+1 \) atoms and \( \nu \) is the maximum interaction order (typically 3 or 4). The update rule is
\begin{equation}
\mathbf{h}_i^{(l+1)} = f_{\mathrm{update}}\left( \mathbf{h}_i^{(l)}, \mathbf{S}_{i\mathcal{N}(i)}^{(l)} \right).
\end{equation}

Readout is performed using a layerwise hierarchical sum over the scalar (\( \ell = 0 \)) channels
\begin{equation}
E_i = \sum_{l=0}^{L} \mathcal{R}^{(l)}\left( \mathbf{h}_i^{(l, \ell=0)} \right)
\end{equation}
where each \( \mathcal{R}^{(l)} \) is a layer-specific dense readout head.

MACE achieves high accuracy and efficiency in modeling systems with strong angular dependencies, such as solvated molecules, interfaces, and complex condensed-phase materials. 
The graph ACE (grACE), which generalizes ACE by employing graph-based, cluster basis functions to capture many-body interactions beyond the local atomic decomposition, exhibits superior accuracy to MACE while retaining a close formal connection to quantum-mechanical Hamiltonian expansions.~\cite{bochkarev2024graph}

\paragraph*{Allegro}

Allegro~\cite{musaelian2023learning} is a local $E(3)$-equivariant MP-GNN that bypasses atom-centered message passing by operating directly on pairwise interactions within a fixed radial cutoff. It encodes many-body correlations via tensor contractions over spherical harmonics and equivariant features, enabling high accuracy with excellent scalability.

Each atom pair \( (i,j) \) is initialized with
\begin{eqnarray}
x_{ij}^{(0)} &\in& \mathbb{R}^{C_0} \quad \text{(scalar feature)}, \\
\mathbf{V}_{ij}^{(0)} &\in& \bigoplus_{\ell=1}^{\ell_{\max}} \mathbb{R}^{C_\ell} \otimes \mathcal{D}^\ell \quad \text{(tensor features)},
\end{eqnarray}
where \( \mathcal{D}^\ell \) is the irreducible representation of the rotation group \( O(3) \) at degree \( \ell \), and \( C_\ell \) is the number of channels.

Over \( L \) layers, Allegro updates pair features using local directional N-msgs over neighboring triplets \( (i,k,j) \). The scalar and equivariant messages are
\begin{align}
\mathbf{S}_{ij}^{\mathrm{scal},(l)} &= \left\{ f_{\mathrm{msg}}^{\mathrm{scal}}(x_{ik}^{(l)}, \hat{\mathbf{r}}_{ik}, w_{ik}^{(l)}) \right\}_{k \in \mathcal{N}(i)}, \\
\mathbf{S}_{ij}^{\mathrm{eq},(l)} &= \left\{ f_{\mathrm{msg}}^{\mathrm{eq}}(\mathbf{V}_{ik}^{(l)}, Y_\ell(\hat{\mathbf{r}}_{ik}), w_{ik}^{(l)}) \right\}_{k \in \mathcal{N}(i)},
\end{align}
where \( Y_\ell(\hat{\mathbf{r}}_{ik}) \) are real spherical harmonics and \( w_{ik}^{(l)} \) are environment-dependent filters. These messages are aggregated and passed through equivariant update functions:
\begin{align}
x_{ij}^{(l+1)} &= f_{\mathrm{update}}^{\mathrm{scal}}(x_{ij}^{(l)}, \mathbf{S}_{ij}^{\mathrm{scal},(l)}), \\
\mathbf{V}_{ij}^{(l+1)} &= f_{\mathrm{update}}^{\mathrm{eq}}(\mathbf{V}_{ij}^{(l)}, \mathbf{S}_{ij}^{\mathrm{eq},(l)}).
\end{align}

After message passing, pairwise scalar energies are predicted from final-layer pair embeddings
\begin{equation}
E_{ij} = \mathcal{R}(x_{ij}^{(L)}), \quad
E_{\mathrm{tot}} = \sum_{i,j \in \mathcal{N}(i)} \sigma_{Z_i,Z_j} E_{ij},
\end{equation}
where \( \sigma_{Z_i,Z_j} \) is a learned scaling factor. Forces are computed as gradients of \( E_{\mathrm{tot}} \) with respect to atomic positions.

Allegro avoids global message propagation and operates only on static, pairwise neighborhoods. This architecture enables linear scaling with system size and has been shown to scale to 100M-atom systems, while remaining competitive with NequIP in accuracy on small-molecule datasets.

\paragraph*{SO3krates: Spherical harmonic coordinate equivariant network}

SO3krates~\cite{frank2022so3krates,frank2024euclidean} is an $SO(3)$-equivariant MP-GNN that models both local and nonlocal quantum interactions via spherical harmonic coordinates (SPHCs). It decouples geometric and atomic features by encoding directionality in basis functions \( Y^{(l)} \), enabling symmetry-preserving attention over arbitrary length scales.

Each atom \( i \) is initialized with
a scalar invariant feature \( f_i^{(0)} \in \mathbb{R}^F \),
 a set of equivariant SPHC tensors \( \chi_i^{(l)} \in \mathbb{R}^{2l+1} \), for \( l = 0, \dots, \ell_{\max} \), which transform via Wigner–\( D \)-matrices under \( SO(3) \).

SPHCs are initialized by aggregating local geometric context
\begin{equation}
\chi_i^{(l)} = \sum_{j \in \mathcal{N}(i)} \tilde{\phi}_{\mathrm{cut}}(r_{ij}) \, Y^{(l)}(\hat{\mathbf{r}}_{ij}),
\end{equation}
where \( Y^{(l)} \) are real spherical harmonics and 
\( \tilde{\phi}_{\mathrm{cut}}(r_{ij}) = \phi_{\mathrm{cut}}(r_{ij}) / \sum_j \phi_{\mathrm{cut}}(r_{ij}) \)
is a normalized cutoff function.

Over \( T \) message-passing layers, invariant and equivariant features are updated using local and long-range directional information. The neighbor-informed local structural messages (N-msgs) for scalars and tensors are
\begin{align}
\mathbf{S}_{i\mathcal{N}(i)}^{(t)} &= \left\{ f_{\mathrm{msg}}^{\mathrm{inv}}(f_j^{(t)}, w_{ij}) \right\}_{j \in \mathcal{N}(i)}, \\
\mathbf{S}_{i\mathcal{N}_\chi(i)}^{(l,t)} &= \left\{ f_{\mathrm{msg}}^{(l)}(Y^{(l)}(\hat{\mathbf{r}}_{ij}), w_{ij}) \right\}_{j \in \mathcal{N}(i) \cup \mathcal{N}_\chi(i)},
\end{align}
where the geometry-aware filter is
\begin{equation}
w_{ij} = \phi_r(r_{ij}) + \phi_s\left( \{ \| \chi_i^{(l)} - \chi_j^{(l)} \|_2 \}_l \right).
\end{equation}

Updates are then computed as
\begin{align}
f_i^{(t+1)} &= f_{\mathrm{update}}^{\mathrm{inv}}(f_i^{(t)}, \mathbf{S}_{i\mathcal{N}(i)}^{(t)}), \\
\chi_i^{(l,t+1)} &= f_{\mathrm{update}}^{(l)}(\chi_i^{(l,t)}, \mathbf{S}_{i\mathcal{N}_\chi(i)}^{(l,t)}).
\end{align}

After message passing, feature–geometry coupling is applied to allow invariant–equivariant interactions
\begin{align}
f_i^{(t+1)} &\mathrel{+}= \psi_1(f_i^{(t+1)}, \{\|\chi_i^{(l)}\|\}_l, \{\|\tilde{\chi}_i^{(l)}\|\}_l), \\
\chi_i^{(l,t+1)} &\mathrel{+}= \psi_2^{(l)}(f_i^{(t+1)}) \cdot \chi_i^{(l,t+1)} + \psi_3^{(l)}(\ldots) \cdot \tilde{\chi}_i^{(l)},
\end{align}
where \( \tilde{\chi}_i^{(l)} \) are cross-coupled SPHCs computed via Clebsch–Gordan contractions.

\paragraph*{ViSNet: Vector–scalar interactive equivariant network}

ViSNet~\cite{wang2024enhancing}  is an $SO(3)$-equivariant MP-GNN that uses a vector–scalar interactive message-passing mechanism (ViS-MP) to update invariant and equivariant features jointly. It replaces Clebsch–Gordan tensor contractions with efficient geometric encodings based on spherical harmonics and incorporates a runtime geometry calculation (RGC) module for bonded angular terms.

Each atom \( i \) is initialized with
a scalar invariant feature \( f_i^{(0)} \in \mathbb{R}^F \),
and equivariant SPHC tensors \( \chi_i^{(l)} \in \mathbb{R}^{2l+1} \) for \( l = 0, \dots, \ell_{\max} \), which transform as irreps of \( SO(3) \).

SPHCs are initialized using directional basis functions over neighbors
\begin{equation}
\chi_i^{(l)} = \sum_{j \in \mathcal{N}(i)} \phi_{\mathrm{cut}}(r_{ij}) Y^{(l)}(\hat{\mathbf{r}}_{ij}),
\end{equation}
where \( Y^{(l)} \) are real spherical harmonics and \( \phi_{\mathrm{cut}} \) is a radial cutoff function.

At each layer, scalar and equivariant features are updated using both local geometry and long-range bonded motifs (e.g., angles, dihedrals) computed via RGC. The N-msgs are defined as
\begin{align}
\mathbf{S}_{i\mathcal{N}(i)}^{(t)} &= \left\{ f_{\mathrm{msg}}^{\mathrm{inv}}(f_j^{(t)}, w_{ij}) \right\}_{j \in \mathcal{N}(i)}, \\
\mathbf{S}_{i\mathcal{N}_\chi(i)}^{(l,t)} &= \left\{ f_{\mathrm{msg}}^{(l)}(Y^{(l)}(\hat{\mathbf{r}}_{ij}), w_{ij}) \right\}_{j \in \mathcal{N}(i) \cup \mathcal{N}_\chi(i)},
\end{align}
where the message weights \( w_{ij} \) are learned functions of distances and bonded geometric context computed on-the-fly.
The update equations are
\begin{align}
f_i^{(t+1)} &= f_{\mathrm{update}}^{\mathrm{inv}}(f_i^{(t)}, \mathbf{S}_{i\mathcal{N}(i)}^{(t)}), \\
\chi_i^{(l,t+1)} &= f_{\mathrm{update}}^{(l)}(\chi_i^{(l,t)}, \mathbf{S}_{i\mathcal{N}_\chi(i)}^{(l,t)}).
\end{align}

ViSNet introduces vector–scalar interaction (ViS-MP) modules to exchange information between scalar and equivariant channels via tensor fusion and residual coupling, preserving equivariance while enhancing expressivity.

\vspace{1em}
\noindent
To conclude this section, we note that while several graph-based NN models have been developed and successfully employed to predict solvent-aware molecular properties~\cite{gimadiev2018assessment,sangala2025graph,shilpa2023recent,xiao2024integrated,ahmad2024ggas2sn}, they operate outside the force/energy learning paradigm and are not designed for PES-consistent dynamics. The properties predicted in these studies are scalar-valued and often exhibit a well-defined structure–property relationship. In principle, models trained on total energies can also be adapted to predict such observables, provided that a mapping from molecular structure or graph to the target property exists. These alternatives will be revisited in Section~\ref{sec:catalogue}, where we categorize applications of MLPs based on whether they model the potential energy surface explicitly or directly predict solvation-dependent properties without PES construction.


\subsection{Kernel models}

Kernel-based potentials use non-parametric methods such as kernel ridge regression (KRR) and Gaussian process regression (GPR) to interpolate molecular energies and forces from quantum mechanical reference data. These models rely on a kernel function that quantifies similarity between molecular or atomic environments, typically using symmetry-preserving descriptors.

Kernel models are strictly data-driven. While they scale poorly with training set size compared to neural networks, they offer high accuracy in the low-data regime. Further, they can enforce desirable physical symmetries (e.g., permutational, rotational, and translational invariance) by construction. Force predictions are obtained via analytic derivatives of the energy, yielding energy-conserving and fully consistent gradients. These features make kernel-based potentials particularly attractive for solvation modeling tasks involving vibrational spectroscopy, conformer ranking, microsolvated clusters, or accurate force-driven dynamics in small to medium-sized systems.

\paragraph*{Kernel ridge regression with physics-inspired descriptors}

Several kernel-based models apply kernel ridge regression (KRR) to molecular descriptors that encode geometric and chemical information in a symmetry-preserving way. Common descriptors include the Coulomb Matrix (CM)\cite{rupp2012fast}, Bag-of-Bonds (BoB)\cite{hansen2015machine}, SLATM (Spectrum of London and Axilrod--Teller--Muto potentials)\cite{huang2020quantum}, MBTR (Many-Body Tensor Representation)~\cite{huo2022unified}, and SOAP. These descriptors range from simple pairwise matrices (e.g., CM, BoB) to more sophisticated many-body or density-based encodings (e.g., MBTR, SOAP). These descriptors map molecules or atomic environments to fixed-length, continuous feature vectors that preserve key physical symmetries (e.g., rotation, translation, permutation). Kernels are functions that define a similarity measure between such vectors~\cite{von2015fourier}.

The kernelized version of the Atomic Cluster Expansion (ACE)~\cite{drautz2019atomic,lysogorskiy2021performant} encodes local atomic environments using systematically constructed orthonormal basis functions that capture geometric correlations among clusters of atoms. These descriptors are body-ordered (e.g., two-body, three-body, etc.), making ACE a principled generalization of classical cluster expansions used in physics-based interatomic potentials. 

All kernel-based models can be paired with kernel functions, such as Gaussian, Laplacian, linear, or polynomial, that define a similarity measure between molecular or atomic environments. The energy is then predicted as a weighted sum over training samples
\begin{equation}
E^{\mathrm{ML}}(\mathbf{d}_q) = \sum_{t=1}^{N_t} \alpha_t \, k(\mathbf{d}_q, \mathbf{d}_t),
\end{equation}
where \( \mathbf{d}_q \) and \( \mathbf{d}_t \) denote the descriptors of the query and training molecules, respectively, and \( k \) is the kernel function.

Two widely used kernel functions are the Laplacian (also known as exponential) and the Gaussian kernels, defined as
\begin{eqnarray}\label{eq_kernels}
k^{\rm Laplacian}(\mathbf{d}_i, \mathbf{d}_j)&  = &\exp\left(-\|\mathbf{d}_i - \mathbf{d}_j\|_1 / \sigma\right), \, {\rm and} \nonumber \\
k^{\rm Gaussian}(\mathbf{d}_i, \mathbf{d}_j) & = & \exp\left(-\|\mathbf{d}_i - \mathbf{d}_j\|_2^2 / 2\sigma^2\right).
\end{eqnarray}
The Laplacian kernel depends on the \( L_1 \) norm (taxicab norm), defined as \( \|{\bf x}\|_1 = \sum_i |x_i| \), while the Gaussian kernel uses the Euclidean norm, defined as \( \|{\bf x}\|_2 = \sqrt{ \sum_i x_i^2} \). Other kernel definitions are also possible, such as those based on pairwise inner products~\cite{scholkopf2002learning}.

The kernel width \( \sigma \) governs the spread of the radial basis functions (RBFs) and thereby the extent to which training examples influence the prediction. To determine \( \sigma \) in a property-independent manner, Ramakrishnan {\it et al.}~\cite{ramakrishnan2015many} proposed a heuristic based on setting the minimum of the off-diagonal elements of the kernel matrix \( {\bf K} \) to 1/2, i.e., \( K_{ij}^{\rm min} = 1/2 \), as a means of conditioning \( {\bf K} \). Accordingly, \( \sigma \) can be estimated from the maximum descriptor difference
\begin{equation}
D_{ij}^{\rm max} = \max\left(\|\mathbf{d}_i - \mathbf{d}_j\|_1\right) \text{ or } \max\left(\|\mathbf{d}_i - \mathbf{d}_j\|_2\right)
\end{equation}
depending on whether Laplacian or Gaussian kernels are used. This leads to the following expressions~\cite{ramakrishnan2015many,ramakrishnan2017machine}
\begin{equation}\label{eq:sigma_opt_max}
\sigma_{\rm opt}^{\rm Laplacian} = \frac{D_{ij}^{\rm max}}{ \log 2}; \quad
\sigma_{\rm opt}^{\rm Gaussian} = \frac{D_{ij}^{\rm max}}{\sqrt{2 \log 2}}.
\end{equation}

However, since descriptor differences typically follow non-trivial distributions, alternative strategies have been employed in specific applications. For instance, in KRR modeling of NMR shielding constants in the QM9-NMR dataset~\cite{gupta2021revving}, improved performance was observed when using the \emph{median} of the descriptor differences in place of \( D_{ij}^{\rm max} \).

The recipe was validated through a more general approach for kernel-based modeling of core-electron binding energies~\cite{tripathy2024chemical}. In this method, the threshold \( K_{ij}^{\rm min} \) is treated as a tunable parameter \( \tau \in (0, 1) \), and the kernel width is set as
\begin{equation}\label{eq:sigma_opt_tau}
\sigma_{\rm opt}^{\rm Laplacian} = \frac{D_{ij}^{\rm max}}{ \log (1/\tau)}; \quad
\sigma_{\rm opt}^{\rm Gaussian} = \frac{D_{ij}^{\rm max}}{\sqrt{2 \log (1/\tau)}}.
\end{equation}
By scanning \( \tau \) in the range 0.03 to 0.99 in increments of 0.03 across various target properties, it was found that the optimal value of \( \tau \) often lies near 1/2.

The FCHL (Faber--Christensen--Huang--Lilienfeld) framework~\cite{faber2018alchemical,christensen2020fchl} integrates both descriptor construction and kernel definition into a unified formulation. It encodes atomic environments using smooth many-body correlation functions (1-, 2-, and 3-body terms), and defines similarity through hierarchical Gaussian kernels that reflect physical interaction scales. Unlike models using fixed descriptors, FCHL builds the kernel into the physics-aware structure of the representation by incorporating physical distance decay, atomic type differentiation, and body-order scaling to reflect the hierarchy of chemical interactions. The learned energy model takes the same functional form
\begin{equation}
E^{\mathrm{ML}} = \sum_j \alpha_j \, k_{\text{FCHL}}(\mathbf{d}, \mathbf{d}_j),
\end{equation}
but here \( k_{\text{FCHL}} \) is a tailored kernel derived from atom-pairwise correlations and spatial decay functions. Forces are obtained via analytic differentiation of the kernel. This approach has been used extensively in solvation energy corrections, spectroscopic modeling, and force-learning tasks where conservative vector fields and physically faithful gradients are essential.

The regression coefficients \( \boldsymbol{\alpha} = (\alpha_1, \alpha_2, \dots, \alpha_{N_t})^\top \) are obtained by minimizing a regularized least-squares loss over training energies
\begin{equation}
\mathcal{L}_{\mathrm{energy}} = \left\| \mathbf{K} \boldsymbol{\alpha} - \mathbf{y} \right\|^2 + \lambda \left\| \boldsymbol{\alpha} \right\|^2,
\end{equation}
which yields the closed-form solution
\begin{equation}
\boldsymbol{\alpha} = (\mathbf{K} + \lambda \mathbf{I})^{-1} \mathbf{y},
\label{eq_krr_train}
\end{equation}
where \( \mathbf{K} \in \mathbb{R}^{N_t \times N_t} \) is the kernel matrix and \( \mathbf{y} \in \mathbb{R}^{N_t} \) is the vector of training energies. 

The regularization strength \( \lambda \) is typically set to zero when the training data are noise-free and the kernel matrix is analytically well-behaved~\cite{ramakrishnan2017machine}. In general, \( \lambda \) is a non-negative real-valued parameter that serves two key purposes. First, when assigned a positive value, it introduces a penalty that regularizes the magnitude of the regression coefficients, thereby reducing the influence of outliers on model performance. Second, and more commonly, regularization becomes essential when the training set contains redundant examples and the kernel matrix \( \textbf{K} \) approaches singularity (i.e., \( K_{ij} \to 1 \)), which can make \( \textbf{K} \) nearly non-invertible. In such scenarios, the addition of the regularization term \( \lambda \mathbf{I} \) conditions the linear system used to obtain the regression coefficients (see Eq.~\ref{eq_krr_train}). It is important to note that such redundancies may arise due to non-unique or poorly discriminating representations, even if only a few training examples are affected. To anticipate and mitigate potential linear dependencies, a small but non-zero value of \( \lambda \), such as \( 10^{-4} \), is commonly adopted.

KRR can be extended to incorporate both energy and force labels in a unified training framework~\cite{christensen2020role,kaeser2020machine}. The central idea is to define a block-structured kernel matrix that couples scalar energies and vector-valued forces through gradient relationships
\begin{equation}
\mathbf{K}_{\mathrm{comb}} =
\begin{bmatrix}
\mathbf{K}_E & \mathbf{K}_{EF} \\
\mathbf{K}_{FE} & \mathbf{K}_{FF}
\end{bmatrix}, \quad
\mathbf{y}_{\mathrm{comb}} =
\begin{bmatrix}
\mathbf{y}_E \\
\mathbf{y}_F
\end{bmatrix},
\label{eq:bigK1}
\end{equation}
where:
\begin{itemize}
    \item \( \mathbf{K}_E \in \mathbb{R}^{N_t \times N_t} \): energy–energy kernel block with entries \( k(\mathbf{d}_i, \mathbf{d}_j) \),
    \item \( \mathbf{K}_{EF} = -\frac{\partial \mathbf{K}_E}{\partial \mathbf{R}^\mathsf{T}} \), \( \mathbf{K}_{FE} = -\frac{\partial \mathbf{K}_E}{\partial \mathbf{R}} \): energy–force and force–energy couplings,
    \item \( \mathbf{K}_{FF} = \frac{\partial^2 \mathbf{K}_E}{\partial \mathbf{R} \partial \mathbf{R}^\mathsf{T}} \): force–force block (Hessian of the kernel),
    \item \( \mathbf{y}_E \in \mathbb{R}^{N_t} \), \( \mathbf{y}_F \in \mathbb{R}^{M_t} \): target vectors for energies and forces.
\end{itemize}
In this case, the regression coefficients are obtained by minimizing a regularized least-squares loss over training energies and forces
\begin{equation}
\mathcal{L}_{\mathrm{energy+forces}} = \left\| \mathbf{K}_{\mathrm{comb}} \boldsymbol{\alpha} - \mathbf{y}_{\mathrm{comb}} \right\|^2 + \lambda \left\| \boldsymbol{\alpha} \right\|^2,
\label{eq:bigK2}
\end{equation}
which yields the closed-form solution
\begin{equation}
\boldsymbol{\alpha} = (\mathbf{K}_{\mathrm{comb}} + \lambda \mathbf{I})^{-1} \mathbf{y}_{\mathrm{comb}},
\label{eq:bigK3}
\end{equation}
where \( \mathbf{K}_{\mathrm{comb}} \in \mathbb{R}^{(N_t+M_t) \times (N_t+M_t)} \) is the combined kernel matrix, \( \mathbf{y}_{\mathrm{comb}} \in \mathbb{R}^{N_t+M_t} \) is the combined vector of training energies and forces.

The energy–force combined KRR problem can be reformulated using an approximated rectangular kernel system\cite{kaeser2020machine}
\begin{equation}
\begin{bmatrix}
\mathbf{y}_E \\
\mathbf{y}_F
\end{bmatrix}
=
\begin{bmatrix}
\mathbf{K}_E \\
\mathbf{K}_{FE}
\end{bmatrix}
\boldsymbol{\alpha} 
\end{equation}
with
$\mathbf{K}_E \in \mathbb{R}^{N_t \times N_t}, \ \mathbf{K}_{FE} \in \mathbb{R}^{M_t \times N_t}$.
This system is overdetermined when \( M_t > N_t \), and the coefficients \( \boldsymbol{\alpha} \in \mathbb{R}^{N_t} \) can be obtained by solving the regularized least-squares problem
\begin{equation}
\boldsymbol{\alpha} = \left( \mathbf{K}_{\mathrm{rect}}^\mathsf{T} \mathbf{K}_{\mathrm{rect}} + \lambda \mathbf{I} \right)^{-1} \mathbf{K}_{\mathrm{rect}}^\mathsf{T} \mathbf{y}_{\mathrm{comb}},
\end{equation}
where \( \mathbf{K}_{\mathrm{rect}} = \begin{bmatrix} \mathbf{K}_E \\ \mathbf{K}_{FE} \end{bmatrix} \in \mathbb{R}^{(N_t + M_t) \times N_t} \), and \( \mathbf{y}_{\mathrm{comb}} = \begin{bmatrix} \mathbf{y}_E \\ \mathbf{y}_F \end{bmatrix} \).

This formulation provides an efficient approximation to the full block-KRR system, avoids constructing the large force–force kernel, and yields energy-conservative force predictions as long as the kernel is differentiable.

\paragraph*{GAP-SOAP: Gaussian approximation potential with SOAP}
The Gaussian approximation potential (GAP)~\cite{bartok2010gaussian,bartok2015gaussian} uses Gaussian process regression (GPR) over SOAP descriptors~\cite{bartok2013representing} to model the total energy as a sum of atomic contributions
\begin{equation}
E_i^{\mathrm{ML}} = \sum_j \alpha_j \, k(\mathbf{d}_i, \mathbf{d}_j),
\end{equation}
where \( \mathbf{d}_i \) is the SOAP descriptor for atom \( i \), and \( k \) is typically a Gaussian kernel that quantifies similarity between local atomic environments. The regression coefficients \( \alpha_j \) are fitted using the Gaussian process formalism, which provides not only predictions but also associated uncertainty estimates for each prediction.

While similar in functional form to kernel ridge regression (KRR), GPR models the output as a distribution over functions and optimizes hyperparameters by maximizing the marginal likelihood~\cite{rasmussen2006gaussian}. This probabilistic treatment contrasts with the deterministic nature of KRR and has been widely adopted in the GAP framework for molecular and materials systems.

Although computationally intensive for large datasets, GAP-SOAP excels in low-data regimes where physical interpretability, gradient consistency, and accuracy are paramount.

\begin{table*}[!htpb]
\centering
\caption{Summary of machine-learned potentials (MLPs) suitable for solvation modeling. Models are distinguished by how they represent atomic environments, support long-range (LR) interactions, and compute forces. ``Grad'' indicates that forces are computed as gradients of a learned scalar energy. The final column distinguishes whether the force is derived from atomic energy contributions, the total energy, or is not defined due to direct force training.
\label{tab:mlp_summary}}
\begin{tabular} {lllll}
\hline 
Model (MLP Category) & Descriptor & LR & Force & Force source \\
\hline 
BPNN~\cite{behler2007generalized} (NN) & Fixed-form & No & Grad & Atomic Energy \\
ANI~\cite{smith2017ani} (NN) & Fixed-form & No & Grad & Atomic Energy \\
CENT~\cite{ghasemi2015interatomic}/QeqNN~\cite{ko2021fourth}/QRNN~\cite{jacobson2022transferable} (NN) & Fixed-form & Yes & Grad & Atomic Energy \\
\hline 
DeePMD~\cite{zhang2018deep} (NN) & Learned & Partial & Grad & Total Energy \\
HIPNN~\cite{lubbers2018hierarchical} (NN) & Learned & Partial & Grad & Total Energy \\
\hline 
SchNet~\cite{schutt2018schnet} (NN) & Learned & No & Grad & Total Energy \\
MGCN~\cite{lu2019molecular} (NN) & Learned  & No & Grad & Total Energy \\
DimeNet~\cite{gasteiger2020directional}/DimeNet++~\cite{gasteiger2020fast} (NN) & Learned & No & Grad & Total Energy \\
MGNN~\cite{chang2025mgnn} (NN) & Learned & Partial & Grad & Total Energy \\
PhysNet~\cite{unke2019physnet} (NN) & Mixed  & Yes & Grad & Total Energy \\
GemNet~\cite{gasteiger2021gemnet} (NN) & Learned & Partial & Grad & Total Energy \\
PaiNN~\cite{schutt2021equivariant} (NN) & Learned & Partial & Grad & Total Energy \\
ENINet~\cite{mao2024molecule} (NN) & Learned & No & Grad & Total Energy \\
SpookyNet~\cite{unke2021spookynet} (NN) & Learned & Yes & Grad & Total Energy \\
NequIP~\cite{batzner20223} (NN) & Learned & Yes & Grad & Total Energy \\
MACE~\cite{batatia2022mace} (NN) & Learned & Yes & Grad & Total Energy \\
Allegro~\cite{musaelian2023learning} (NN) & Learned & No & Grad & Total Energy \\
SO3krates~\cite{frank2022so3krates} (NN) & Learned & Yes & Grad & Total Energy \\
ViSNet~\cite{wang2024enhancing} (NN) & Learned & Yes & Grad & Total Energy \\
\hline 
KRR~\cite{rupp2012fast} (Kernel) & Fixed-form & No & Grad & Total Energy \\
FCHL~\cite{faber2018alchemical} (Kernel) & Fixed-form & Yes & Grad & Total Energy \\
GAP-SOAP~\cite{bartok2010gaussian} (Kernel) & Fixed-form & Partial & Grad & Total Energy \\
ACE (kernelized)~\cite{drautz2019atomic} (Kernel) & Fixed-form & Yes & Grad & Atomic Energy \\
\hline 
MTP~\cite{shapeev2016moment} (Linear) & Fixed-form & Partial & Grad & Atomic Energy \\
SNAP~\cite{thompson2015spectral} (Linear) & Fixed-form & Partial & Grad & Atomic Energy \\
ACE (linear)~\cite{drautz2019atomic} (Linear) & Fixed-form & Yes & Grad & Atomic Energy \\
\hline 
NewtonNet~\cite{haghighatlari2022newtonnet} (NN) & Learned & Partial & Force-only & N/A \\
GDML~\cite{chmiela2017machine}/sGDML~\cite{chmiela2019sgdml} (Kernel) & Fixed-form & No & Force-only & N/A \\
\hline 
KRR with local-PCA~\cite{rupp2015machine} (Kernel) & Fixed-form & Partial & Force-only & N/A \\
ForceNet~\cite{hu2021forcenet} (NN) & Learned & Partial & Force-only & N/A \\
\hline 
\end{tabular}
\end{table*}

\subsection{Linear models}

Linear-MLPs are linear and polynomial descriptor-based MLPs that use fixed, symmetry-adapted basis functions to approximate the potential energy surface. These methods rely on linear regression over carefully constructed features, such as tensor contractions, bispectrum components, or orthonormal polynomial expansions, to capture atomic environments in a manner consistent with physical symmetries. Unlike kernel methods, these models offer scalable training and faster predictions while maintaining systematic improvability and interpretability.

\paragraph*{MTP: Moment tensor potentials}
MTP represents atomic environments using basis functions built from tensor contractions of relative position vectors~\cite{shapeev2016moment}
\begin{equation}
E_i^{\mathrm{ML}} = \sum_\mu \theta_\mu B_\mu(\mathbf{r}_i),
\end{equation}
where \( B_\mu \) are symmetry-invariant tensor descriptors. 

\paragraph*{SNAP: Spectral neighbor analysis potential}
SNAP models~\cite{thompson2015spectral} atomic energies using a linear combination of bispectrum components derived from the spherical harmonics expansion of the atomic neighbor density
\begin{equation}
E_i^{\mathrm{ML}} = \sum_k \beta_k B_k^{(i)},
\end{equation}
where \( B_k^{(i)} \) are rotationally invariant bispectrum coefficients and \( \beta_k \) are linear regression weights. 

\paragraph*{Linear ACE: Linear atomic cluster expansion}
Linear ACE expresses the atomic energy as a linear combination of orthonormal, symmetry-adapted basis functions that encode many-body interactions up to a given body order~\cite{drautz2019atomic}. The total energy is decomposed into per-atom contributions
\begin{equation}
E_i^{\mathrm{ML}} = \sum_\mu c_\mu B_\mu(\mathbf{r}_i),
\end{equation}
where \( B_\mu(\mathbf{r}_i) \) are basis functions constructed from products of radial functions and spherical harmonics over neighbor clusters, and \( c_\mu \) are regression coefficients fitted using linear least-squares. 

Table~\ref{tab:mlp_summary} provides a comparative overview of the major \gls*{mlp} architectures used in solvation modeling. The models are categorized by learning paradigm: neural network–based potentials (NN-MLPs), kernel-based potentials (Kernel-MLPs), and linear models (Linear-MLPs). They are further distinguished by their descriptor types and capacity to model long-range interactions.

Long-range interactions are a key consideration in solvation modeling, as they underpin solvent organization, dielectric screening, and polarization phenomena beyond the first solvation shell. To reflect this, we include a dedicated column in our classification tables indicating whether a given model accounts for long-range effects, either explicitly (e.g., via charge equilibration, Coulomb terms, or dispersion corrections) or implicitly (e.g., through deep, equivariant, or attention-based architectures).

All models listed support analytic force computation via gradients of the learned energy function, making them suitable for MD simulations and geometry optimization tasks in both gas-phase and condensed-phase solvation environments.

\subsection{Automatic differentiation}

Modern neural network potentials such as SchNet, DimeNet, and PhysNet compute atomic forces analytically as the negative gradient of the total energy with respect to atomic positions as per Eq.\ref{eq:force_energybased_ML}. These gradients are obtained via \emph{automatic differentiation} (autodiff), a core feature of deep learning libraries such as PyTorch~\cite{paszke2019pytorch}, TensorFlow~\cite{abadi2016tensorflow}, and JAX~\cite{bradbury2018jax}. 
In practice, autodiff frameworks trace the computational graph of the scalar energy prediction and compute derivatives with respect to atomic coordinates using built-in routines such as 
\texttt{torch.autograd.grad} (PyTorch), 
\texttt{tf.GradientTape} (TensorFlow), or 
\texttt{jax.grad} (JAX). General-purpose libraries such as EagerPy~\cite{rauber2020eagerpy} also support such operations across multiple backends.
This allows energy-based MLPs to seamlessly return physically consistent forces without requiring manual gradient derivations.

In contrast, kernel-based models (e.g., GAP, sGDML, FCHL) often fall outside the scope of standard autodiff frameworks due to their functional complexity. These models traditionally rely on \emph{manually derived analytical gradients}, particularly when forces are treated as primary learning targets.

To address this gap, Schmitz \textit{et al.}~\cite{schmitz2022algorithmic} propose a general framework that employs \emph{algorithmic differentiation} (AD) to construct physically consistent machine-learned force fields. Their method derives gradients and higher-order derivatives (e.g., Hessians) directly from scalar energy models, enforcing conservation laws and ensuring that the resulting force field is curl-free. Specifically, the kernel function \( k \) in a Gaussian Process (GP) model is redefined as
\begin{equation}
k_{\text{force}}(\mathbf{x}, \mathbf{x}') = \nabla_{\mathbf{x}} \nabla_{\mathbf{x}'}^\top k(\mathbf{x}, \mathbf{x}'),
\end{equation}
guaranteeing a conservative force field by construction.

The framework also supports composite kernels with 
descriptors \( D(\cdot) \) based on deep NN architectures
\begin{equation}
k(\mathbf{x}, \mathbf{x}') = k(D(\mathbf{x}), D(\mathbf{x}')),
\end{equation}
enabling hybrid models that integrate GNN-derived features into kernel-based force fields. This provides a scalable and interpretable route to building accurate, energy-conserving MLPs beyond standard neural networks.

\subsection{Gradient-domain models\label{subsec:GDMLPs}}

While many MLPs are trained to reproduce energies and derive forces via differentiation, a distinct class of models operates directly in the gradient domain, i.e., learning force fields from force data. These are called gradient-domain MLPs (GD-MLPs) and can be further classified based on whether they preserve energy conservation:
\begin{itemize}
    \item Strictly conservative GD-MLPs (e.g., sGDML, NewtonNet) construct forces as exact gradients of a learned scalar potential, ensuring thermodynamic consistency.
\item Non-conservative MLPs (e.g., ForceNet in its unconstrained form) learn force fields directly (F-MLPs) from data without enforcing that the forces are derivable from a scalar energy function; such models are discussed further in Section~\ref{subsec:forceconly}.
\end{itemize}
The distinction between conservative and non-conservative models is critical for tasks such as long-time MD and free energy simulations, where consistency between forces and PESs is required.

Early kernel-based models, such as those based on GPR~\cite{li2015molecular} and KRR~\cite{rupp2015machine}, predicted forces directly (i.e., these are F-MLPs) but did not guarantee integrability, i.e., the existence of a corresponding energy function. This challenge was addressed by sGDML~\cite{chmiela2019sgdml}, which enforces symmetry and conservation analytically to produce high-fidelity force fields suitable for vibrational and thermodynamic modeling.

\paragraph*{NewtonNet: A Newtonian equivariant message-passing network}

NewtonNet~\cite{haghighatlari2022newtonnet} is an $O(3)$-equivariant message-passing neural network inspired by Newton's equations of motion. It introduces physically motivated modules, such as force, displacement, and energy calculators, that build latent equivariant features throughout the network, resulting in interpretable many-body interactions and improved data efficiency.

Each atom \( i \) is initialized with scalar features \( \mathbf{s}_i^{(0)} \in \mathbb{R}^F \) based on an embedding of its atomic number, and pairwise distances \( r_{ij} \) are encoded using Bessel basis functions with cutoff filtering. These define invariant edge features \( \mathbf{e}_{ij} \in \mathbb{R}^E \), used throughout message passing.

Atomic features are updated at each layer \( l \) using symmetric message passing
\begin{align}
\mathbf{m}_{ij}^{(l)} &= f_{\mathrm{msg}}\left( f_{\mathrm{atom}}(\mathbf{s}_i^{(l)}), f_{\mathrm{atom}}(\mathbf{s}_j^{(l)}), f_{\mathrm{edge}}(r_{ij}) \right), \\
\mathbf{s}_i^{(l+1)} &= \mathbf{s}_i^{(l)} + \sum_{j \in \mathcal{N}(i)} \mathbf{m}_{ij}^{(l)},
\end{align}
where \( f_{\mathrm{atom}} \), \( f_{\mathrm{edge}} \), and \( f_{\mathrm{msg}} \) are MLPs. This message construction is symmetric under \( i \leftrightarrow j \), enabling latent force consistency via Newton’s third law.

The latent force between atoms is defined as
\begin{equation}
\mathbf{F}_{ij}^{(l)} = f_{\mathrm{force}}(\mathbf{m}_{ij}^{(l)}) \cdot \hat{\mathbf{r}}_{ij}, \quad \mathbf{F}_i^{(l)} = \sum_{j \in \mathcal{N}(i)} \mathbf{F}_{ij}^{(l)},
\end{equation}
ensuring antisymmetry and equivariance. These forces are passed into a displacement module
\begin{equation}
\Delta \mathbf{r}_i^{(l)} = f_{\mathrm{disp}}(\mathbf{s}_i^{(l)}) \cdot \mathbf{F}_i^{(l)},
\end{equation}
where \( f_{\mathrm{disp}} \) is a learned function and \( \Delta \mathbf{r}_i^{(l)} \) accumulates virtual displacements.

Energy contributions are computed via a projection of equivariant vectors to invariant scalars
\begin{equation}
\Delta E_i^{(l)} = -f_{\mathrm{energy}}(\mathbf{s}_i^{(l)}) \cdot (\mathbf{F}_i^{(l)} \cdot \Delta \mathbf{r}_i^{(l)}),
\end{equation}
and added to the scalar features for the next layer
\begin{equation}
\mathbf{s}_i^{(l+1)} \leftarrow \mathbf{s}_i^{(l+1)} + \Delta E_i^{(l)}.
\end{equation}

The final atomic energies are predicted from the last scalar features
\begin{equation}
E_i = \mathcal{R}(\mathbf{s}_i^{(L)}), \quad E_{\text{tot}} = \sum_i E_i,
\end{equation}
and forces are obtained via gradients
\begin{equation}
\mathbf{F}_i = -\nabla_{\mathbf{r}_i} E_{\text{tot}}.
\end{equation}

It is helpful to contrast NewtonNet with direct force-only models such as ForceNet, which is discussed in Section~\ref{subsec:forceconly}. While ForceNet predicts atomic forces directly without learning an energy potential, NewtonNet retains a conservative formulation by predicting energies and obtaining forces as analytical gradients. Moreover, NewtonNet introduces interpretable latent force vectors that obey Newton’s third law (action–reaction symmetry), in contrast to ForceNet, which enforces geometric consistency through conditional filter convolutions and rotational data augmentation. 
By integrating learned displacement and energy modules, NewtonNet enables force prediction to emerge naturally from equivariant geometric features, without requiring force labels as the primary form of supervision.

\paragraph*{Gradient-domain KRR}

The combined energy and force formulation of KRR is particularly effective because force data provides a much richer supervision signal than energy alone: each configuration contributes a single scalar energy label, but \(3N\) force components~\cite{christensen2020role}. The number of force labels \(M_t\), therefore, exceeds the number of energy labels \(N_t\) by an order of magnitude or more in most datasets. Leveraging this imbalance improves data efficiency and enables more accurate learning of local PES curvature.

To emphasize force accuracy, one may train solely on forces (see Eq.~\ref{eq:bigK1}, Eq.~\ref{eq:bigK2}, and Eq.~\ref{eq:bigK3}) by minimizing
\begin{equation}
\mathcal{L}_{\mathrm{force}} = \left\| \mathbf{K}_{FF} \boldsymbol{\alpha} - \mathbf{y}_F \right\|^2 + \lambda \left\| \boldsymbol{\alpha} \right\|^2,
\end{equation}
which yields a closed-form solution
\begin{equation}
\boldsymbol{\alpha} = \left( \mathbf{K}_{FF} + \lambda \mathbf{I} \right)^{-1} \mathbf{y}_F.
\end{equation}
When \( \lambda = 0 \), we have
\begin{equation}
\mathbf{y}_F = \mathbf{K}_{FF} \boldsymbol{\alpha}, \quad 
\Rightarrow \quad
\mathbf{y}_E = \mathbf{K}_{EF} \boldsymbol{\alpha} + \mathrm{const},
\end{equation}
allowing energy reconstruction up to an additive constant. This is the setting in which force-only learning remains conservative and physically meaningful.

For out-of-sample predictions, the force on a query geometry \( \mathbf{r}_q \) is given by
\begin{equation}
\mathbf{F}_q^{\mathrm{ML}} = \sum_{t=1}^{N_t} \boldsymbol{\alpha}_t \cdot \frac{\partial^2 k(\mathbf{r}_q, \mathbf{r}_t)}{\partial \mathbf{r}_q \partial \mathbf{r}_t},
\end{equation}
where \( \mathbf{r}_q \in \mathbb{R}^{3N} \) is the vector of Cartesian coordinates for the query configuration. If the model is conservative, the energy at \( \mathbf{r}_q \) can be reconstructed as
\begin{equation}
E_q^{\mathrm{ML}} = \sum_{t=1}^{N_t} \boldsymbol{\alpha}_t \cdot \frac{\partial k(\mathbf{r}_q, \mathbf{r}_t)}{\partial \mathbf{r}_t} + \mathrm{const}.
\end{equation}
This expression highlights the ability of gradient-domain KRR models to generalize force predictions to unseen geometries, while retaining energy consistency and physical interpretability.

\paragraph*{GDML and sGDML}
A notable and widely used example of gradient-domain kernel methods is the gradient-domain ML (GDML) model~\cite{chmiela2017machine}, which directly learns conservative force fields from force-only data. Unlike descriptor-based models, GDML operates directly on full molecular geometries \( \{ \mathbf{r}_i \} \in \mathbb{R}^{3N} \) and ensures that the predicted force field is the gradient of a scalar energy function. This is achieved by expressing forces as
\begin{equation}
{F}^{\mathrm{ML}}_{\mu}({\bf r}_i) = \sum_{j=1}^{N_t} {\alpha}_{\mu j} \cdot \nabla_{r_{\mu,i}} k({\bf r}_i, {\bf r}_j),
\end{equation}
where \( k({\bf r}_i, \bf{r}_j) \) is a scalar-valued kernel function, typically a Gaussian kernel of the form
\begin{equation}
k({\bf r}_i, {\bf r}_j) = \exp\left(- \frac{\| {\bf r}_i - {\bf r}_j \|^2}{\sigma^2} \right),
\end{equation}
with \( \sigma \) controlling the kernel width. The resulting force–force kernel block is constructed via second derivatives
\begin{equation}
\left[ \mathbf{K}_{FF} \right]_{(\mu i)(\nu j)} = \frac{\partial^2 k(\mathbf{r}_i, \mathbf{r}_j)}{\partial r_{i}^{\mu} \partial r_{j}^{\nu}},
\end{equation}
where \( r_i^\mu \) and \( r_j^\nu \) are Cartesian components and \( \mu, \nu \in \{x, y, z\} \).

Because GDML constructs forces directly from gradients of a scalar-valued kernel, the model is strictly conservative by design. The corresponding energy can be reconstructed (up to an additive constant) as
\begin{equation}
E^{\mathrm{ML}}(\mathbf{r}_t) = \sum_{t=1}^{N_t} \boldsymbol{\alpha}_t \cdot k(\mathbf{r}_q, \mathbf{r}_t) + \mathrm{const}.
\end{equation}

The symmetry-adapted variant, sGDML~\cite{chmiela2019sgdml}, extends this framework by explicitly enforcing permutation invariance of chemically equivalent atoms. This is accomplished by symmetrizing the kernel over all valid atomic index permutations \( \pi \in \mathcal{P} \)
\begin{equation}
k^{\mathrm{sGDML}}(\mathbf{R}_i, \mathbf{R}_j) = \frac{1}{|\mathcal{P}|} \sum_{\pi \in \mathcal{P}} k(\pi \cdot \mathbf{R}_i, \mathbf{R}_j),
\end{equation}
where \( \pi \cdot \mathbf{R}_i \) denotes the permuted geometry. This guarantees that both the predicted forces and the reconstructed energy surface obey the correct exchange symmetry.

Thanks to their analytic structure, strict conservativity, and symmetry enforcement, GDML and sGDML are exceptionally data-efficient. They have been successfully applied to a wide range of small molecular \cite{raghunathan2023intramolecular} and microsolvated systems, offering accurate force predictions and energy-conserving trajectories from as few as 100 reference geometries.

\vspace{0.75em}
\noindent
In summary, gradient-domain models offer data-efficient, physically grounded alternatives to energy-based MLPs. They are particularly advantageous for solvation applications involving local force perturbations, vibrational analysis, or AIMD-derived datasets. However, for reliable long-time dynamics or free energy calculations, conservative models such as sGDML or jointly trained energy–force MLPs are generally preferred.

In contrast to gradient-domain models, force-only methods bypass the energy altogether and regress atomic forces directly. The following models exemplify this approach.

\subsection{Force-only models\label{subsec:forceconly}}

Force-only MLPs (F-MLPs) represent a direct approach, bypassing the energy function entirely and predicting atomic forces as vector-valued targets. Unlike gradient-domain models that (at least implicitly) construct a \gls*{pes}, force-only models are not required to be conservative and are often trained with vector losses alone.

This design choice can offer improved force accuracy and reduced training cost, especially in high-dimensional systems or when reference energies are unavailable. However, the lack of an underlying energy functional introduces risks:
\begin{itemize}
    \item The resulting force field may violate conservation laws.
\item Thermodynamic properties derived from force-only simulations may be inconsistent.
\item Long-time dynamics may suffer from artifacts such as energy drift or nonphysical trajectories.
\end{itemize}
To mitigate these issues, some models incorporate architectural or training constraints that approximate conservative behavior. For example, ForceNet regresses forces directly but uses rotational augmentation and architectural regularization to encourage geometric consistency.

\paragraph*{ForceNet: Force-centric GNN}

ForceNet~\cite{hu2021forcenet} is a force-centric message-passing neural network designed to predict atomic forces directly, without relying on energy gradients. In contrast to energy-based models such as SchNet or PaiNN that derive forces as gradients of scalar energy, ForceNet learns atomic forces as target quantities, offering increased flexibility and computational efficiency, especially for large-scale applications.

The model inputs are atomic numbers \( Z_i \) and positions \( \mathbf{r}_i \in \mathbb{R}^3 \), which define the node features and directional edge features. 
Each atom \( i \) is initialized with a scalar embedding \( \mathbf{s}_i^{(0)} = \mathbf{a}_{Z_i} \in \mathbb{R}^F \), and edge features \( \mathbf{e}_{ij} \in \mathbb{R}^E \) are constructed from pairwise distances and directions.

At each layer \( l \), scalar node features are updated using residual message passing
\begin{equation}
\mathbf{s}_i^{(l+1)} = f_{\mathrm{update}}\left( 
\mathbf{s}_i^{(l)} + \sum_{j \in \mathcal{N}(i)} 
f_{\mathrm{msg}}(\mathbf{s}_j^{(l)}, \mathbf{s}_i^{(l)}, \mathbf{e}_{ij})
\right),
\end{equation}
where the message function uses conditional filter convolutions (cfconv), extending SchNet’s framework to depend on both sender and receiver embeddings.

After \( L \) layers, the predicted atomic force for atom \( i \) is obtained via
\begin{equation}
\mathbf{F}_i = \mathcal{R}(\mathbf{s}_i^{(L)}),
\end{equation}
where \( \mathcal{R} \) is a decoder dense layer that maps scalar node embeddings to 3D vector outputs.

Because ForceNet does not derive forces from a learned energy potential, it does not enforce energy conservation or exact rotational equivariance. Instead, it uses rotation data augmentation during training to approximate equivariant behavior
\begin{equation}
(\mathbf{R} \mathbf{r}, \mathbf{R} \mathbf{F}) \mapsto \text{training pair},
\end{equation}
where \( \mathbf{R} \in SO(3) \) is a random rotation matrix.

On the OC20 dataset, ForceNet demonstrates state-of-the-art accuracy in force prediction, outperforming energy-gradient-based models like DimeNet++ and GNS while requiring significantly fewer GPU days and enabling faster inference. Its success highlights the potential of force-only training for scalable atomistic modeling.

\paragraph*{Equivariant force prediction using KRR with local-PCA}
The early study by Rupp \textit{et al.}~\cite{rupp2015machine} highlighted the importance of symmetry-preserving representations for predicting quantum mechanical properties of atoms in molecules in the context of gradient-domain KRR. Although the terminology of \textit{invariance} and {\it equivariance} is not used explicitly, the authors implement these principles through their design of descriptors and learning targets.

To predict scalar-valued properties such as core-level ionization energies and NMR chemical shifts, the authors use a modified version of the Coulomb matrix as a local atomic descriptor. Specifically, they construct local Coulomb matrices centered on each atom, consisting of the atom's own nuclear charge and the Coulomb interactions with its \( k \) nearest neighbors. These descriptors are explicitly made {\it invariant} to translation, rotation, and permutation of atom indices, which is a key requirement for learning scalar observables that do not depend on molecular orientation or labeling.

For vector-valued targets such as atomic forces, the model enforces {\it equivariance} under 3D rotations. 
Each force vector is first projected onto a local atom-centered coordinate frame, and the machine learning model predicts its components \( (F_{x'}, F_{y'}, F_{z'}) \) in that frame. The force vector in the global frame is then reconstructed via a \( 3 \times 3 \) transformation matrix \( \mathbf{T}_i \)
\begin{equation}
\mathbf{F}_i = \mathbf{T}_i \mathbf{F}_i^{\mathrm{local}}.
\label{eq:localtrans}
\end{equation}

A local atom-centered coordinate frame is associated with each atom \( i \) using 
a set of weighted relative position vectors to its neighboring atoms \( j \), defined as
\begin{equation}
\mathbf{x}_j = \frac{Z_j}{\|\mathbf{r}_j - \mathbf{r}_i\|^3} (\mathbf{r}_j - \mathbf{r}_i), \quad j \ne i,  \quad r_{ij} \leq \tau,
\end{equation}
where \( \tau \) is a fixed cutoff radius that enforces locality of the atomic environment, \( \mathbf{r}_j \) and \( \mathbf{r}_i \) are the Cartesian coordinates of atoms \( j \) and \( i \), and \( Z_j \) is the atomic number of atom \( j \). This weighting scheme emphasizes the contribution of nearby and heavy atoms while down-weighting distant and lighter ones.

These weighted vectors are used to build a covariance matrix of the local atomic environment
\begin{equation}
\mathbf{T}_i = \sum_{j \ne i} \mathbf{x}_j \mathbf{x}_j^\top,
\end{equation}
whose eigenvectors define a canonical local reference frame. The corresponding rotation matrix \( \mathbf{T}_i \in \mathbb{R}^{3 \times 3} \) aligns the global Cartesian frame with the principal axes of the local neighborhood.

This local frame enables {\it rotational equivariance} for vector-valued targets. Specifically, force vectors \( \mathbf{F}_i \) are projected into the local frame during training
\begin{equation}
\mathbf{F}_i^{\mathrm{local}} = \mathbf{T}_i^\top \mathbf{F}_i,
\end{equation}
and the model is trained to regress these local components. At inference time, predicted forces of a query atom, $q$, are rotated back to the global frame
\begin{equation}
\mathbf{F}_q = \mathbf{T}_q \mathbf{F}_q^{\mathrm{local}}.
\label{eq:localtrans}
\end{equation}

To describe the local chemical environment of atom \( i \), a descriptor matrix \( \mathbf{M}_i \in \mathbb{R}^{m \times 4} \) is constructed as
\begin{equation}
\mathbf{M}_i = 
\begin{bmatrix}
Z_{j_1} & x'_{j_1} - x'_i  & y'_{j_1} - y'_i  & z'_{j_1} - z'_i  \\
Z_{j_2} & x'_{j_2} - x'_i   & y'_{j_2}- y'_i   & z'_{j_2} - z'_i  \\
\vdots & \vdots & \vdots & \vdots \\
Z_{j_m} & x'_{j_m} - x'_i   & y'_{j_m} - y'_i  & z'_{j_m} - z'_i 
\end{bmatrix},
\end{equation}
where \( (x'_j -  x'_i, y'_j -  y'_i, z'_j -  z'_i) \) are the relative coordinates of neighbor atom \( j \) projected into the local PCA-derived frame. The rows are sorted by increasing distance from the central atom \( i \), enforcing {\it permutation invariance} with respect to atom indexing. The descriptor \( \mathbf{M}_i \) includes only atoms \( j \) within the cutoff sphere \( r_{ij} \leq \tau \), where \( m \) is the number of such neighbors.

Because all atomic positions are expressed relative to \( \mathbf{r}_i \), the descriptor \( \mathbf{M}_i \) is also {\it invariant under translation}. A rigid shift \( \mathbf{r}_j \rightarrow \mathbf{r}_j + \boldsymbol{\epsilon} \) for all \( j \) does not change the relative vectors, the covariance matrix, or the resulting descriptor.

This PCA-based procedure thus ensures that the learned model respects the physical symmetries of the system. Although developed before modern GNNs, this approach anticipates many of the principles underlying equivariant architectures, including rotation- and translation-invariant descriptors and frame-consistent force prediction.

While the local PCA framework enables equivariant force prediction, the approach does not guarantee that the learned force field is conservative. 
As a result, the predicted force field may not be \emph{integrable}, and hence may violate energy conservation in MD simulations. This arises because the model is trained directly on force components without enforcing that they are derived from a differentiable scalar field.

\begin{table*}[!htpb]
\centering
\caption{Summary of geometric symmetry groups relevant to equivariant MLPs.}
\label{tab:groups}
\begin{tabular}{@{}ll@{}}
\hline
Group & Description \\
\hline
$E(3)$  & Euclidean group: includes translations, rotations, and reflections \\
$SE(3)$ & Special Euclidean group: includes translations and proper (orientation-preserving) rotations \\
$O(3)$  & Orthogonal group: includes all 3D rotations and reflections (but no translations) \\
$SO(3)$ & Special Orthogonal group: includes only proper 3D rotations \\
\hline
\end{tabular}
\end{table*}

\subsection{Symmetry and equivariance}
\label{symmMPNN}
Modern MP-GNN-MLPs for atomistic systems incorporate geometric symmetries as architectural priors 
for feature aggregation to ensure physical consistency across diverse atomistic configurations. In these models, the relevant symmetry group is the Euclidean group $E(3)$, which comprises all rigid-body transformations in 3D space, including translations, rotations, and reflections~\cite{marsden2013introduction}. These symmetries are encoded by the \textit{invariance} of scalar quantities (e.g., total energy) and the \textit{equivariance} (or covariance) of vector and tensor quantities (e.g., forces, dipole moments), meaning such quantities transform consistently with the geometry under operations from $E(3)$.

More generally, a function $f$ is \emph{invariant} if $f(Tx) = f(x)$ and \emph{equivariant} if $f(Tx) = T f(x)$, where $T$ denotes a spatial transformation (e.g., rotation, translation, or reflection) acting on the input $x$. For out-of-sample predictions, MLPs should respect these symmetries by predicting invariant energies and equivariant forces, ensuring physically consistent behavior under arbitrary spatial configurations. This is achieved not by explicitly identifying transformations, but by embedding symmetry constraints into the model architecture, so that the outputs transform correctly by design, even when the specific transformation was not encountered during training.

\paragraph*{Equivariance via group representations}
Let $G$ be a symmetry group (e.g., $E(3)$, see Table~\ref{tab:groups} for more examples), and let $f : X \to Y$ be a function between vector spaces. The function $f$ is said to be \emph{equivariant} with respect to $G$ if
\begin{equation}
    f(D_X[g]\,x) = D_Y[g]\,f(x), \quad \forall g \in G,
    \label{eq:symmeq1}
\end{equation}
where $D_X[g]$ and $D_Y[g]$ are representations of the group element $g$ acting on the input and output spaces, respectively~\cite{cohen2016group,kondor2018clebsch,pitts2013nominal,auslander2014groups}. 

Eq.~\ref{eq:symmeq1} formalizes the notion of equivariance in terms of group actions. It states that applying a group transformation $g \in G$ to the input $x$, and then evaluating the function $f$, is equivalent to first evaluating $f(x)$ and then applying the corresponding transformation to the output. The maps $D_X[g]$ and $D_Y[g]$ are linear representations of the group $G$ on the input and output spaces, capturing how elements of $G$ act on vectors in $X$ and $Y$. This condition ensures that the function $f$ respects the symmetry structure defined by $G$ and transforms outputs in a consistent way under group actions.
If $D_Y[g] = \mathbb{I}$ (identity element) for all $g$, then $f$ is \emph{invariant} under $G$
\begin{equation}
    f(D_X[g]\,x) = f(x).
\end{equation}

\begin{figure*}[ht]
\centering
\includegraphics[width=\linewidth]{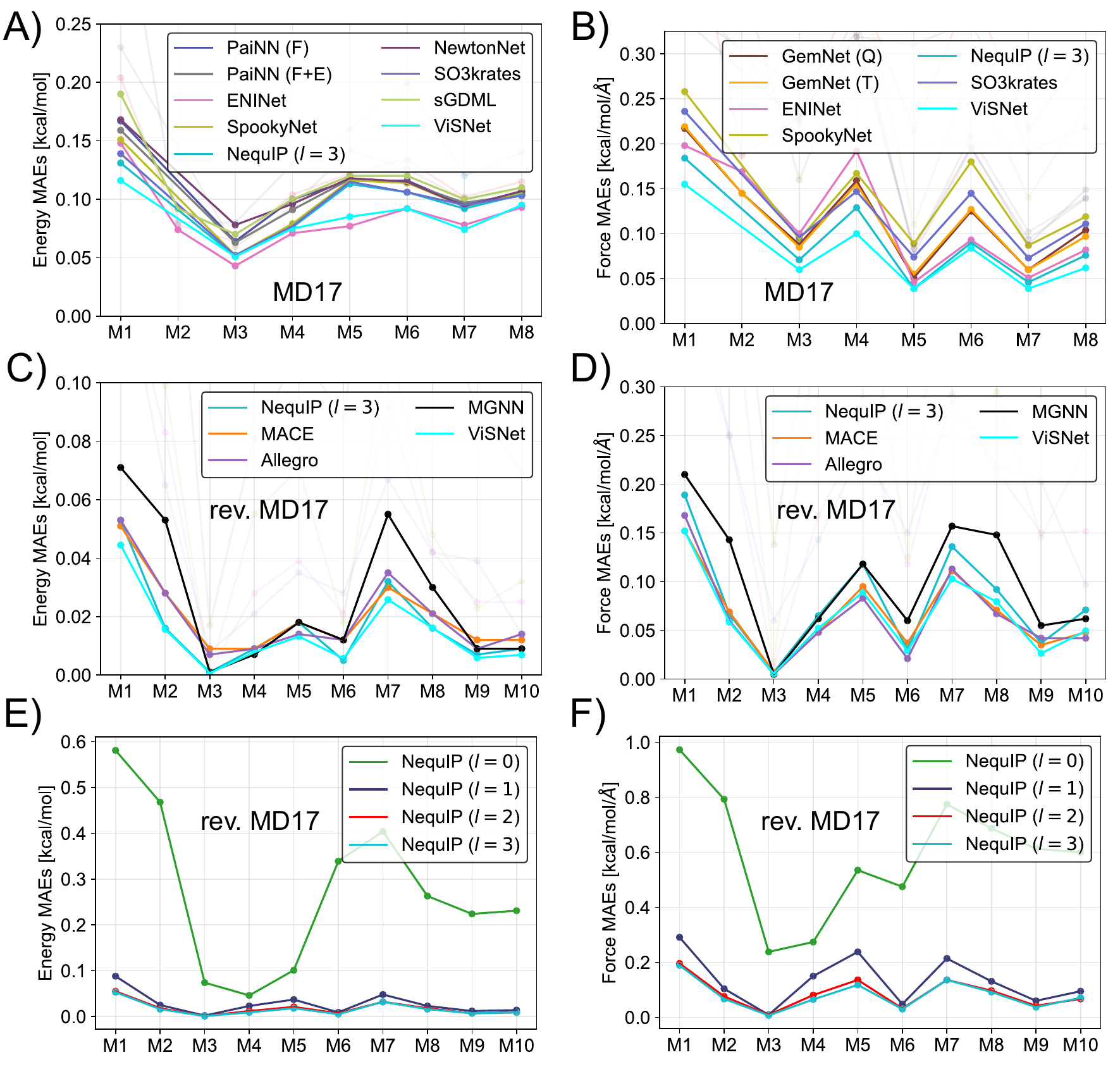}
\caption{
Mean absolute errors (MAEs) from 50 out-of-sample predictions (training: 950 samples; test: 50) for energies (left panels) and forces (right panels), evaluated on the MD17 (A, B) and revised MD17 (C--F) datasets using various MP-GNN-MLPs. 
Panels A and B: MD17 molecules M1--M8: aspirin, benzene, ethanol, malonaldehyde, naphthalene, salicylic acid, toluene, and uracil. 
Panels C--F: revised MD17 molecules M1--M10: aspirin, azobenzene, benzene, ethanol, malonaldehyde, naphthalene, paracetamol, salicylic acid, toluene, and uracil. 
\textit{Methods (Panels A--B):} SchNet, DimeNet, PhysNet, GemNet-Q, GemNet-T, PaiNN (force-only and joint energy/force variants), ENINet, SpookyNet, NequIP ($\ell = 3$), NewtonNet, SO3krates, sGDML, ViSNet. 
\textit{Methods (Panels C--D):} NequIP ($\ell = 0$--$3$), MACE, Allegro, FCHL19, GAP (SOAP), sGDML, ACE, MGNN, ViSNet. 
\textit{Panels E--F:} Performance of NequIP for $\ell = 0, 1, 2, 3$. 
Only the top-performing models are highlighted in bold for clarity. 
The original data used to create this figure and other details are provided in the Supplementary Material. 
}
\label{fig:md17}
\end{figure*}

These expressions describe how physical quantities behave under symmetry transformations applied to the molecular configuration $\mathcal{R}=\{ {\bf r}_i \}$.

\begin{itemize}
    \item \textbf{Energy is invariant:} $E(\mathcal{R}) = E(g \cdot \mathcal{R})$ \\
    This means that the total energy of the system does not change under a transformation $g$ from the symmetry group. Here, $g \cdot \mathcal{R}$ denotes the transformed configuration. This invariance reflects a core physical principle:  scalar observables such as energy remain unchanged under coordinate transformations.

\item \textbf{Force on an atom transforms equivariantly:} $\mathbf{F}_i(\mathcal{R}) \to \mathbf{R} \mathbf{F}_i(\mathcal{R})$ under rotation $\mathbf{R}$ \\
The force $\mathbf{F}_i$ is a local vector quantity acting on atom $i$ and depends on the atomic configuration $\mathcal{R}$. When the entire system is rotated by a matrix $\mathbf{R}$, the force on each atom must also rotate accordingly: the vector itself changes, but the physical relation between geometry and force remains consistent. This transformation property, preserving directional relationships under symmetry operations, is the defining feature of equivariance.

\item \textbf{Dipole moment (and other global vectors) transform equivariantly:} \\
The dipole moment $\boldsymbol{\mu}$, defined as the weighted sum of atomic positions $\mathbf{r}_i$ with partial charges $q_i$,  $\boldsymbol{\mu} = \sum_i q_i \mathbf{r}_i$: transforms as a vector under rotations: $\boldsymbol{\mu} \to \mathbf{R} \boldsymbol{\mu}$. In the continuum limit, it is expressed as an integral over the charge density: $\boldsymbol{\mu} = \int \rho(\mathbf{r}) \mathbf{r}\, d^3r$. More broadly, any global vector constructed from position-dependent quantities, such as total angular momentum or net force, exhibits the same equivariant behavior under spatial transformations.

\end{itemize}

\paragraph*{Equivariant MP-GNN-MLPs}
Traditional message-passing neural networks maintain invariance by encoding only scalar features based on distances or angles. In contrast, \emph{equivariant neural networks} (e.g., NequIP~\cite{batzner20223}, Allegro~\cite{musaelian2023learning}, tensor field networks~\cite{thomas2018tensor}) are designed so that intermediate representations transform under the irreducible representations (irreps) of rotation groups such as $SO(3)$ or $O(3)$ (see Table~\ref{tab:mpgnn-summary-readout} for a comparison across MP-GNN-MLPs).
These irreps are indexed by angular momentum order $\ell \geq 0$ and parity $p \in \{-1, +1\}$. Examples are
\begin{itemize}
    \item Scalars: $(\ell=0, p=+1)$ (e.g., energy),
    \item Vectors: $(\ell=1, p=-1)$ (e.g., forces),
    \item Pseudovectors: $(\ell=1, p=+1)$ (e.g., angular momentum).
\end{itemize}
Angular momentum is a pseudovector (or axial vector) with positive parity because it is defined as the cross product of position and momentum vectors
\begin{equation}
    \mathbf{L} = \mathbf{r} \times \mathbf{p},
\end{equation}
and under a parity (spatial inversion) transformation, both $\mathbf{r}$ and $\mathbf{p}$ change sign, i.e., $\mathbf{r}, \mathbf{p} \to -\mathbf{r}, -\mathbf{p}$, but their cross product ($\mathbf{L}$) remains unchanged.

Equivariant GNNs maintain symmetry by composing features through tensor products that respect the underlying group structure. In the context of $O(3)$ or $E(3)$ symmetry, features are labeled by irreps indexed by angular momentum index $\ell$ and parity $p \in \{-1, +1\}$. A tensor product between two features $T_{\ell_1 p_1}$ and $T_{\ell_2 p_2}$ yields a direct sum of irreps $T_{\ell_{\text{out}} p_{\text{out}}}$ satisfying
\begin{equation}
    |\ell_1 - \ell_2| \leq \ell_{\text{out}} \leq \ell_1 + \ell_2, \quad
    p_{\text{out}} = p_1 p_2.
\end{equation}

This rule mirrors angular momentum coupling in quantum mechanics~\cite{sakurai2020modern}: the composition of a degree-$\ell_1$ and degree-$\ell_2$ feature yields a range of angular components $\ell_{\text{out}}$ from $|\ell_1 - \ell_2|$ up to $\ell_1 + \ell_2$. Parity is multiplicative, so two even-parity features ($p = +1$) yield an even-parity output, whereas combining even and odd gives odd ($+1 \cdot -1 = -1$).

Such tensor products allow the network to build complex geometric interactions while preserving equivariance. For instance
\begin{itemize}
    \item Scalar $\otimes$ scalar $\to$ scalar: $(\ell=0, p=+1) \otimes (\ell=0, p=+1) \to (\ell=0, p=+1)$,
    \item Vector $\cdot$ vector $\to$ scalar: $(\ell=1, p=-1) \otimes (\ell=1, p=-1) \to (\ell=0, p=+1)$,
    \item Vector $\times$ vector $\to$ pseudovector: $(\ell=1, p=-1) \otimes (\ell=1, p=-1) \to (\ell=1, p=+1)$.
\end{itemize}

These operations form the backbone of message-passing layers in equivariant models such as NequIP, Allegro, and MACE, enabling them to learn rich geometric representations while strictly enforcing symmetry constraints.

In NequIP, for example, messages from atom $j$ to atom $i$ are formed as tensor products between spherical harmonic–expanded edge vectors and equivariant node features, ensuring $E(3)$-equivariance at every layer. This design yields smoother potential energy surfaces and accurate directional response without requiring hand-crafted symmetry functions.

\begin{table*}[!htpb]
\centering
\caption{ Reported error metrics for machine-learned interatomic potentials (MLPs) used in atomistic modeling applications. Mean absolute deviation (MAD) and root mean square deviation (RMSD) are the most commonly reported measures of model accuracy for total energies, forces, and solvation free energies.  All acronyms are defined in the main text.  }
\label{tab:mlp_errors}
\begin{tabular}{l cc ccc}
\hline 
{Property}       & Metric & {Reported error} & {Unit} & {Reference level} & {Source} \\
\hline 
Total energy            & MAD       & 0.1           & kcal/mol      & classical, PB implicit solvation & \cite{liao2024calculation}\\
                        & MAD       & 0.4-0.8       & meV/atom      & DFT       & \cite{yang2023neural}\\
                        & MAD       & 0.04          & kJ/mol        & CCSD(T)   & \cite{schran2018high}\\
                        & MAD       & 0.1           & eV/atom       & DFT       & \cite{chen2023accelerating} \\
                        & RMSD      & 0.1           & kcal/mol/atom & DFT       & \cite{celerse2024capturing}\\
                        & RMSD      & 0.9           & kcal/mol      & DFT       & \cite{yao2023machine} \\
                        & RMSD      & 4$\times10^{-4}$ & eV/atom    & DFT       & \cite{anmol2024unveiling} \\
                        & RMSD      & 0.3-0.5        & meV/atom     & DFT       & \cite{brezina2024elucidating} \\
&&&&\\
Solvation free energy   & MAD       & 0.09          & kcal/mol      & DFT/Exp.   & \cite{vermeire2021transfer} \\
                        & MAD       & 0.37          & kcal/mol      & Exp.       & \cite{yu2023solvbert} \\
&&&&\\
Forces                   & MAD       & 12.6-16.3          & meV/\AA{}      & DFT       & \cite{yang2023neural} \\
                        & MAD       & 0.06               & eV/\AA{}       & DFT       & \cite{chen2023accelerating} \\
                        & RMSD      & 100                & meV/\AA{}      & DFT       & \cite{celerse2024capturing}\\
                        & RMSD      & 0.4                & kcal/mol/\AA{} & DFT       & \cite{yao2023machine} \\
                        & RMSD      & 7.2$\times10^{-2}$ & eV/\AA{}/atom  & DFT       & \cite{anmol2024unveiling} \\
                        & RMSD      & 33.1-69.7          & meV/\AA{}/atom & DFT       & \cite{brezina2024elucidating} \\
\hline 
\end{tabular}
\end{table*}
\subsection{Model performances}

Many of the models discussed in this section have been benchmarked in the MLP studies 
on the QM9, MD17, and revised MD17 datasets. To provide a comparative evaluation of their performance, we have compiled mean absolute errors (MAEs) for energies and forces on the MD17 and revised MD17 datasets from the original studies and plotted them in Figure~\ref{fig:md17}.

In the original MD17 dataset, ENINet and ViSNet report the lowest MAEs for energies, while ViSNet and NequIP (with $\ell = 3$) are most accurate for MD17 forces. On the revised MD17 dataset, ViSNet, MACE, and Allegro perform well across most molecules. 

The performance of NequIP across tensor orders $\ell=0$ to $\ell=3$ (Panels E--F) shows that force MAEs drop significantly as $\ell$ increases from 0 to 2, with diminishing changes between $\ell = 2$ and $\ell = 3$. This illustrates the importance of high-rank tensor channels for accurately capturing directional interactions.

\begin{figure}[ht]
\centering
\includegraphics[width=\linewidth]{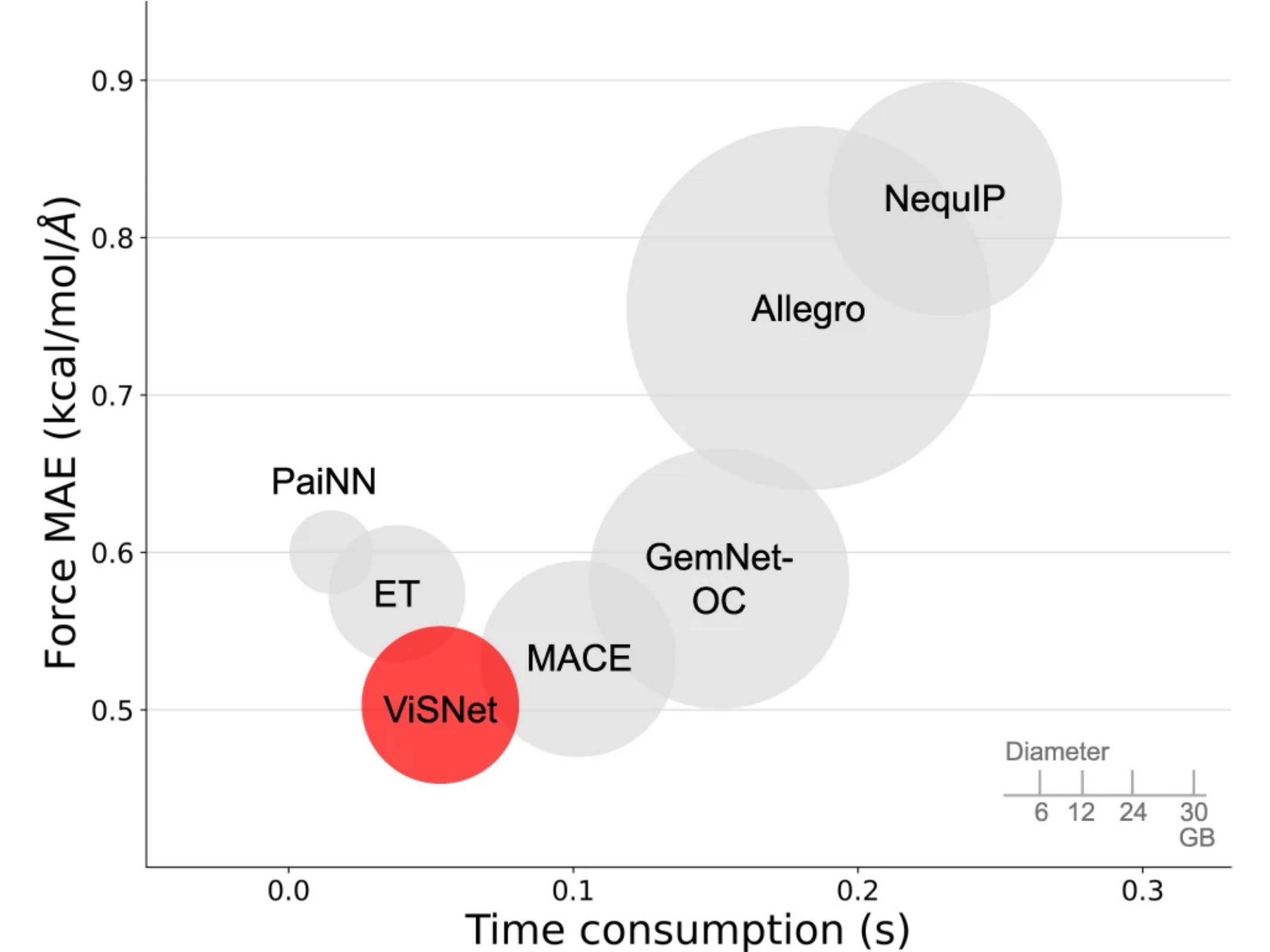}
\caption{
Performance comparison of ViSNet (red) against PaiNN, ET (Equivariant Transformer)~\cite{tholkeequivariant}, MACE, GemNet-OC (a version of GemNet~\cite{gasteigergemnet} pretrained on the Open Catalyst 2020 dataset), Allegro, and NequIP (grey) for the Chignolin dataset. Axes represent accuracy (vertical) and training time (horizontal); memory usage is depicted as bubble volume. 
Figure from~\RRef{wang2024enhancing}, used under a Creative Commons license.
}
\label{fig:mae_time}
\end{figure}

Although we do not systematically evaluate computational efficiency in this review (e.g., scaling, memory usage, or GPU performance), we highlight results from the ViSNet study~\cite{wang2024enhancing} that benchmark runtime and memory usage on a larger system. 
Figure~\ref{fig:mae_time} summarizes accuracy, training time, and memory consumption for multiple MLPs on the 166-atom Chignolin mini-protein dataset. This benchmark includes about 10,000 conformations and reflects large-scale molecular settings.

Among the tested models shown in Figure~\ref{fig:mae_time}, ViSNet achieved the best accuracy, followed by MACE. However, PaiNN demonstrated the fastest training and lowest memory usage, highlighting its suitability for large-scale simulations. Allegro, while one of the top performers on MD17, was found to be memory-intensive and slower for Chignolin. These results highlight the trade-off between accuracy and computational efficiency across architectures.

\subsection{Error metrics}

Evaluating the accuracy of MLPs for solvation modeling requires careful consideration of both raw and normalized error metrics. Table~\ref{tab:mlp_errors} summarizes reported error values for key physical properties across various MLP frameworks.

Energies are typically benchmarked against quantum chemical methods (e.g., DFT, MP2, or CCSD(T)), with mean absolute deviations (MADs) reported in meV/atom or kcal/mol. MAD below 1~kcal/mol (about \(k_BT\) at room temperature) is generally considered chemically accurate. For dynamical simulations, however, force accuracy is more critical. Root mean square deviations (RMSDs) below 0.05~eV/\AA{} (about 0.00097 hartree/bohr) are typically sufficient for MD simulations, vibrational spectra, and solvation structures. Since each configuration contributes $3N$ force components, force-based training provides a dense supervisory signal.

Beyond direct error metrics, models are evaluated based on their ability to reproduce experimentally observable or trajectory-derived quantities, such as solvation free energies (\(\Delta G_{\text{solv}}\)), RDFs, hydration numbers, or diffusivities. Both energy and force accuracies contribute to the reliability of these higher-level properties, with force fidelity crucial for trajectory-based observables.

While RMSD and MAD offer absolute measures of prediction error, their usefulness in multi-property contexts is limited due to the different scales of physical quantities. Morawietz {\it et al.}~\cite{morawietz2012neural} addressed this using the normalized RMSD (nRMSD), defined as
\begin{equation}
    \textrm{nRMSD} = \frac{\textrm{RMSD}}{\sigma},
\end{equation}
where \(\sigma\) is the standard deviation of the reference values. Their NN-MLPs for water monomers and dimers yielded nRMSDs below 2\% for energies, forces, and charges, highlighting the benefit of contextualizing errors relative to the data distribution.

In multi-property learning, each target property can be normalized by dividing by its maximum value to ensure uniform scaling~\cite{ramakrishnan2015many}. A more general alternative is to standardize each property using its mean and standard deviation~\cite{igual2017introduction}
\begin{equation}
    {\bf z} = \frac{{\bf p} - \mu}{\sigma},
    \label{eq:normscore}
\end{equation}
where \(\mu\) and \(\sigma\) are computed from the training set. This standard score transformation, which is commonly implemented in data science toolkits such as \texttt{sklearn.preprocessing.StandardScaler}~\cite{pedregosa2011scikit}, makes prediction errors comparable across heterogeneous properties such as energies, forces, dipole moments, and solvation free energies.

Such normalization is crucial in solvation modeling, where the magnitudes of relevant properties span multiple orders: covalent bond changes (about 100~kcal/mol), hydrogen bonding (about 10~kcal/mol), and conformational fluctuations (<1~kcal/mol). 
Although 1~kcal/mol is often cited as a chemically meaningful threshold, standardized or relative error metrics, normalized to the spread of the target property, offer a more consistent and reliable basis for comparing model performance across diverse solvation tasks.

Normalization strategies are also relevant for function-valued targets, such as absorption spectra, RDFs, or SDFs. In such cases, bin-wise normalization of the target values ensures dimensionless error metrics~\cite{kayastha2022resolution}, enabling consistent evaluation across spectral or structural observables.

\subsection{Deployment in solvation modeling}

The integration of MLPs into solvation modeling workflows relies on synergistic coordination between software infrastructure, reproducible datasets, deployment strategies, and model comparison practices. This section outlines the key components enabling MLP-based simulations of solvated systems, ranging from training frameworks and benchmarking efforts to hybrid modeling strategies and deployment in AIMD.

A variety of open-source tools support the training and deployment of MLPs in solvation modeling, some of which are collected in Table~\ref{tab:software_frameworks}. The recent editorial survey~\cite{rupp2024guest} highlights community-wide efforts to develop software ecosystems for atomistic ML, with a specific focus on integrating MLPs into the current tooling landscape, which includes simulation engines, workflow automation, and dataset infrastructure. 

\begin{table*}[!htpb]
    \centering
    \caption{Some open-source software frameworks for MLP development.}
    \label{tab:software_frameworks}
    \begin{tabular}{llll}
        \hline 
        Software & Primary interfaces & Backends & Application \\
        \hline 
        DeePMD-kit~\cite{wang2018deepmd}        & LAMMPS, i-PI, GROMACS        & Tensorflow & DeepMD training and large-scale MD \\
        NequIP~\cite{batzner20223}              & LAMMPS, ASE    & PyTorch & Equivariant MLPs for directional interactions \\
        SchNetPack~\cite{schutt2018schnet}      & LAMMPS       & PyTorch & MLP training, atom-wise embeddings \\
        i-PI~\cite{kapil2019pi}                 &  LAMMPS, etc.       & Python& Classical and path-integral MD with MLPs \\
        GAP and QUIP~\cite{bartok2010gaussian}  & LAMMPS, ASE, CASTEP (LOTF, AIMD)   & & Gaussian approximation potentials \\
        Dscribe~\cite{himanen2020dscribe}       &          &Python& Descriptor generation (SOAP, etc.) \\
        QML~\cite{QML}                          &         & Python& Descriptor generation (SLATM, etc.), kernel models \\
        Autosolvate~\cite{hruska2022autosolvate}&        &  Python& Setup for explicit solvation environments \\
         TorChANI~\cite{gao2020torchani}&  & PyTorch & Training  of ANI \\ 
        \hline 
    \end{tabular}
\end{table*}

Reproducibility in MLP-aided solvation modeling depends on detailed reporting of training data, model architectures, simulation parameters, and validation metrics. Beyond ML benchmarks, rigorous quantum chemical benchmarking also plays a key role in validating interaction energies in solvated systems. Chen {\it et al.}~\cite{chen2020accurate} systematically assessed DFT, composite, and DLPNO-CCSD(T)/CBS methods on glycine–water clusters up to 40 waters, showing that range-separated hybrids like B3LYP-D3(BJ) and $\omega$B97M-V achieve MADs of 2–4 kJ/mol relative to high-level wavefunction theory. 
These error margins serve as a reference for validating MLP predictions on microsolvated clusters.
Transfer learning from quantum chemistry to experiment has also emerged as a strategy for mitigating data scarcity and experimental noise in solvation modeling. Vermeire and Green~\cite{vermeire2021transfer} demonstrated that pretraining MP-NN-MLPs on COSMO-RS–derived solvation free energies (CombiSolv-QM) and fine-tuning on curated experimental data (CombiSolv-Exp) can reduce test MAEs from ~0.47 to ~0.21 kcal/mol, and as low as 0.09 kcal/mol on high-accuracy subsets. 

Benchmark datasets such as MD17~\cite{chmiela2017machine}, revised-MD17~\cite{christensen2020role}, QM9~\cite{ramakrishnan2014quantum,qm9pack}, and Open Catalyst~\cite{chanussot2021open} have established standard protocols for evaluating the performance of MLPs across diverse chemical systems. These datasets have supported the development of widely used MLPs (e.g., ANI, PhysNet, SchNet) and have also been extended to enable the training of reactive MLPs suitable for solvation modeling such as CH-GAP~\cite{ibragimova2025unifying}.
Recent benchmarking efforts~\cite{zheng2024revisiting} in aqueous solubility prediction further emphasize the role of dataset quality, descriptor design, and model robustness. Gao {\it et al.}\cite{gao2022accurate} demonstrated that DNNs trained on curated descriptors offer improved transferability across multiple test sets, compared to GCN models trained on the same solubility data. 
In a related effort, Gao {\it et al.}~\cite{gao2020accurate} used MGCN and SchNet architectures to predict aqueous solubility directly from 3D molecular graphs, bypassing handcrafted descriptors. SchNet achieved the best performance across multiple benchmarks, indicating the strength of end-to-end, structure-based models for solvation properties.

MLPs can be deployed to predict solvation-relevant properties such as hydration free energies, solvent-induced reactivity, and noncovalent interactions. For example, Temel {\it et al.}~\cite{temel2023performance} showed that ANI-based MLPs trained on explicit solvent systems can reproduce DFT-level hydration free energies via simple linear scaling. 
An alternative to direct force field training is the free energy machine learning (FML) framework~\cite{weinreich2021machine}, which learns solvation free energies of the QM9 molecules as ensemble properties. FML uses Boltzmann-weighted averages of FCHL representations derived from short MD trajectories, enabling systematic improvement of prediction accuracy with training set size.
Recent work by Jindal {\it et al.}~\cite{jindal2025computing} demonstrates how atomic polar tensor (APT)–based equivariant NN-MLPs, trained exclusively on finite water clusters, can accurately reproduce the IR spectrum of bulk liquid water. Their results highlight the ability of localized models, when properly embedded (e.g., via continuum solvation), to bridge finite-size calculations and condensed-phase observables. 

Descriptor engineering remains a critical and evolving area, with growing evidence that carefully designed representations can significantly enhance MLP sensitivity in reactive systems~\cite{lange2024comparative,zhang2024modelling,ibragimova2025unifying}.
The MolMerger algorithm~\cite{ramani2024graph} introduces a GNN-based model that explicitly encodes solute–solvent interactions via charge-guided graph construction, enabling robust generalization to unseen solvents and highlighting the importance of interaction-aware featurization for ML deployment in chemically diverse solvent environments.

MLPs enable DFT-quality MD simulations at a lower cost, particularly for solvated systems where solvent reorganization and H-bond dynamics are critical. When used with explicit solvent, MLP-based MD simulations have been applied to study pK$_\text{a}$ shifts, solvent–solute structuring, and thermodynamic fluctuations~\cite{steinmann2016solvation,zhang2024modelling}. Microsolvated cluster models and cluster–continuum schemes provide tractable approximations for systems such as protonated solutes~\cite{celerse2024capturing,tavzrtscher2025automated}, while solvent placement via grid-based free energy solvation theory~\cite{steiner2021quantum} further improves structural fidelity.

Model selection in solvation contexts must consider trade-offs among accuracy, interpretability, scalability, and computational overhead. Kernel-based models (e.g., GAP, ACE) provide excellent performance on small data regimes, but scale poorly with dataset size. Neural architectures (e.g., SchNet, PhysNet, NequIP) scale well and capture complex many-body interactions. Equivariant models such as NequIP and MACE excel at solvent shell structure and polarization modeling, though they require GPU-accelerated training and careful tuning. The AIMNet2 model~\cite{anstine2025aimnet2} exemplifies a transferable MLP trained on a chemically diverse dataset spanning 14 elements and both neutral and charged species. It couples message-passing–based local environments with physics-informed long-range electrostatics and dispersion terms, allowing general-purpose deployment without system-specific retraining. Benchmarking efforts such as the Solubility Challenge~\cite{conn2023blinded} or other data curation efforts~\cite{llompart2024will,marques2025self} help standardize comparison.

Hybrid solvation models combine explicit or MLP-treated solute and solvent shells with continuum representations. PCM and COSMO~\cite{chen2023accelerating,ehlert2021robust} are commonly used for long-range effects, while the first solvation shell is modeled with MLPs or {\it ab initio} methods. In more complex schemes, QM/MM-style partitions treat the chemically active region with DFT, the immediate solvent environment with MLPs, and the remainder classically. 
R\"ocken {\it et al.}~\cite{rocken2024predicting} introduced ReSolv, a NequIP-based MLP trained in two stages: first on vacuum DFT data from QM7-X~\cite{hoja2021qm7}, and then reweighted using free energy perturbation to match experimental hydration free energies from FreeSolv~\cite{riquelme2018hydration}. 
Hybrid MLIP–classical schemes are also explored in metallic or organic–inorganic interfaces~\cite{wan2024construction}.

Recent efforts have also focused on developing solvation-aware MLPs that combine physical interpretability with deployment in dynamic environments. The PairF-Net model~\cite{ramzan2022machine} introduced a pairwise decomposition of atomic forces based on interatomic vectors that ensures rotational and translational invariance. Kalayan {\it et al.}~\cite{kalayan2024neural} extended this to an energy-conserving framework capable of force prediction in solvated geometries, despite being trained only on gas-phase data. Their PairF-Net implementation supports ML/MM simulations in OpenMM and captures solvent-induced conformational shifts, with the rMD17-aq dataset serving as a benchmark for aqueous-phase performance. Further, Williams {\it et al.}~\cite{williams2024stable} proposed PairFE-Net, trained on metadynamics-enhanced datasets, to enable long-timescale stable MD for flexible molecules. Their work highlights the need for conformational completeness to ensure reliable free energy predictions under solvated conditions.

These strategies enable MLPs to be used flexibly across the solvation modeling landscape, from small molecules in water to complex biomolecular and interfacial environments.

\subsection{Sampling of configurational space\label{subsec:sampling}}

Efficient and accurate sampling of configurational space is a recurring theme across the development, training, and deployment of MLPs for solvation modeling. Case studies surveyed in Sections~\ref{sec:A1} and~\ref{sec:A2} illustrate how sampling-enhancing strategies, including active learning, metadynamics, replica exchange, and umbrella sampling, improve configurational diversity and ensure reliable prediction of solvation thermodynamics and reactivity.

Although the general methodology for building MLPs has been widely reviewed~\cite{rupp2015machinenut,pan2024training,dral2022quantum,miksch2021strategies,artrith2021best,handley2010potential,yang2024machine,behler2015constructing,witkoskie2005neural,jacobs2025practical,chen2025application,wang2025design,zhang2023artificial}, the adaptation of these strategies to solvation is non-trivial. Solvent fluctuations, hydrogen bonding, and long-range polarization introduce significant configurational complexity. In bulk systems, dynamic fluctuations decay with configurational sampling, but in microsolvated clusters, statistical uncertainties are amplified due to reduced degrees of freedom and lack of ensemble averaging.

\begin{figure*}[htpb]
\centering
\includegraphics[width=1.0\linewidth]{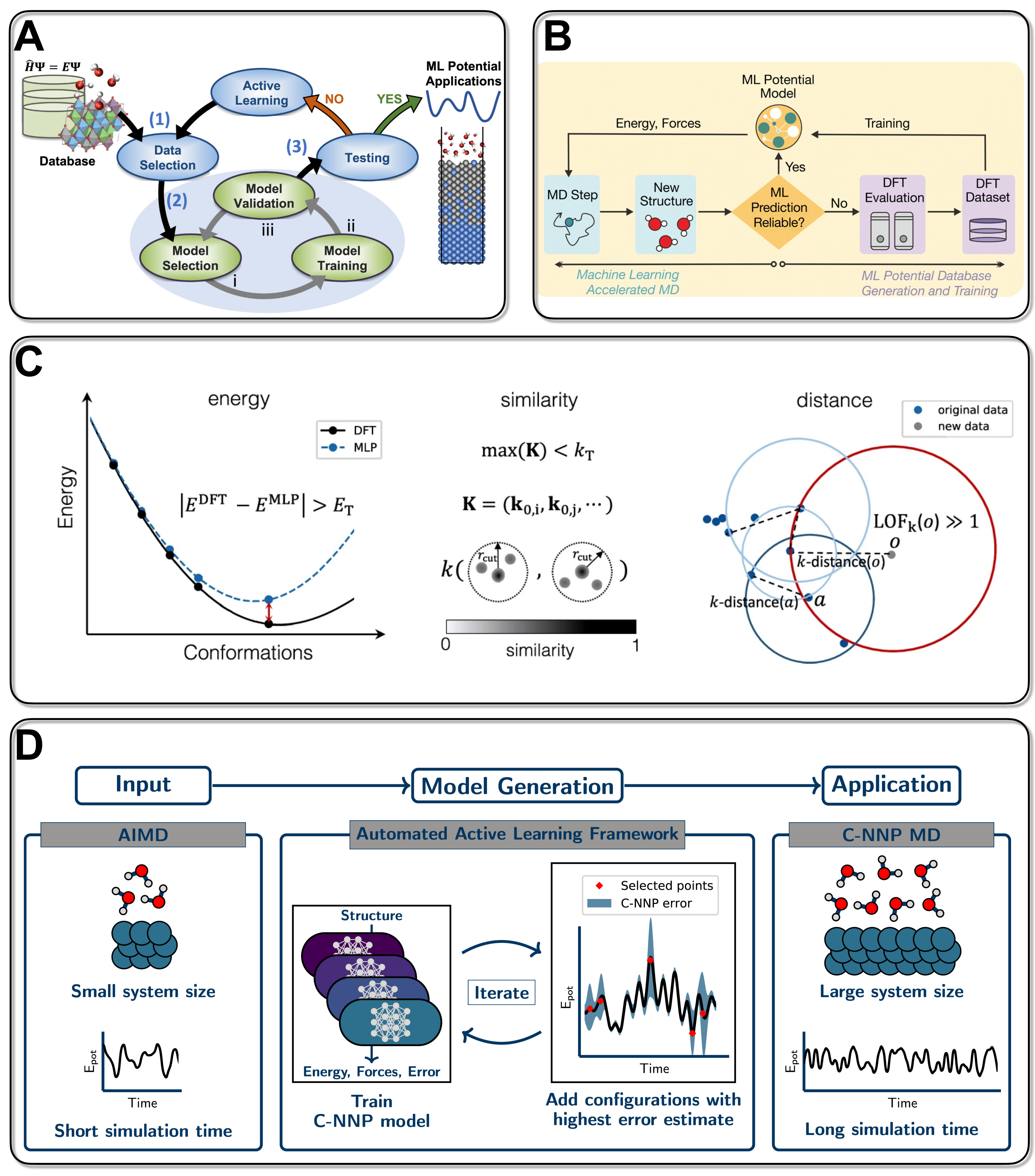}
\caption{
Representative active learning or learning-on-the-fly strategies for MLP-based solvation modeling:
(A) Generic active learning loop (adapted from~\cite{miksch2021strategies}),
(B) ML-accelerated MD workflow for catalytic processes (adapted from~\cite{chen2023accelerating}),
(C) Energy-based structure selection for Diels--Alder reactions in water/methanol (adapted from~\cite{zhang2024modelling}),
(D) C-NNP training at solid--liquid interfaces (adapted from~\cite{schran2021machine}).
All subfigures are adapted from the original works for scholarly synthesis and commentary. Original copyrights remain with the respective publishers.
}
\label{fig:activelearning}
\end{figure*}

Sampling-enhancing techniques, namely \gls*{qbc}~\cite{schran2021machine,schran2020committee}, energy-based selection~\cite{zhang2024modelling}, and extrapolation-grade filtering~\cite{chen2023accelerating}, are increasingly employed in both training and validation workflows. 
A recent contribution by Pan et al.~\cite{pan2025fast} demonstrates how pairing large-scale conformer enumeration with deep learning enables accurate tautomer ratio predictions in aqueous solution.
Active learning or \gls*{lotf} is particularly useful for dynamically expanding datasets during MD simulations~\cite{csanyi2004learn}. Methods like DP-GEN~\cite{zhang2020dp}, FLARE~\cite{yu2021machine}, and AL4GAP~\cite{guo2023al4gap} rely on uncertainty estimation and targeted retraining to refine force fields during simulations systematically. C-NNP models~\cite{schran2021machine} and DeepPot workflows~\cite{celerse2024capturing} exemplify successful LOTF integration.

Figure~\ref{fig:activelearning} illustrates representative workflows that integrate active learning 
or on-the-fly strategies into MLP-based solvation modeling. 

Recent work by Pattanaik et al.~\cite{pattanaik2023confsolv} exemplifies data-driven approaches to configurational sampling by generating a large-scale dataset of $\sim$9 million solute conformers across 41 solvents and training an MP-GNN-MLP to predict relative free energies in solution. Their model enables rapid identification of low-energy conformers relevant to solvation thermodynamics, significantly reducing the need for exhaustive first-principles calculations. 
These examples highlight how different studies have addressed the challenge of sampling configurational space during model training and deployment.

Metadynamics, umbrella sampling, and replica exchange techniques bias sampling toward rare events such as hydrogen-bond rearrangements, proton transfer, and solvation shell exchange. The differentiability and low cost of MLPs make them ideal for coupling with these strategies. \gls*{pimd} simulations further enable accurate modeling of \gls*{nqes}~\cite{liu2022toward}. MLPs can be integrated into umbrella sampling schemes, where configurations are sampled under harmonic restraints along collective variables (CVs) and later combined using weighted histogram analysis methods (WHAM)~\cite{kumar1992weighted} or multistate Bennett acceptance ratio (MBAR)~\cite{bennett1976efficient,shirts2008statistically}. When multiple system copies are simulated at varying conditions and swapped, \gls*{remd} further improves ensemble convergence~\cite{fabregat2020hamiltonian,sugita1999replica}. Accelerated MD techniques also benefit from MLP efficiency, and CV-based approaches expand configurational reach~\cite{goetz2020insights}. A conceptually distinct approach is offered by Boltzmann generators~\cite{noe2019boltzmann}, which use invertible NNs to learn a direct mapping between a simple latent distribution (e.g., Gaussian) and the Boltzmann-weighted configuration space, showing promise for overcoming rare event bottlenecks in solvation modeling.

A recent study by Mendible-Barreto et al.~\cite{mendible2025considerations} offers a detailed analysis of how the choice and distribution of CVs in training datasets affect the ability of equivariant NN-MLPs to recover \gls*{fes}s. By systematically comparing training strategies based on classical and {\it ab initio} data for systems like butane and alanine dipeptide, the authors show that even sparse or biased CV sampling can yield accurate FES predictions if high-quality reference energies are used, thereby highlighting a practical route for sampling-aware MLP training.

Studies using DeepPot-SE, ACE, and NequIP potentials demonstrate that combining MLPs with enhanced sampling enables the robust reconstruction of the \gls*{fes} and stable long-time MD in challenging solvent environments. Validation via RDFs,  vibrational density of states (VDOS), t-distributed stochastic neighbor embedding (t-SNE) trajectory comparisons, and free energy benchmarks confirm the effectiveness of these approaches~\cite{chen2023accelerating,rocken2024predicting,celerse2024capturing,yao2023machine}.

Wieder et al.~\cite{wieder2021fitting} used ANI-1ccx with MBAR to predict tautomer solvation free energies.
Free energy perturbation (FEP) and thermodynamic integration (TI) 
\begin{equation}
\Delta G_{\text{solv}} = \int_{0}^{1} \left\langle \frac{\partial E(\lambda)}{\partial \lambda} \right\rangle_{\lambda} \mathrm{d}\lambda
\end{equation}
are also feasible with MLPs, using ensemble-averaged quantities, 
where $E(\lambda)$ is the potential energy along an alchemical path interpolating between the gas-phase and solvated system.

Efficient and diverse sampling is essential not only during training data generation and uncertainty quantification, but also in achieving accurate predictions of solvation thermodynamics, reactivity, and interfacial behavior. These strategies enable broader configurational coverage and improved convergence of key observables in chemically and dynamically complex solvation environments. PIMD with DP-MP2~\cite{liu2022toward} captures NQEs in RDFs and diffusion. However, non-conservative force-only MLPs may produce unphysical energy differences, biasing TI/FEP unless trained on both energy and force to ensure consistency:
Inconsistent total energy, $U$, leads to unreliable free-energy, $F$, estimates as $F=U-TS$. Hence, conservative MLPs are essential for rigorous thermodynamic modeling.

Recent work by Axelrod and Gómez-Bombarelli~\cite{axelrod2023molecular} systematically examines the role of conformer ensembles in molecular property prediction using deep learning, emphasizing the importance of physically realistic conformer sampling, offering practical guidance for incorporating ensemble information into ML workflow. Global structure generators (e.g., ABCluster~\cite{elm2020toward}) provide diverse initial states. Smith et al.~\cite{smith2018less} used \gls*{qbc} to efficiently train ANI-1x using only 25\% of the data, while Kulichenko et al.~\cite{kulichenko2024data} emphasized uncertainty-guided selection. The importance of chemically meaningful and thermodynamically diverse configurations is further emphasized by the use of force diversity metrics~\cite{botu2015learning}, electrostatic screening~\cite{behler2021machine}, dimensionality reduction~\cite{friederich2021machine}, and benchmark diagnostics~\cite{pinheiro2021choosing}. For spectroscopically sensitive solvation (e.g., hydrogen bonding), the centroid-based self-consistent-charge density-functional tight-binding method (SCC-DFTB) has also been proposed~\cite{brezina2024elucidating}.

In summary, diverse sampling strategies, from enhanced CV-based sampling and LOTF to force-conservative learning, are crucial for accurate and transferable MLP-based solvation modeling.

\section{Case Studies: Trends and Themes\label{sec:catalogue}}

The use of MLPs in solvation modeling has rapidly expanded, enabling first-principles simulations at reduced computational cost across a wide spectrum of systems and tasks. To clarify this growing landscape, we organize representative studies into a structured catalogue based on three criteria:  
(1) the application type,  
(2) the solvation paradigm, and  
(3) the MLP architecture.

This classification facilitates cross-comparison of methodologies and highlights promising directions for model development. It also serves as a diagnostic framework for identifying underexplored combinations of solvent treatment and model design.

We distinguish two primary application types, which form the two main subsections of this catalogue of case studies:
\begin{enumerate}
    \item \textbf{A1: Solvation-aware PES Construction with MLPs for dynamics and thermodynamics.}  
Here, MLPs are trained to reproduce the PES in solvated systems and are subsequently used to perform \gls*{md} simulations, free energy sampling, or reaction path analysis. These studies emphasize generating and using an accurate PES under solvent effects.
    \item \textbf{A2: Direct prediction of solvation-dependent properties without PES construction.}  
In these studies, MLPs are not used to model full PESs. Instead, they are trained to predict solvation-dependent properties, such as dipole moments, redox potentials, or solvation energies, directly from the molecular structure or other structural descriptors. These approaches bypass full force field construction and focus on property-specific inference.
\end{enumerate}

Each case study is further cross-classified as a combination of the solvation modeling type (as discussed in \ref{ssec:traditional_paradigms}) and MLP type (as classified in \ref{subsec:classifyMLPs}):

\begin{itemize} 
\item S1: Explicit solvation 
\item S2: Implicit solvation 
\item S3: Hybrid (microsolvation / cluster–continuum) 
\end{itemize}

\begin{itemize} 
\item M1: NN-MLP (energy/force, neural networks)
\item M2: Kernel-MLP (energy/force, kernel)
\item M3: Linear-MLP (energy/force, linear)
\item M4: GD-MLP (gradient-domain)
\item M5: F-MLP (force-only)
\end{itemize}

A summary of all case studies organized under this classification is provided in Table~\ref{tab:case_catalogue}.

\subsection{Solvation-aware PES modeling\label{sec:A1}}  
\subsubsection{Water contact layer at solid/liquid interfaces modeled by NN-MLPs}

\textbf{S1:} Explicit solvation (liquid water on solid surfaces) \\
\textbf{M1:} Descriptor-based NN-MLP (BPNN)

G{\"a}ding {\it et al.}~\cite{gading2024role} applied MD simulations powered by \gls*{cnnps} to investigate the structure of interfacial water at six different solid surfaces: graphene, MoS$_2$, Au(111), Au(100), Pt(111), and Pt(100). The models were built on the BPNN architecture, using \gls*{acsfs} as descriptors, and were trained iteratively on reference {\it ab initio} data using an active learning protocol enhanced by a \gls*{qbc} approach.
To monitor model reliability during training, the authors evaluated the \gls*{rmse} in the potential 
energy per atom. 

The resulting NN-MLPs achieved near {\it ab initio} accuracy in reproducing both forces and energies. Simulated structural observables, such as water density profiles and \gls*{2dpcfs}, were in close agreement with DFT results obtained using the optB88-vdW functional, which effectively captures dispersion interactions. 
Minor discrepancies between ensemble members due to finite-size effects
were revealed in \gls*{rdfs} calculated with smaller simulation boxes.
However, \gls*{rdfs} calculated with larger simulation boxes matched well with \gls*{aimd} trajectories, indicating the absence of such finite-size artifacts. 

This modeling framework revealed that the structure of the water contact layer is sensitive to the chemical composition and crystallographic symmetry of the substrate. Homogeneous, isotropic layers formed on surfaces such as graphene and Au(111), while MoS$_2$, Au(100), and the Pt facets exhibited pronounced anisotropy, spatial inhomogeneity, and even chemisorbed sublayers of interfacial water. Overall, the study highlights how descriptor-based NN-MLPs can not only accelerate MD simulations but also provide accurate, detailed insights into complex aqueous interfacial phenomena, relevant to wetting, hydration, and nanoscale transport.

\subsubsection{Solvation forces for efficient MD}

\textbf{S2:} Implicit solvation using \gls*{pb} electrostatics \\
\textbf{M1:} Descriptor-based NN-MLP (Deep neural network trained on \gls*{pb}-derived energy and force labels using internal coordinates as descriptors)

Liao et al.~\cite{liao2024calculation} developed a \gls*{dnn} model to predict atom-specific solvation free energies and forces, serving as an efficient surrogate to traditional \gls*{pb} electrostatics in implicit solvent models. The model was trained on internal coordinates of small biomolecules (Ala dipeptide and Met-enkephalin), with reference labels derived from GPU-accelerated \gls*{pb} calculations using the SurfPB solver.

To ensure translational and rotational invariance, atomic Cartesian coordinates were transformed into internal coordinates, and force vectors were rotated into a canonical molecular frame. This allowed rapid inference of solvation effects across long MD trajectories. When applied to enhanced sampling MD, the free energy surfaces reconstructed using \gls*{dnn}-predicted solvation forces closely matched those obtained from explicit solvent simulations.

Compared to the PB method, the trained \gls*{dnn} model achieved about $100$-fold speedups for inference on 1000–2000 snapshots, while maintaining Pearson correlation coefficients near 1.0 against reference energies and forces. This approach is particularly well-suited for solvent-inclusive MD of small molecules and may be extended in the future via \gls*{gnn}s for broader transferability.

\subsubsection{Metadynamics of oxygen reduction at Au–water interfaces}

\textbf{S1:} Explicit solvation (Au(100)–water interface with reactive intermediates) \\ 
\textbf{M1:} End-to-end NN-MLP (equivariant GNN: PaiNN)

Yang et al.~\cite{yang2023neural} developed an active-learning-based equivariant GNN potential based on the PaiNN architecture to perform long-timescale reactive \gls*{metad} simulations of the \gls*{orr} at a solvated Au(100) interface. 
The model was trained on configurations sampled from AIMD and enhanced sampling trajectories, encompassing both equilibrium structures and rare, high-energy intermediates. 
Reference energies and forces were obtained using the PBE+D3 functional.

To ensure broad coverage of chemical space, CUR matrix decomposition, which is a low-rank approximation technique to decompose a matrix $\mathbf{A}$ into a compressed matrix product \( \mathbf{C} \times \mathbf{U} \times \mathbf{R} \), was employed during dataset curation, with Euclidean distances in structural space used to identify and retain the most diverse and informative samples. 
The resulting model achieved energy MAEs below 1 meV/atom and force MAEs between 12.6 and 16.3 meV/\AA{}. 

When used in production \gls*{metad} simulations, the model was able to capture the full thermodynamic and structural landscape of the solvated Au(100)–water interface. While the NN-MLP-MD density profiles for bulk water were somewhat smoother than those from AIMD, the authors attributed this to incomplete equilibration and variability in the AIMD data. In contrast, the energy profiles averaged over long trajectories starting from different initial states (e.g., Au(100) with four adsorbed *OH groups) showed excellent convergence, highlighting the robustness of the model over extended timescales.

Notably, the trained potential also demonstrated strong transferability. Even without being explicitly trained on Au(110)–water and Au(111)–water configurations, it was able to predict energies and forces for these systems with low uncertainty, matching the reliability observed in the original Au(100)-water system. This study highlights how symmetry-aware GNN models, when combined with careful dataset construction and active learning, can provide accurate and transferable force fields for chemically reactive, solvent-inclusive interfacial systems.

\subsubsection{Neural network potential for protonated water clusters in helium droplets}

\textbf{S3:} Microsolvation (from a few He atoms to bulk superfluid helium) \\ 
\textbf{M1:} Descriptor-based NN-MLP (BPNN)

Schran et al.~\cite{schran2018high} trained an BPNN-MLP model to simulate helium–solute interactions in protonated water clusters (H$_3$O$^+$ and H$_5$O$_2^+$), spanning both microsolvated and bulk regimes. The NN-MLP was initially trained on two-body interaction energies computed at the \gls*{ccsd_t} level using counterpoise-corrected aug-cc-pVTZ basis sets, ensuring high fidelity to electronic structure data. Radial and angular symmetry functions were employed to preserve translational, rotational, and permutational invariances. When validated against \gls*{ccsd_t} energy grids for fixed solute geometries, the NN-MLP achieved a small \gls*{mad} of about 0.04 kJ/mol.

To incorporate \gls*{nqes}, the trained NNP was coupled to \gls*{aipimd} and \gls*{pimc} simulations at 1.67 K. The \gls*{aipimd} simulations employed the PBE exchange–correlation functional with D3 dispersion corrections for up to two-body terms. Simulations with the NN-MLP accurately captured structural observables such as \gls*{sdfs} and \gls*{rdfs} of helium atoms surrounding the solutes, both for microsolvated states and extended liquid helium environments.

To improve coverage of the \gls*{pes}, a refinement step was introduced by analyzing the symmetry function values from NN-MLP-driven \gls*{pimd} trajectories and identifying structural regions underrepresented in the original training set. For H$_3$O$^+$ and H$_5$O$_2^+$, 3545 and 4101 such structures were retrieved, respectively. These were recomputed at the \gls*{ccsd_t} level and added to the training dataset, further refining the potential and enhancing its accuracy.

Minimal deviations between the retrained NN-MLP and  \gls*{ccsd_t} benchmarks were observed. However, for hydronium configurations microsolvated by 14 helium atoms, minor artifacts appeared in the energy landscape near shallow minima, particularly at larger isovalues between the \gls*{vad} and the reference grid. These discrepancies diminished at smaller isovalues, indicating the shallow nature of the NN-MLP predicted \gls*{pes} in those regions.

Interestingly, for the Zundel cation (H$_5$O$_2^+$) with asymmetric proton sharing, the \gls*{sdfs} showed nontrivial anomalies in helium placement. The helium atoms were displaced due to competing minima on the NN-MLP grid, leading to a slightly shallower interaction potential near the excess proton. Nonetheless, the model captured key structural motifs and reproduced solvation features across a wide range of system sizes, from isolated complexes to bulk-like conditions, demonstrating its versatility and transferability.

\subsubsection{Solvent-driven ring opening of N-enoxyphthalimide}

\textbf{S1:} Explicit solvation (bulk methanol and TFE) \\
\textbf{M1:} Descriptor-based NN-MLP (DeepPot-SE with transfer learning)

C{\'e}lerse et al.~\cite{celerse2024capturing} developed a DeepMD MLP using the DeepPot-SE framework and 
data at the hybrid-DFT-level with transfer learning to investigate solvent-mediated reactivity in bulk-phase environments. Their workflow combined active learning with \gls*{wtmtd} to construct fully converged \gls*{fes}s for the ring-opening reaction of N-enoxyphthalimide in methanol and trifluoroethanol (TFE). These simulations uncovered a diverse ensemble of \gls*{ts} configurations, often stabilized by up to five explicit solvent molecules, reflecting the inherently dynamic nature of solute–solvent interactions.

To balance computational efficiency and accuracy, initial MD trajectories were generated at the GGA-DFT level (PBE), and transfer learning was used to fine-tune the potential to PBE0 hybrid-GGA-DFT quality. This approach avoided the cost of direct hybrid-level AIMD while retaining chemical accuracy, particularly for modeling TS stabilization and solvation effects.

Interestingly, the computed free energy barriers differed notably from past static DFT microsolvation studies (see References 22 and 43 in \RRef{celerse2024capturing}), which tended to overestimate the barrier heights. The authors suggest that these discrepancies stem from two main factors: first, the limited treatment of entropy in static cluster models; and second, the inherent bias introduced by manual placement of solvent molecules, which can lead to large variations in predicted barrier heights. In contrast, their ML-accelerated MetaD approach captures the realistic, thermally accessible ensemble of solute–solvent configurations.

The predicted FESs not only matched experimental trends in solvent selectivity between MeOH and TFE but also highlighted how both nucleophilic interactions and solvent-induced entropic effects shape reactivity. This work demonstrates how ML-enhanced simulations can access complex reaction landscapes and \gls*{ts}s that would be challenging to characterize with traditional static or continuum solvation models.

\subsubsection{Explicit solvent modeling of adsorption and reactivity at metal–water interfaces}

\textbf{S1:} Explicit solvation (water flms on Cu and Pd surfaces) \\
\textbf{M3:} Linear-MLP (MTP with active learning)

Chen et al.~\cite{chen2023accelerating} demonstrated the power of \gls*{mlamd} with active learning to model heterogeneous catalysis under explicit solvation conditions. They employed an MTP-based linear-MLP, trained via Maxvol-based active learning. The resulting model retained DFT-level accuracy (MADs of ~1 meV/atom and 0.04 eV/\AA{}). MTP outperformed SchNet and ANI in predicting water energetics and RDFs, achieving better accuracies even with small training sets (800 DFT configurations) and enabling long-timescale MLaMD simulations of up to 500 ps compared to the AIMD simulations (50 ps).

The study focused on adsorption processes involving CO*, OH*, COH*, HCO*, and OCCHO* on Cu and Pd surfaces in the presence of 32 explicit water molecules (where * denotes an adsorbed species). MLaMD simulations captured structural and energetic solvation effects inaccessible via implicit solvent models or short-time AIMD runs. For instance, the binding energy of OH* on Cu(111) was found to be 0.34 eV lower than in prior AIMD results, likely due to enhanced hydrogen bonding and longer sampling. Structured water layers also influenced adsorption site preferences with OH* favoring top sites in the explicit solvent, which differs from predictions in vacuum or continuum models.

The authors also employed MLaMD-driven \gls*{wtmtd} simulations to compute free energy barriers for C–H bond breaking in ethylene glycol on Cu(111) and Pd(111). The calculated barriers (1.28 eV and 0.93 eV, respectively) were consistent with trends previously obtained using QM/MM potential of mean force approaches.

While effective for neutral systems, most MLPs, such as those used here, struggle to capture long-range electrostatics in charged environments such as electrolytes, where weak dielectric screening makes Coulomb interactions especially important.

\subsubsection{ML-Based implicit solvent forces for alanine dipeptide}

\textbf{S2:} Thermodynamic solvation via average solvent environment configuration (ASEC) \\
\textbf{M1:} NN-MLP (DeepPot-SE representation trained on ASEC forces)

Yao et al.~\cite{yao2023machine} developed a DeepMD-MLP using DeepPot-SE to model solvent-induced forces on alanine dipeptide, trained directly on data from explicit solvent MD simulations. Their approach used an \gls*{asec} built from 5000 MD frames for each of 1296 solute conformations. To guide the learning process, the training also incorporated dihedral angle constraints ($\phi$, $\psi$), ensuring physically relevant sampling of solute conformations.

Using a force-matching strategy, the model achieved chemical accuracy, with RMSEs below 0.5 kcal/mol/\AA{} for solute–solvent forces at the MM level and below 1 kcal/mol/\AA{} at the QM/MM level. When used in umbrella sampling, the trained MLP reproduced the 2D \gls*{fes} (Ramachandran plot) of alanine dipeptide with high fidelity, achieving RMSD less than 0.9 kcal/mol relative to explicit solvent references. The resulting map closely resembled those reported for solvent-free environments~\cite{maier2015ff14sb}. Notably, the workflow was efficient enough to be extended to QM-level simulations, with the ASEC-based implicit model contributing negligible overhead relative to QM solute calculations.

\subsubsection{Explicit solvent dynamics of the Menshutkin reaction via DeepPot-SE}

\textbf{S1:} Explicit solvation (bulk water with 106 H$_2$O molecules) \\
\textbf{M1:} NN-MLP (DeepPot-SE with on-the-fly probability-based enhanced sampling)

Karmakar {\it et al.}~\cite{anmol2024unveiling} developed a DeepPot-SE MLP for the Menshutkin reaction (a prototypical  S$_\text{N}$2 methyl transfer reaction) in explicit water using active learning and enhanced sampling to achieve DFT-level accuracy over nanosecond simulation timescales. Starting from AIMD metadynamics at multiple temperatures, the model was iteratively refined using DP-GEN and trained on over 500,000 configurations, achieving RMSE below 0.1 eV/\AA{} in force prediction.

WT-MetaD simulations using the trained  NN-MLP allowed reconstruction of converged \gls*{fes}, yielding a barrier of 29.7 kcal/mol and a reaction free energy of $-39 \pm 2$ kcal/mol, consistent with experimental trends and outperforming SMD-based implicit solvation. FES analysis revealed significant changes in solvation structure along the reaction coordinate, particularly enhanced stabilization of the charged products via hydrogen bonding. The study highlights how MLPs trained with explicit water environments can capture both the energetic and dynamic solvation effects that are missing in continuum models.

\subsubsection{Modeling  of$\pi$-hydrogen bonding in benzene–water and benzene–ammonia solutions}

\textbf{S1:} Explicit solvation (liquid water and ammonia) \\
\textbf{M1:} Descriptor-based NN-MLP (C-NNP ensemble of BPNN)

Brezina et al.~\cite{brezina2024elucidating} performed thermostatted ring polymer molecular dynamics (TRPMD) simulations using MLPs trained on ab initio data to study $\pi$-hydrogen bonding between benzene and solvent molecules (H$_2$O, NH$_3$). The authors developed a committee of BPNN MLPs (C-NNPs) trained on hybrid DFT (revPBE0-D3) AIMD trajectories to accurately reproduce the ab initio PES while reducing the computational cost. The study systematically analyzed solvation structures, bond lifetimes, orientational distributions, and vibrational signatures of $\pi$-hydrogen bonding in both solvents.

Spatial distribution functions revealed two prominent solvation caps where $\pi$-hydrogen bonds form, with comparable structures in both solvents. A time-resolved existence criterion was used to determine bond populations and lifetimes (1.7--1.8 ps). Vibrational spectroscopy (via TRPMD) captured blue-shifted OH stretching features in water, consistent with experimental Raman SC spectra. Importantly, water showed significant $\pi$-hydrogen bond anticooperativity between the two binding sites on benzene, while ammonia did not, due to weaker solvent–solvent H-bonding. Despite accurate reproduction of most properties, the C-NNP model underestimated the anticooperativity effect, highlighting limitations of local descriptors and standard training protocols. The authors suggest using MP-NN-MLPs architectures such as  NequIP with customized active learning to address such long-range effects.

\subsubsection{Water at solid–liquid and confined interfaces}

\textbf{S1:} Explicit Solvation (bulk water, water on surfaces, nanoconfinement) \\
\textbf{M1:} Descriptor-based NN-MLP (BPNN, \gls*{cnnps})

Schran et al.~\cite{schran2021machine} developed a data-driven and automated \gls*{ml} pipeline based on \gls*{cnnps} to enable efficient and accurate \gls*{md} simulations of complex aqueous systems. The approach involves generating BPNN-based models for specific thermodynamic state points using an active learning strategy (\gls*{qbc}) applied to short ab initio MD trajectories. Six systems were studied: 
(i) fluoride ion in water, 
(ii) sulfate ion in water, 
(iii) water in carbon nanotubes, 
(iv) water in hBN nanotubes, 
(v) water confined between MoS$_2$ sheets, and 
(vi) water on rutile TiO$_2$(110). 

The C-NNPs achieved high accuracy with compact training sets ($\sim$300 configurations) and were validated using an automated protocol analyzing RDFs, VDOS, and force RMSEs. For the water–TiO$_2$ interface, long (5 ns) simulations enabled insight into interfacial layering, diffusion anisotropy, and free energy profiles, revealing firmly structured and immobile water contact layers. The framework leverages symmetry functions and fixed hyperparameters, requiring minimal user tuning, and is implemented in the open-source packages. This study exemplifies a scalable workflow for solvation-aware MLP development and application.


\begin{table*}[htbp]
\centering
\caption{Summary of case studies in MLP-aided solvation modeling}
\label{tab:case_catalogue}
\renewcommand{\arraystretch}{1.1}
\begin{tabular}{lll lll}
\hline 
Study & S1 & S2 & S3 & MLP Type (M1--M4) \\
\hline 
\multicolumn{5}{l}{\textit{A1: PES-based applications}} \\ 
Interfacial water on solids~\cite{gading2024role} & \checkmark & & & M1 (BPNN) \\
Solvation forces via PB-DNN~\cite{liao2024calculation} & & \checkmark & & M1 (DeepNN, internal coords) \\
ORR on Au(100)–water~\cite{yang2023neural} & \checkmark & & & M1 (PaiNN) \\
Protonated water in He droplets~\cite{schran2018high} & & & \checkmark & M1 (BPNN) \\
Solvent-driven ring-opening~\cite{celerse2024capturing} & \checkmark & & & M1 (DeepPot-SE) \\
Adsorption energies/barriers~\cite{chen2023accelerating} & \checkmark & & & M3 (MTP vs SchNet, ANI) \\
Alanine dipeptide~\cite{yao2023machine} &  & \checkmark &  & M1 (DeepPot-SE) \\
Menshutkin reaction in water~\cite{anmol2024unveiling} & \checkmark & & & M1 (DeepPot-SE + LOTF) \\
Benzene $\pi$-H bonding~\cite{brezina2024elucidating} & \checkmark & & & M1 (C-NNP: ensemble of BPNN) \\
Water at Interfaces and Confinement~\cite{schran2021machine} & \checkmark & & & M1 (C-NNP: ensemble of BPNN) \\
\hline 
\multicolumn{5}{l}{\textit{A2: Direct property modeling}} \\ 
Aquamarine $\Delta$-ML~\cite{hilfiker2024leveraging} & & \checkmark & & M1 (MACE) \\
Redox potentials~\cite{jinnouchi2025absolute} & \checkmark & \checkmark & & M2 (GAP-SOAP) \\
Ion–ligand screening~\cite{wang2020active} & & \checkmark & & M1 (GNN) \\
Size-transferable mbGDML~\cite{maldonado2023modeling} & \checkmark & & & M4 (mbGDML vs. SchNet, GAP) \\
Redox/acidity free energies~\cite{wang2022automated} & & \checkmark & & M1 (DeepPot-SE) \\
FieldSchNet~\cite{gastegger2021machine} &  & \checkmark & \checkmark & M2 (Field-aware NN: FieldSchNet) \\
Solvent effects on reaction rates~\cite{chung2024machine} & & \checkmark & & M1 (CGR GCNN ) \\
Transferable GNN implicit solvent~\cite{katzberger2024general} & & \checkmark & & M1 (GNN correction to GB) \\
Hybrid solvation free energies~\cite{bonnet2024solvation} & & & \checkmark & M1 (NequIP GNN) \\
Ionic media and DNA~\cite{coste2023developing} & & \checkmark & & M1 (Allegro, $\Delta$-ML) \\
\hline 
\end{tabular}
\end{table*}

\vspace{0.5em}
\subsection{Direct modeling of solvation properties\label{sec:A2}}
\subsubsection{$\Delta$-ML modeling of solvated properties of the Aquamarine dataset}

\textbf{S2:} Implicit solvation ({MPB model of water) \\
\textbf{M1:} Neural network MLP (MPNN: MACE)

In \RRef{hilfiker2024leveraging}, 
MACE was applied to predict solvent-phase dipole moments and many-body dispersion energies (\( E_{\text{MBD}} \)) of drug-like molecules from the Aquamarine benchmark dataset, which includes 40 physicochemical properties for approximately 60{,}000 conformers of 1{,}653 molecules containing up to 92 atoms and composed of H, C, N, O, F, P, S, and Cl elements~\cite{medrano2024dataset}.

Conformational sampling was performed in both gas phase and implicit water using a combination of tools, including CREST, Maestro, Omega, and RDKit. Geometries of selected conformers were relaxed at the DFTB3+MBD level, while implicit solvation was modeled using the size-modified Poisson--Boltzmann (MPB) formalism. CREST employed the GFN2-xTB semiempirical method with a GBSA solvation model to generate accurate 3D structures in solution.

To improve prediction accuracy, a \(\Delta\)-ML approach was adopted, where MACE was trained to learn the difference between solvated and gas-phase property values, \(\Delta P = P_{\text{sol}} - P_{\text{gas}}\), instead of directly predicting \(P_{\text{sol}}\)~\cite{hilfiker2024leveraging}. This strategy yielded notable gains, especially for solvation-sensitive properties. For example, solvation was found to significantly affect computed Hirshfeld charges, particularly for the heavier atoms, P, S, and Cl. 

Accuracy assessments showed that the model performed well for small to medium-sized molecules. 
For systems with fewer than 20 atoms, RMSE in atomic positions was below \(0.1\,\text{\AA}\), even when 
solvent effects were included. In contrast, systems with more than 40 atoms exhibited deviations larger 
than \(2.0\,\text{\AA}\), highlighting the increasing complexity with conformational flexibility.

\subsubsection{ML-aided computation of absolute redox potentials using hybrid-DFT and ML-based FFs}

\textbf{S2:} Implicit/thermodynamic solvation via MD (explicit water with bulk-like sampling) \\
\textbf{M2:} Kernel-MLP ($\Delta$-ML with GAP-SOAP)

Jinnouchi et al.~\cite{jinnouchi2025absolute} developed an ML–aided first-principles framework for computing absolute standard hydrogen electrode potentials (ASHEP) and redox potentials of various atomic and molecular species, using thermodynamic integration (TI) and thermodynamic perturbation theory (TPT). Their approach combines hybrid DFT (PBE0+D3 and HSE06) with a kernel-based MLP, denoted in the study as machine-learned force fields (MLFF), enabling statistically robust sampling while reducing the computational cost of first-principles simulations. The framework, implemented within VASP, supports active learning and on-the-fly dataset refinement via uncertainty-driven sampling~\cite{jinnouchi2019fly}.

Simulations were performed with periodic explicit solvent models containing 64 water molecules. Importantly, including Grotthuss-type proton transfer events through MLFF sampling significantly accelerated convergence of free energy estimates. Proton insertion free energies were computed without artificial constraints, yielding an ASHEP of $-4.52 \pm 0.09$~V and a real proton potential of $-11.12 \pm 0.09$~eV, aligning experimental and benchmark theoretical values.

To further assess the method's generality, redox potentials for seven aqueous redox couples (e.g., Fe$^{3+}$/Fe$^{2+}$, Cu$^{2+}$/Cu$^{+}$, O$_2$/O$_2^-$) were computed, achieving an RMSE of 140~mV. However, some discrepancies were observed in hydrogen bonding features, particularly in the RDFs of neutral and reduced O$_2$ interacting with water. These were mitigated by coupling TI simulations driven by MLFFs with corrections from reference first-principles calculations. Upon uncertainty evaluations,
reference data were incorporated into the MLFF training set. This hybrid correction scheme enabled accurate modeling of proton solvation and redox processes across a range of oxidation states and molecular configurations.

\subsubsection{Screening of ether-based solvate ionic liquids}

\textbf{S2:} Implicit/thermodynamic solvation via ion–ligand configurational sampling \\
\textbf{M1:} NN-MLP (GNNs trained via active learning)

Wang et al.~\cite{wang2020active} developed an active learning workflow that couples NN-MLPs with DFT calculations to efficiently screen over 1{,}000 oligoether ligands for their potential use in forming solvate ionic liquids (SILs) with Li\textsuperscript{+}, Mg\textsuperscript{2+}, and Na\textsuperscript{+}. Initial ligand conformers were generated using RDKit and then optimized at the BP86-D3/def2-SVP level of theory. The NN-MLPs, trained on force and energy data from these DFT calculations, were iteratively improved via uncertainty-aware sampling of MD trajectories, allowing the model to improve its predictions with each generation progressively.

The resulting workflow enabled rapid exploration of ligand–ion configuration space and identified promising ligands with stronger binding affinities and improved oxidative stabilities compared to standard tri- and tetraglyme benchmarks. DFT refinements validated the NN-predicted ligand–ion structures, and Pareto front analysis highlighted the tradeoff between binding strength and chemical stability. 

For Mg\textsuperscript{2+}, ligands with extended alkyl spacers and favorable coordination environments were found to offer high binding affinities, though high melting points limit their utility. Na\textsuperscript{+} coordination showed smaller improvements due to inherently weaker ion–dipole interactions. Interestingly, the inclusion of implicit solvent effects revealed that solvation typically reduces binding energies relative to gas-phase DFT. However, the authors noted that continuum solvation models may fall short in capturing the subtle solvation environments characteristic of SILs, making it difficult to assess SIL-forming capability based on implicit solvent data alone reliably.

Overall, this study demonstrates the scope of ML-accelerated atomistic modeling for the rational design of ion-coordinating ligands, highlighting the importance of balancing coordination strength, solvation effects, and chemical stability in ligand screening.

\subsubsection{Size-transferable force fields via many-body GDML}

\textbf{S2:} Explicit solvation and liquid-phase modeling using truncated many-body expansions \\
\textbf{M4:} Force-only Kernel-MLP (GDML) compared with GAP and NN-MLP (SchNet)

Maldonado et al.~\cite{maldonado2023modeling} introduced a data-efficient framework for training GDML-based FFs using a many-body expansion (MBE) approach, termed mbGDML. Rather than training a global potential, mbGDML learns separate FFs for 1-, 2-, and 3-body interactions, each trained on  1{,}000 MP2/def2-TZVP data points. This modular strategy enables accurate modeling of molecular clusters and bulk phases for solvents such as water, acetonitrile, and methanol.

Unlike conventional GDML models, which are limited in system size due to their global nature, mbGDML achieves size transferability by summing low-order contributions. 
The model accurately handles systems with up to 20 solvent molecules, yielding MADs below 0.38 kcal/mol per monomer in energy and 0.06 kcal/(mol\,\AA) in forces. 
Comparison with other MBE-based potentials such as mbGAP and mbSchNet, as well as size-transferable GNN models such as NequIP, showed that mbGDML offers competitive, often superior, accuracy while using considerably less training data.

The authors also demonstrated that mbGDML reproduces bulk-phase RDFs in good agreement with experiment, showcasing its ability to capture key structural features of liquid solvents. While only 1- to 3-body interactions were included in the present study, the authors noted that higher-order terms (4-body and beyond) could be incorporated using electrostatic or general quantum embedding approximations. These approximations may offer a practical path forward for scaling to more complex condensed-phase systems without the prohibitive cost of explicit high-order reference calculations.

\subsubsection{ML-aided thermodynamics via DeepPot-SE}

\textbf{S1:} Explicit solvation via AIMD + ML acceleration \\
\textbf{M1:} NN-MLP (DeepPot-SE with iterative concurrent learning)

Wang and Cheng~\cite{wang2022automated} developed an automated workflow that combines free energy perturbation (FEP) and thermodynamic integration (TI) with concurrent learning of MLPs using the DeepPot-SE framework. The workflow iteratively trains separate MLPs for the initial and final states of redox or acid–base reactions, enabling accurate trajectories from \gls*{mlamd}. 

The method was applied to a range of aqueous species, including redox-active molecules such as hydroquinones. It achieved high chemical accuracy: redox potentials within $\sim$0.2 V and p$K_\text{a}$ values within one unit of experiment. Training required fewer than 2000 configurations per system, and the MLPs showed excellent transferability across alchemical coupling parameters.

To improve accuracy, the method accounted for systematic DFT energy errors (e.g., Hartree potential shifts) by combining PBE-D3 trajectories with single-point HSE06 evaluations. This enabled AIMD-level accuracy at a fraction of the cost and dramatically improved the convergence of thermodynamic integrals.

The MLaMD-based FEP-TI framework provides a scalable route for computing acidity constants, redox potentials, and solvation free energies in explicit solvent, and is suitable for high-throughput screening in complex electrochemical environments.

\subsubsection{FieldSchNet: Solvent effects on spectra and reactions}

\textbf{S2--S3:} Implicit and explicit solvation effects in spectroscopy and reactivity \\
\textbf{M2:} Field-aware equivariant NN (FieldSchNet)

Gastegger et al.~\cite{gastegger2021machine} introduced FieldSchNet, a field-aware NN-MLP designed to predict molecular energies and response properties, such as dipole moments, polarizabilities, and NMR shielding tensors, in the presence of external electric fields. The model accounts for solvent effects through a dual 
strategy, where implicit solvation is captured via a learned Onsager-like polarizable continuum, while explicit solvent interactions are handled using ML/MM electrostatic embedding.

FieldSchNet demonstrated reliable predictions of spectroscopic observables derived from IR, Raman, 
and NMR spectra for solvated ethanol and methanol. It also successfully modeled solvent-mediated reactivity, such as shifts in activation barriers in a Claisen rearrangement reaction. Notably, ML/MM simulations revealed that hydrogen bonding between the ether oxygen and surrounding water molecules played a key role in lowering the reaction barrier, as evidenced by a pronounced RDF peak at $2.0\,\text{\AA}$ corresponding to the transition state of hydrogen bond formation/breaking between solvent and solute. 

Beyond accuracy, the model delivered impressive computational efficiency with a speedup of nearly 30{,}000-fold compared to full quantum mechanical simulations. This performance enabled the exploration of large configuration spaces and facilitated inverse design tasks, such as identifying solvent environments that stabilize specific transition states. Together, these results highlight the promise of FieldSchNet for rational solvent engineering and fast, field-aware molecular modeling in complex environments.

\subsubsection{Solvent effects on reaction rates}

\textbf{S2:} Implicit solvation (COSMO-RS-derived data) \\
\textbf{M1:} \gls*{gcnn}

Chung and Green~\cite{chung2024machine} developed a \gls*{gcnn} model to predict solvent effects on activation energies and enthalpies, specifically, solvent-induced changes in activation free energy ($\Delta\Delta G^\ddagger_\mathrm{solv}$) and activation enthalpy ($\Delta\Delta H^\ddagger_\mathrm{solv}$). 
The model was trained on a large dataset of 28{,}000 reactions and 295 solvents using COSMO-RS–computed values as reference. 
The model requires only atom-mapped reaction SMILES and solvent SMILES as input, without the need for structural descriptors.

To improve predictive performance, the authors applied transfer learning, fine-tuning the model on a smaller dataset that included experimental activation barriers for 165 reaction–solvent pairs. The fine-tuned model achieved sub-kcal/mol accuracy and reliably estimated solvent-induced shifts in reaction rates, offering a practical alternative to costly quantum chemistry calculations for high-throughput solvent screening.

However, during fine-tuning, the model exhibited overfitting to the RP-solv features (solvation free energies and solvation enthalpies of reactants and products). While this did not significantly compromise generalizability, it points out the importance of regularization strategies when adapting to small experimental datasets.

\subsubsection{Transferable GNN-based implicit solvent model for organic molecules}

\textbf{S2:} Implicit solvation (ML correction to continuum model) \\
\textbf{M1:} Graph neural network (GNN-based correction to classical GB)

Katzberger and Riniker~\cite{katzberger2024general} developed a transferable GNN–based implicit solvent model for organic molecules in water. The GNN augments the classical continuum model, GB-Neck2 (a variant of the generalized Born model), by learning short-range corrections to solute–solvent mean forces, using a multi-task architecture that predicts per-atom scaling factors for polar and non-polar solvation components. The model was trained on a dataset of over 3 million force vectors derived from explicit-solvent MD simulations of 369,000 diverse organic molecules.

The GNN achieves up to 18-fold effective speedup over explicit solvent simulations (4.4$\times$ faster MD and 4$\times$  improved sampling) while matching reference free-energy profiles and intramolecular hydrogen-bond statistics. Prospective simulations reproduced conformational ensembles, torsional distributions, and solvation-driven energy barriers with high fidelity, outperforming standard implicit models. This study establishes a scalable and general-purpose ML framework for efficient solvation-aware MD in water.

\subsubsection{ML-accelerated hybrid solvation free energies for ions}

\textbf{S3:} Hybrid solvation (explicit inner shell, implicit outer dielectric) \\
\textbf{M1:} E(3)-equivariant GNN (NequIP-based ML force field)

Bonnet and Marzari~\cite{bonnet2024solvation} extended a hybrid solvation framework for computing ion solvation free energies by integrating \gls*{mlamd} with quasi-chemical theory (QCT). The model treats the first two solvation shells (18 water molecules) explicitly and replaces expensive first-principles MD with a NequIP-based GNN trained on DFT data using an active learning workflow. This enabled 200 ps simulations that converged solvation free energies within chemical accuracy (0.04 eV).

The \gls*{mlamd} model, trained using two-stage refinement and cross-validation with DFT energies and forces, was deployed in LAMMPS for large-scale MD. The hybrid solvation environment includes a spherical wall to contain explicit waters and applies analytical corrections for long-range dielectric effects using Born-type screening. Solvation energies for 10 alkali and alkaline-earth cations (Li$^+$ to Ba$^{2+}$) showed improved agreement with experimental values over earlier AIMD approaches, with mean relative errors reduced to 1.8\%. This study illustrates the viability of combining physically motivated hybrid solvation models with ML-based energy functions for efficient and accurate thermodynamic property prediction.

\subsubsection{Delta-learned equivariant GNN for ionic media with implicit solvation}

\textbf{S2:} Implicit solvation (water coarse-grained out) \\
\textbf{M1:} Allegro, $\Delta$-ML

Coste et al.~\cite{coste2023developing} developed a deep implicit solvation (DIS) model for ionic aqueous systems using a $\Delta$-learned equivariant NN architecture. In this model, sodium and chloride ions are modeled explicitly, while water is treated implicitly, allowing accurate many-body potentials of mean force (PMFs) to be learned efficiently across varying salt concentrations (0.15–2.0 mol L$^{-1}$). The ML model (Allegro) is trained using a force-matching loss to minimize the difference between all-atom and a prior potential, which models van der Waals and electrostatic interactions, enabling accurate structural reproduction even in sparse ionic regimes.

For systems involving DNA, a coarse-grained representation of the molecule is combined with explicit ion treatment. The DIS model captures context-dependent ion distributions, including near phosphate backbones and DNA grooves, and significantly improves over prior and classical CG potentials. Although the model lacks explicit long-range electrostatics beyond Wolf summation and DNA flexibility, it achieves good agreement with RDFs, ion residence times, and 3D density maps. Significant simulation speedups over the all-atom model (0.5 vs.\ 84 ns/day for salt; 0.1 vs.\ 1.5 ns/day for DNA) highlight the DIS model’s efficiency for large-scale biomolecular simulations in ionic media.


\vspace{1em}
\noindent
\textbf{Summary:}
Table~\ref{tab:case_catalogue} summarizes case studies in MLP-aided solvation modeling, categorized by application type (A1 vs.\ A2), solvation paradigm (S1--S3), and MLP architecture (M1--M5). Two broad trends emerge. First, A1-type simulations, which involve learning the PES for configurational sampling, are predominantly coupled with explicit solvation (S1), especially in bulk or interfacial systems. In contrast, A2-type applications that target direct prediction of solvation-dependent properties, such as redox potentials or free energies, frequently adopt implicit solvation models (S2), reflecting their focus on energetic observables rather than full configurational sampling.

Second, A1-style efforts are almost exclusively based on NN-MLPs (M1), typically leveraging end-to-end or equivariant architectures for learning PES and forces. By comparison, A2 tasks show greater architectural diversity, including kernel-based models (M2, M4) and hybrid $\Delta$-ML approaches that incorporate prior physical knowledge.

While microsolvated systems (S3) remain less common overall, they are almost entirely studied via A1-type simulations with force-based learning. A2-style property predictions for S3 remain rare, with notable exceptions such as FieldSchNet~\cite{gastegger2021machine} and the GNN-based hybrid solvation model~\cite{bonnet2024solvation}.

Surveying ongoing efforts based on modeling objective (PES vs.\ property prediction) and solvation representation (explicit, implicit, or hybrid) is crucial for guiding both method development and benchmarking. Unambiguous classification is particularly important. For example, A2 tasks involving forces or energies over microsolvated ensembles fall under A1.

\section{Open Challenges and Avenues\label{sec:challenge}}

Despite remarkable progress, solvation-aware MLPs face critical open challenges across physical fidelity, architecture scalability, and workflow deployment.

MLPs must balance locality-driven efficiency aspects with the nonlocal interactions physically inherent to solvation. While accurate for short-time dynamics, force-only models lack a consistent energy reference and thus fail to support ensemble-based thermodynamic observables or enhanced sampling methods~\cite{fu2022forces}. 
The lack of an absolute energy reference in force-only models impedes Boltzmann-weighted ensemble calculations, while non-conservative forces introduce path-dependent integration errors. Ensemble observables such as free energies, chemical potentials, and partition functions become ill-defined, preventing reliable thermodynamic predictions or enhanced sampling applications like metadynamics or umbrella sampling.
Even energy-conserving MLPs suffer from truncated descriptors, limited cutoff schemes, or force drift, especially under periodic boundary conditions. For instance, ANI-2x systematically overstabilizes intramolecular hydrogen bonding in solvated environments due to missing long-range polarization~\cite{morado2023does}.

Electrostatic screening, dielectric response, and long-range charge transfer remain difficult to capture with standard local models. Periodic systems and interfaces introduce additional complexities such as spatial heterogeneity, solvent structuring, and directional noncovalent forces that challenge conventional descriptors and training protocols. For example, the molecular arrangement at the bulk is altered when liquids meet an interface~\cite{hayes2015structure}, introducing orientation-dependent polarization, and long-range structural correlations that are difficult to capture with local MLP descriptors.
The TEA Challenge 2023~\cite{poltavsky2025crash} highlighted significant model degradation under such conditions, especially for weakly interacting or rotationally mobile atoms (e.g., Pb in MAPbI$_3$). Encoding physical priors such as many-body dispersion or nonlocal electrostatics is essential for accurate solvation-phase predictions. Equivariant architectures such as NequIP and MACE achieve high accuracy on training distributions, but often produce significant maximum errors (up to 88 kcal mol$^{-1}$ \AA$^{-1}$) in sparse or complex regions~\cite{poltavsky2025crash}. Atom-wise force errors can vary 5–6$\times$ within a single molecule, complicating \gls*{ff} reliability. Unlike kernel methods (e.g., sGDML, GAP), current DNNs lack intrinsic uncertainty estimates, limiting their use in active learning and error-aware simulations.

$\Delta$-ML frameworks enable efficient correction of low-level MLPs using higher-level quantum or experimental data~\cite{ramakrishnan2015big}. This improves accuracy in energetics and free energies without retraining from scratch. Notable examples include 
extending ANI-2x with dispersion- and polarization-aware corrections~\cite{morado2023does}, 
improving the BuRNN framework~\cite{wan2024construction}, and 
predicting CCSD(T)/CBS energies from HF-derived features via MOB-ML~\cite{jacobson2024machine}.
Domain adaptation via $\Delta$-ML also enables solvent transferability using 
minimal data~\cite{meng2023something}, suggesting a path toward universal solvation-aware potentials.

Multiscale and hybrid models offer promising solutions. For instance, BuRNN~\cite{wan2024construction} demonstrates how polarizable ML/MM hybrids can treat local accuracy and long-range dielectric response simultaneously. Fragmentation-based MP2 models with two-body truncation and electrostatic embedding extend solvation simulations beyond DFT accuracy and scale~\cite{liu2022toward}.

Recent efforts by Schran et al.~\cite{schran2021machine} embed experimental calibration into DFT-MD workflows, improving agreement with thermal observables. Expanding this paradigm to include spectroscopic and thermodynamic data (e.g., IR, NMR, solubility) in model training and validation will be crucial for 
predictive solvation modeling where MLP predictions are faithfully connected to experimental observations.

The practical deployment of MLPs is hampered by fragmented software ecosystems. Libraries for various NN-MLPs and kernel models use different APIs and formats, complicating integration. Unified toolkits, pretrained model repositories, and user-friendly interfaces, such as ColabFit~\cite{kulichenko2024data,vita2023colabfit}, can significantly lower adoption barriers. Consistent documentation, open benchmarks, and curated metadata for solvation systems are needed to support reproducibility and community development.

Local descriptor-based MLPs face the so-called ``short-blanket dilemma''~\cite{zhai2023short,zhai2024many,muniz2023neural}, i.e., improving the fit to short-range properties (e.g., RDFs) often worsens long-range physics (e.g., dielectric response). Models that incorporate explicit nonlocal terms~\cite{ko2021general,wu2024simple,ple2023force} offer a way forward, enabling solvation models to capture polarization, charge delocalization, and directional interactions with higher fidelity. 

The emerging “foundation” ML models that are pre-trained on large, diverse datasets and then fine-tuned on smaller, task-specific ones, may shape the next generation of transferable and data-efficient MLPs for solvation modeling. Similar efforts have already emerged as a prominent trend in the field of inorganic materials.~\cite{batatia2023foundation,lysogorskiy2025graph}.
Extending the foundation model paradigm to solvation modeling poses distinct challenges due to the dynamic, heterogeneous nature of liquid environments. Capturing solute-solvent interactions, long-range polarization, and composition-dependent effects will require large, chemically diverse liquid-phase datasets and careful fine-tuning strategies to maintain accuracy across solute-solvent combinations and thermodynamic conditions.

Addressing these challenges will require coordinated advances in model architecture, training data design, hybrid physical–ML modeling, and community infrastructure. By improving the physical expressiveness, transferability, and usability of solvation-aware MLPs, future models can enable reliable simulations across chemical domains ranging from microsolvation to bulk liquids and complex interfacial systems.

\section{Perspectives and Outlook\label{sec:outlook}}

Despite the growing maturity of MLP-based solvation modeling, several promising avenues remain underexplored. Four directions in particular could significantly broaden the impact of solvation-aware MLPs. Promising directions include: (i) integrating property prediction with hybrid solvation (A2+S3) for tractable yet accurate modeling, (ii) benchmarking MLPs across solvent-aware properties like IR spectra and solubilities, (iii) developing transferable models using diverse datasets and architectures (e.g., transformers, $\Delta$-learning), and (iv) applying solvation-aware MLPs in biomolecular modeling and photocatalysis to capture conformational and reactive solvent effects.

A notable gap in the current literature is the scarcity of workflows combining property-specific prediction tasks (A2) with hybrid solvation paradigms (S3); see Table~\ref{tab:case_catalogue}. This absence may reflect both conceptual and technical challenges, such as defining physically meaningful property labels across microsolvated clusters or generating consistent training data from cluster–continuum models. These challenges are especially evident for solvent-sensitive properties such as valence and core-level excitations, NMR chemical shifts, or IR vibrational frequencies, where solvation induces spectral shifts that are difficult to decompose into local and bulk contributions~\cite{tripathi2025impact}. Nevertheless, A2+S3 workflows represent a promising frontier. They can enable accurate, tractable predictions for systems where fully explicit solvation is prohibitive, yet pure continuum models oversimplify critical interactions. Future work using equivariant GNNs trained on curated microsolvation ensembles may bridge this gap and unlock new data-driven solvation strategies.

Benchmarking MLPs across solvent-aware observables, beyond energies and forces, remains a critical need. For example, the hydrated electron~\cite{herbert2017hydrated}, a one-electron but spatially diffuse species, poses stringent demands on electronic structure fidelity~\cite{gao2024enhanced}. Its sensitivity to dispersion, charge delocalization, and solvent shell structure makes it a strong candidate for testing coupled QM–ML frameworks. Analogous to Zuo et al.’s comprehensive benchmark of materials properties~\cite{zuo2020performance}, future efforts should extend solvation MLP evaluation to IR spectra, diffusion constants, solubilities, and p$K_\text{a}$ shifts.

Developing transferable, general-purpose MLPs for solvation is feasible only with chemically diverse datasets that span solute–solvent combinations and structural motifs. While small-molecule sets like QM9 offer a starting point, their solvation coverage is limited. Augmenting such datasets with 
varying numbers of explicit solvent molecules and ensuring solvent structure quality (e.g., via improved water models such as TIP4P/TIP5P) is essential. Moreover, capturing multipolar and anisotropic electrostatics via distributed charge model (DCM)~\cite{devereux2014novel} or polarizable multipole potentials, such as the inexpensive atomic multipole optimized energetics for biomolecular applications  (iAMOEBA)~\cite{wang2013systematic}, will be crucial for simulating dynamic environments. This addresses sampling sufficiency and solvent structuring questions, shedding light on bulk liquid structure---the neglected child, the ``Cinderella'' of condensed matter~\cite{tabor1991gases}---which, unlike gases and solids, remains far less well-characterized and understood.

The challenge of model transferability across solvents, temperatures, and compositions requires both architectural and data innovations. Transformer-based pretraining, generative models (e.g., diffusion autoencoders), and latent space alignment offer one route~\cite{zhang2025roadmap}. Elm’s ``clusteromics'' concept~\cite{elm2020toward} offers another route suggesting building structured databases of acid–base clusters and hybrid solvation structures, then unifying them through conservative MLPs and $\Delta$-ML corrections. This approach can scaffold general-purpose models for both microhydration and bulk solvation environments. 
The advent of a foundation model for solvation could greatly reduce our reliance on manual, system-specific force field parameterization, paving the way for automated and reliable prediction of solvent effects even for newly designed or unconventional molecules.

Solvent effects are critical in protein folding, ligand binding, and conformational energetics. Although AlphaFold~\cite{jumper2021highly,evans2021protein,abramson2024accurate} revolutionized protein structure prediction, explicit modeling of solvent-accessible surfaces and flexible side chains remains limited. MLPs incorporating solvation physics could help map protein free energy landscapes more accurately, especially in flexible, solvent-exposed regions. This offers a pathway toward solvent-sensitive peptide therapeutics, protein engineering, and {\it in silico} design of biopolymers.

Solvent-aware MLPs are poised to accelerate research in photocatalytic water treatment and solar fuel design~\cite{sun2022application}. Modeling visible-light-driven mechanisms involving radicals like hydroxyl radicals (${\cdot}$OH), superoxide ions (${\cdot}$O$_2^-$), and peroxide radicals (${\cdot}$OOH) requires both electronic and solvation accuracy. MLPs trained on explicit-solvent QM data for metal oxides (e.g., TiO$_2$, ZnO, WO$_3$) and metal-free photocatalysts (e.g., doped graphene) could enable high-throughput discovery of photocatalytically active sites under solvent-rich conditions~\cite{mohamadpour2024photocatalytic}.


\section*{Supplementary Material \label{sec:SI}}
Original data, references, and plotting scripts used for making Figure~\ref{fig:md17} are available at: \href{https://github.com/raghurama123/Rev-MLP4Sol}{https://github.com/raghurama123/Rev-MLP4Sol}.

\section*{Data Availability}
The data that support the findings of this study are within the article and its supplementary material. 

\section*{Acknowledgments}
RB and SR thank the support of Mahindra University, under Project No. IRP/2024/24.
SD and RR acknowledge the support of the Department of Atomic Energy, Government
of India, under Project Identification No.~RTI~4007.




\bibliography{literature}

\end{document}